\documentclass{article}
\usepackage{jheppub,esint,shuffle,psfrag}
 \usepackage[utf8]{inputenc}
\usepackage{xcolor}
\usepackage{varioref}
\usepackage{amsmath,amsfonts,amsthm,mathrsfs}
\usepackage{amssymb,amsthm}
\usepackage{enumerate}
\usepackage{float}
\usepackage{fancyvrb}
\usepackage{verbatim}
\usepackage{wrapfig}
\usepackage{appendix}
\usepackage{amstext}
\usepackage{amssymb}
\usepackage{braket}
\usepackage{graphicx}
\usepackage{color}
\usepackage{varioref}
\usepackage{multirow,graphics}
\usepackage{epstopdf}

\usepackage{bbm}
\usepackage{slashed}
\usepackage{relsize}

\numberwithin{equation}{section}

\usepackage{tikz}
\usetikzlibrary{plotmarks,calc,decorations, decorations.pathmorphing, patterns}
\usetikzlibrary{arrows}
\tikzstyle dynkin node=[very thick,shape=circle,draw,inner sep=0pt,minimum size=5mm]
\tikzstyle dynkin line=[very thick]
\tikzstyle inverse line=[gray,line width=1.46pt,line cap=round, dash pattern=on 0pt off 2\pgflinewidth]
\tikzstyle red phase=[red,decoration={snake,amplitude=0.1mm,segment length=1.6mm},decorate]
\tikzstyle blue phase=[blue,decoration={snake,amplitude=0.1mm,segment length=0.9mm},decorate]
\tikzstyle green phase=[green,decoration={snake,amplitude=0.1mm,segment length=0.9mm},decorate]
\tikzstyle brown phase=[brown,decoration={snake,amplitude=0.1mm,segment length=0.9mm},decorate]
\newcommand{\boundellipse}[3]
{(#1) ellipse (#2 and #3)
}
\usetikzlibrary{decorations.pathmorphing}
\tikzstyle arrow=[thick,rounded corners=18pt,-latex]
\tikzstyle box=[draw,rounded corners,outer sep=4pt]
\tikzstyle B node=[outer sep=0pt]
\tikzstyle Q node=[inner sep=1pt,outer sep=0pt]
\definecolor{purple_nice}{rgb}{0.4,0.2,0.7}
\definecolor{fuel_blue}{RGB}{42,162,185}

\def\<{\langle}
\def\>{\rangle}

\newcommand{\p}{\partial}

\newcommand{\beq}{\begin{equation}}
\newcommand{\eeq}{\end{equation}}
\newcommand{\bem}{\begin{multline}}
\newcommand{\eem}{\end{multline}}
\newcommand{\bea}{\begin{equation}\begin{aligned}}
\newcommand{\eea}{\end{aligned}\end{equation}}

\newcommand{\e}{\operatorname{e}}
\newcommand{\ii}{\operatorname{i}}
\newcommand{\Qm}{\mathbf{Q}}
\newcommand{\R}{\mathbb{R}}
\newcommand{\Sb}{\mathbb{S}}
\newcommand{\ub}{\mathbf{u}}
\newcommand{\V}{\mathbb{V}}
\newcommand{\tdel}{\tilde{\delta}}
\newcommand{\del}{\delta}

\makeatletter
\@ifundefined{usebibtex}{} {}
\makeatother

\def\Tr{\text{Tr}~}

\title{
\Large Mirror channel eigenvectors of the $d$-dimensional fishnets}
\author{Sergey Derkachov$^a$, Gwenaël Ferrando${}^{b,c}$, Enrico Olivucci$^d$}
\affiliation{a. St. Petersburg Department of the Steklov Mathematical Institute of Russian Academy of Sciences, Fontanka 27, 191023 St. Petersburg, Russia. \\ b. Laboratoire de Physique de l’Ecole Normale Superieure, CNRS, Universite PSL, Sorbonne Universites,
24 rue Lhomond, 75005 Paris, France \\c. Institut de Physique Theorique, Universite Paris-Saclay, CNRS, CEA Saclay, 91191 Gif-sur-Yvette, France
\\d. Perimeter Institute for Theoretical Physics, Waterloo, Ontario N2L2Y5, Canada.}
\begin{document}
\abstract{We present a basis of eigenvectors for the graph building operators acting along the mirror channel of planar fishnet Feynman integrals in $d$-dimensions.
The eigenvectors of a fishnet lattice of length $N$ depend on a set of $N$ quantum numbers $(u_k,l_k)$, each associated with the rapidity and bound-state index of a lattice excitation. Each excitation is a particle in $(1+1)$-dimensions with $O(d)$ internal symmetry, and the wave-functions are formally constructed with a set of creation/annihilation operators that satisfy the corresponding Zamolodchikovs-Faddeev algebra. These properties are proved via the representation, new to our knowledge, of the matrix elements of the fused R-matrix with $O(d)$ symmetry as integral operators on the functions of two spacetime points. The spectral decomposition of a fishnet integral we achieved can be applied to the computation of Basso-Dixon integrals in higher dimensions. }

\maketitle

\section{Introduction}

The fishnet integrals are a class of Feynman diagrams with square lattice topology \cite{Zamolodchikov1980a} of remarkable importance for massless quantum field theory and---especially---for theories with conformal symmetry. Diagrams of fishnet type describe the planar limit of correlators in the strongly-deformed $\mathcal{N}=4$ supersymmetric Yang--Mills theory introduced by V.Kazakov and O.Gürdo\u{g}an \cite{Gurdogan:2015csr}. Moreover, for minimal size of the square lattice (ladder integrals), they form the basis of functions needed for the bootstrap of four-point functions of $\frac{1}{2}$-BPS operators with specific R-symmetry polarisations in the undeformed theory \cite{Coronado_2019, Coronado_2020}. Furthermore, other classes of fishnet integrals---with different lattice topology---describe completely the correlators of other planar conformal field theories, for instance the $3D$ chiral theory obtained as a deformation of ABJM super-conformal theory \cite{Caetano:2016ydc}. Finally, we shall mention that specific fishnets describe the Landau singularity of massless scattering amplitudes at all-loops \cite{Prlina:2018ukf}.

The remarkable properties of the fishnets is the possibility to find algorithms for their computation at any loop order \cite{Basso:2017jwq}. The procedure relies on methods of quantum integrability that map the Feynman integral to (the integral kernel of) a diagonalisable operator---the transfer-matrix of an integrable XXX spin chain with conformal symmetry \cite{Chicherin:2012yn,Gromov:2017cja}. Also, for a square lattice without boundary conditions imposed, these integrals enjoy infinite-dimensional Yangian symmetry \cite{Chicherin:2017frs, Chicherin:2017cns}.

\begin{figure}[H]
\begin{center}
\includegraphics[scale=0.65]{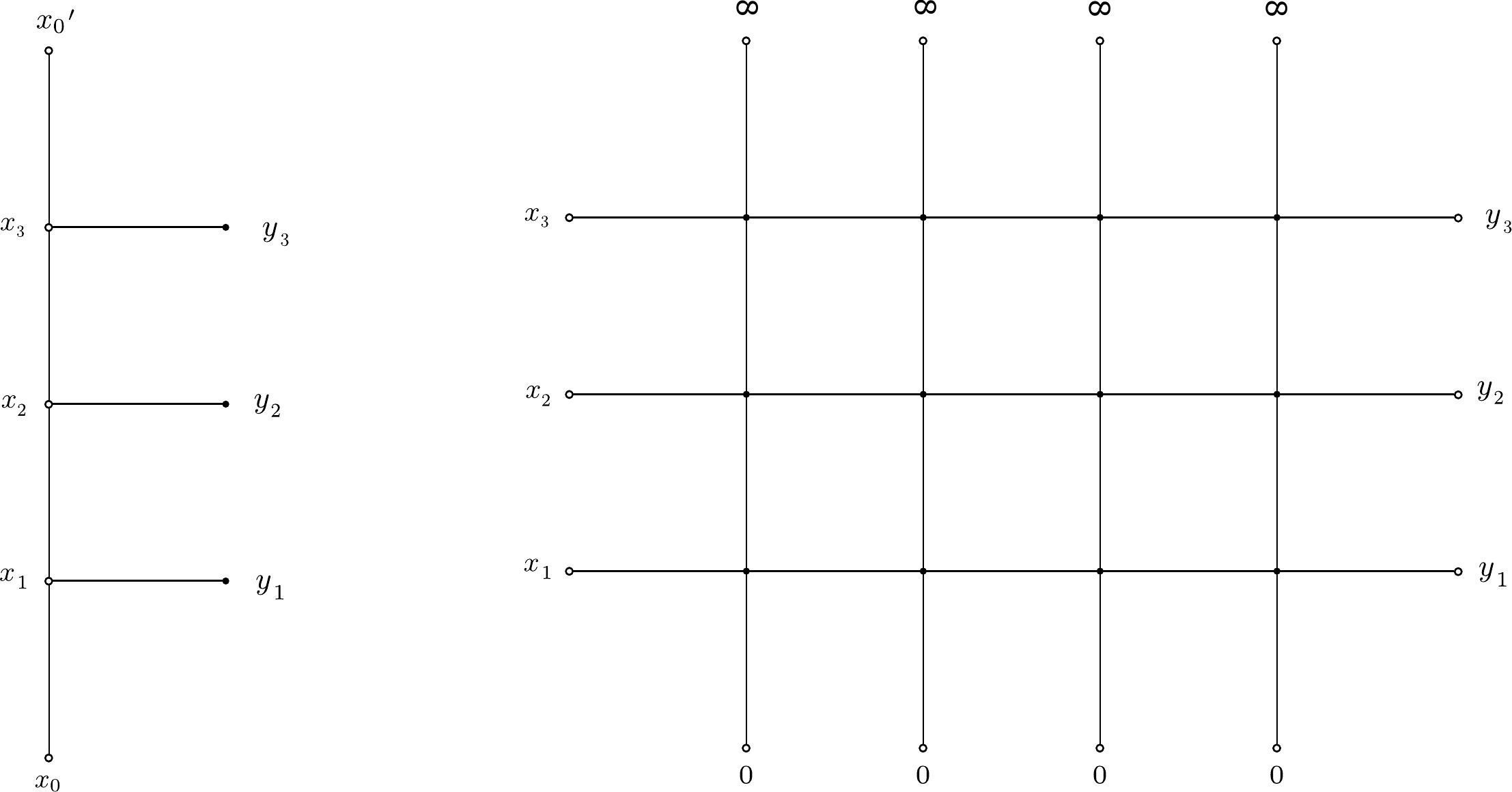}
\end{center}
\caption{\textbf{Left:} Graph-building operator for a fishnet integral of size $N=3$, with fixed boundaries $x_0$ and $x'_0$. Solid lines are bare propagators. \textbf{Right:} Fishnet integral of size $M\times N$ with $M=4$ and $N=3$. The boundaries are fixed to points $x_0=0$ and $x_0'=\infty$ (amputation of upper legs), for which the graph-builder is especially simple to diagonalise. The Basso--Dixon integral would correspond to a reduction $x_k=x$ and $y_k=y$. Black dots are integrated vertices and circles are external points.}
\label{fig:BDintro}
\end{figure}
A successful application of the spin-chain tools to fishnets is the computation of Basso--Dixon (BD) integrals in $d=2,4$ spacetime dimensions. BD integrals \cite{Basso:2017jwq} are specific reductions of a fishnet with open boundaries, which can be constructed using a so-called graph-building operator, see Fig.~\ref{fig:BDintro}. This graph-building operator is said to act in the ``mirror channel" of the fishnet lattice. B.~Basso and L.~Dixon originally obtained \cite{Basso:2017jwq} a nice explicit determinant formula for this general family of Feynman diagrams in $d=4$. Their derivation may be decomposed in two steps. First, the Feynman diagram is rewritten in some specific representation, then, the obtained expression is transformed to a determinant. The BD integrals in two dimensions were computed in \cite{Derkachov2019}, and in that case the transition to the determinant form is straightforward. In $d=4$ the analogue transformation is more complicated and was proven in \cite{Basso:2021omx}.

It was demonstrated in \cite{Derkachov2019,Derkachov:2019tzo,Derkachov:2020zvv} that the representation obtained in the first step is the so-called separated variables representation. Quantum separation of variables (SoV),  introduced by E.~Sklyanin \cite{Sklyanin:1984sb,Sklyanin:1991ss,Sklyanin:1995bm}, is one of the various techniques \cite{Faddeev:1996iy,Faddeev:1980zy,Kulish:1981bi,Sklyanin:1991ss} used to solve quantum integrable models.

Roughly speaking, SoV consists in finding a basis of the quantum space in which the spectral problem simplifies drastically. Thus, it can be understood as some far-reaching generalisation of the usual Fourier transform. From the point of view of quantum mechanics, the Fourier transform is a transition from coordinate representation to momentum representation. It is the simplest example of a canonical transformation, and the generalised eigenvectors of the momentum operator are used as the integral kernel of the Fourier transform. The freedom in using various unitary equivalent representations is typical of quantum mechanics, but there is a natural distinguished representation in the case of integrable systems -- Sklyanin's SoV representation.
The SoV basis is an eigenbasis for a particularly interesting family of commuting operators \cite{Sklyanin:1984sb,Sklyanin:1991ss,Sklyanin:1995bm}.
BD integrals are related to non-compact conformal spin chains \cite{Derkachov:2001yn,Chicherin:2012yn}, and it is indeed possible to define a commuting family of operators $\mathbf Q(u)$, which includes the graph-building operator.

For a long time, application of the SoV method was restricted to the models with
symmetry group of the lower rank \cite{Sklyanin:1991ss,Derkachov:2001yn,Derkachov:2002tf,
Bytsko:2006ut,Kharchev:2001rs}, or to the Toda chain
\cite{Sklyanin:1984sb,Kharchev:1999bh,Kharchev:2000yj,Kharchev:2000ug,Kozlowski:2014jka}.
In the last few years, however, much progress has been made for compact \cite{Gromov:2016itr,Maillet:2018bim,Ryan:2018fyo,Maillet:2018czd,Maillet:2019nsy,Gromov:2019wmz,Maillet:2020ykb,Gromov:2020fwh,Ryan:2020rfk} and non-compact \cite{Cavaglia:2019pow,Gromov:2020fwh,Cavaglia:2021mft} spin chains with higher-rank symmetry.

In the present paper, we generalise the first step in the computation of BD integrals to the case of general $d$, and we construct the corresponding $d$-dimensional SoV representation. Namely, we give an explicit description of a basis of eigenvectors of $\mathbf Q(u)$ and analyse their properties, thus revealing an underlying realisation of the Zamolodchikovs--Faddeev algebra \cite{Zamolodchikov:1978xm, Faddeev:1980zy} with respect to the exchange of their quantum numbers, i.e. the excitations of the lattice. The construction of the eigenvectors for an arbitrary number of sites extends, to any size $N$ of the fishnet, the two-site eigenvector presented in \cite{Basso:2019xay}. However, the investigation of the symmetry properties of the eigenvectors, the calculation of the corresponding inner product and of Sklyanin's measure are based on a new integral interchange relation. The main ingredient of this interchange relation is a particular $O(d)$-invariant R-matrix---more precisely, a solution of the Yang--Baxter equation acting in the tensor product of two arbitrary symmetric traceless representations of $O(d)$.

Applied to the computation of the $M\times N$ BD integral, our results allow us to rewrite it in the following way:
\begin{multline}
\pi^{\frac{Nd}{2}}\bra{x,\dots, x}\left(\prod_{i=1}^N \hat{x}_{i-1,i}^{2\del}\right)\mathbf{B}_{N,\tdel}^{M+1}\ket{y,\dots ,y} \\
= \sum_{\substack{0\leqslant l_1,\dots, l_N\leqslant +\infty \\ 1\leqslant m_i\leqslant d_{l_i}}} \int _{-\infty}^{+\infty}\cdots \int _{-\infty}^{+\infty} \bra{x,\dots ,x} \left(\prod_{i=1}^N \hat{x}_{i-1,i}^{2\del}\right) \ket{\ub_1,\dots ,\ub_N;C_{m_1,l_1}\otimes\dots\otimes C_{m_N,l_N}}\\
\times \langle{\ub_1,\dots ,\ub_N;C_{m_1,l_1}\otimes\dots\otimes C_{m_N,l_N}}\ket{y,\dots ,y} \frac{\mu(\mathbf{u}_1,\dots , \mathbf{u}_N)}{N!} \prod_{k=1}^N Q^{M+1}_{l_k}(u_k)\text{d} u_k\, ,
\end{multline}
where $\mathbf{u}_k = (u_k,l_k)$, and $\lbrace\ket{\ub_1,\dots ,\ub_N;C_{m_1,l_1}\otimes\dots\otimes C_{m_N,l_N}}\rbrace$ is a complete basis of eigenvectors of the graph-building operator $\mathbf{B}_{N,\tdel}$,
\beq
Q_l(u) = \frac{\Gamma(\del)\Gamma\left(\frac{d}{4}-\frac{\del}{2} + \frac{l}{2} - \ii u\right)\Gamma\left(\frac{d}{4}-\frac{\del}{2} + \frac{l}{2} + \ii u\right)}{\Gamma(\tdel)\Gamma\left(\frac{d}{4}+\frac{\del}{2} + \frac{l}{2} + \ii u\right)\Gamma\left(\frac{d}{4}+\frac{\del}{2} + \frac{l}{2} - \ii u\right)}\, ,
\eeq
and Sklyanin's measure is
\beq
{\mu(\mathbf{u}_1,\dots , \mathbf{u}_N) = \prod_{1\leqslant j< k\leqslant N} \left[(u_j-u_k)^2 + \frac{(l_j-l_k)^2}{4}\right]\left[(u_j-u_k)^2 + \frac{(d-2+l_j+l_k)^2}{4}\right]}\, .
\eeq
The explicit construction of the eigenvectors is described in Section \ref{sec:diagonalisation}. The next technical step to be performed is the simplification of $\langle{\ub_1,\dots ,\ub_N;C_{m_1,l_1}\otimes\dots\otimes C_{m_N,l_N}}\ket{y,\dots ,y}$. Though this was quite easy in two and four dimensions \cite{Derkachov2019,Derkachov:2019tzo,Derkachov:2020zvv}, we will show, on the simplest non-trivial example, i.e. the $N=2$ case, that the situation in generic dimension is much more involved.

The rest of the paper is organised as follows. In Section \ref{sec:O(D)Rmatrix}, we work out the properties of a solution $\mathbb{R}_{l_1,l_2}(u)$ of the Yang-Baxter equation acting on the tensor product of spaces of rank-$l_1$ and rank-$l_2$ symmetric traceless tensors. In particular, we shall present two new representations for the matrix elements of the $O(d)$-invariant R-matrix. The first one is obtained by a direct application of the fusion procedure \cite{Kulish:1981gi}, while the second one is an integral representation and is in fact equivalent to the main interchange relation. Along with that, the next section introduces the graphical Feynman diagram notations of lines and vertices that will be used to prove the most cumbersome identities throughout the paper. Section \ref{sec:diagonalisation} contains the explicit construction of the eigenvectors of the fishnet, which is done in an iterative manner as in the $d=2$ case \cite{Derkachov:2001yn,Derkachov:2014gya, Derkachov2019}, the determination of the spectrum, and the analysis of the eigenvectors' properties. An eigenvector for a lattice of size $N$ turns out to be described by a set of $N$ excitations, each characterised by a rapidity $u_k$ and a bound-state index or spin $l_k$, according to the analysis carried out in \cite{Basso:2019xay}. The rearrangement of the excitations inside an eigenvector and the overlap of eigenvectors reveal a picture of factorised scattering of excitations described by the S matrix $\mathbb{R}_{l_1,l_2}(u)$, up to a non-trivial phase. We work out, in Section \ref{sec:BD diagrams}, the simplest examples of application to Basso-Dixon integrals in any $d$.

Appendix \ref{app:integral relations} contains the basic integral identities used throughout the paper, while the proof of the integral representation for the R-matrix and of some related identities are relegated to Appendices \ref{app:main}, \ref{app:Symanzik}, and \ref{app:O}. Appendix \ref{app:spin_basis} presents an alternative basis of the eigenvectors which makes use of auxiliary spinors, in the spirit of the $d=4$ results of \cite{Derkachov:2019tzo, Derkachov:2020zvv}.

\section{$O(d)$-invariant R Matrices}
\label{sec:O(D)Rmatrix}

For $l\in\mathbb{N}$, we denote by $\V_l$ the (complex) vector space of symmetric traceless tensors of rank $l$, in dimension $d$. We shall denote its dimension by $d_l$. We will sometimes refer to $l$ as the spin.

This section contains an explicit construction of the R-matrices $\mathbb{R}_{l_1,l_2}$ acting in the tensor product $\V_{l_1}\otimes\V_{l_2}$ and satisfying the Yang--Baxter relation in $\V_{l_1}\otimes\V_{l_2}\otimes \V_{l_3}$ for arbitrary $l_1,l_2,l_3$. Our starting point will be the R-matrix by A.~Zamolodchikov and Al.~Zamolodchikov \cite{Zamolodchikov:1978xm}
\begin{equation}
\forall \,C\in\V_1\otimes\V_1,\quad\left[\R_{1,1}(u)C\right]^{\mu\nu} = \frac{1}{u+\ii}\left[u C^{\mu\nu} + \ii C^{\nu\mu} - \frac{\ii u}{u + \ii\frac{d-2}{2}} {C}{^{\rho}_{\rho}} \delta^{\mu\nu}\right]\, .
\end{equation}
In the first part of the section, we apply the fusion procedure \cite{Kulish:1981gi,Kulish:1981bi} to the construction of the general R-matrix $\mathbb{R}_{l_1,l_2}$. The fusion procedure was used for
the calculation of R-matrices $\R_{1,2},\R_{1,3}$ and $\R_{2,2}$ by N.~MacKay
\cite{MacKay:1990mp} and for $\R_{1,l}$ by N.~Reshetikhin \cite{Reshetikhin:1985eun,Reshetikhin:1985mhr}, we therefore generalise their results.

In the second part of the section, we shall prove an identity, which we call the interchange relation, and in which $\mathbb{R}_{l_1,l_2}$ plays the key role. This identity will be used extensively in the rest of the paper as it allows to prove symmetry properties of the eigenvectors of the transfer-matrix operators under the exchange of excitations. As a matter of fact, the interchange relation could be considered as the defining relation for the R-matrix $\mathbb{R}_{l_1,l_2}$, since it contains all the information about it. Starting from this identity, it is possible to derive an integral representation for the R-matrix, which allows to prove in a simple way its unitarity and the Yang--Baxter property. Vice versa, from the integral representation for the R-matrix it is possible to derive the interchange relation.

The equivalence of the two expressions for $\mathbb{R}_{l_1,l_2}$---the integral representation and the representation obtained directly by fusion procedure---is far from obvious. Our proof is very technical and we postpone it to Appendix \ref{app:main}. We should note that the spectral decomposition for the general R-matrix $\mathbb{R}_{l_1,l_2}$ was actually obtained thirty years ago by N.~MacKay \cite{MacKay:1990mp, MacKay:1991bj}. The latter result is in some sense complementary to both our expressions and we have checked their equivalence in the case of $\mathbb{R}_{1,l}$.

\subsection{Fusion procedure}

In this subsection we show that the R-matrix acting on $\V_{l_1}\otimes\V_{l_2}$ is defined by the following matrix elements
\begin{multline}
\label{Rll}
x^{\otimes l_1}\otimes y^{\otimes l_2} \cdot \left[\R_{l_1,l_2}(u)\zeta^{\otimes l_1}\otimes \eta^{\otimes l_2}\right] =  \frac{\left(\ii u + \frac{l_2 - l_1}{2}\right)_{l_1}}{\left(\ii u - \frac{l_1 + l_2}{2}\right)_{l_1}} \sum_{\substack{k\geqslant0\, ,n\geqslant 0 \\ k+n \leqslant \min(l_1,l_2)}} \frac{l_1! l_2!}{k! n! (l_1 - k - n)! (l_2 - k - n)!}\\
\times \frac{ \left(x\cdot y\, \zeta\cdot \eta\right)^k \left(x\cdot \eta\, y\cdot \zeta\right)^n}{\left(\ii u + \frac{4 - l_1 - l_2 - d}{2}\right)_k\left(-\ii u + \frac{2-l_1-l_2}{2}\right)_n} (x\cdot \zeta)^{l_1 - k - n} (y\cdot \eta)^{l_2 - k - n}\,,
\end{multline}
where all contractions, represented with a dot, of tensor indices are done using the Euclidean metric $\delta_{\mu\nu}$, and $x,y,\zeta,\eta$ are four null vectors in $\mathbb{C}^d$. One has for instance
\beq
\zeta^2 = \zeta\cdot\zeta = \zeta_{\mu}\zeta^\mu = \zeta^\mu\zeta^\nu\delta_{\mu\nu} = 0\, .
\eeq
We also use the Pochhammer symbol
\beq
(a)_l = \frac{\Gamma(a+l)}{\Gamma(a)} = \prod_{k=0}^{l-1} (a+k)\, .
\eeq

The proof of \eqref{Rll} is done in two steps. We first apply fusion to increase one of the spins, keeping the other equal to 1. In that case, the previous formula contains only three terms and reads
\begin{multline}\label{R1l}
x\otimes y^{\otimes l} \cdot \left[\R_{1,l}(u)\zeta\otimes\eta^{\otimes l}\right] = \frac{1}{u+\ii\frac{l+1}{2}}\Bigg[\left(u - \ii\frac{l-1}{2}\right) x\cdot\zeta (y\cdot\eta)^l \\
+ \ii l\, x\cdot\eta\, y\cdot\zeta (y\cdot\eta)^{l-1} - \ii l\frac{u - \ii\frac{l-1}{2}}{u+\ii\frac{d+l-3}{2}} x\cdot y\, \zeta\cdot\eta (y\cdot\eta)^{l-1}\Bigg]\, .
\end{multline}
Equivalently, we could have written, for $C\in\V_1\otimes\V_l$,
\begin{multline}\label{R1l explicit}
\left[\R_{1,l}(u)C\right]^{\mu\nu_1\cdots\nu_l} = \frac{1}{u+\ii\frac{l+1}{2}}\Bigg[\left(u - \ii\frac{l-1}{2}\right)C^{\mu\nu_1\cdots\nu_l} + \ii \sum_{j=1}^lC^{\nu_j\mu\nu_1\cdots\widehat{\nu_j}\cdots\nu_l}\\
- \ii\frac{u - \ii\frac{l-1}{2}}{u+\ii\frac{d+l-3}{2}}\sum_{j=1}^l\delta^{\mu\nu_j} {C}^{\rho}_{\rho} {}^{\nu_1\cdots\widehat{\nu_j}\cdots\nu_l} + \frac{1}{u+\ii\frac{d+l-3}{2}}\sum_{1\leqslant j<k\leqslant l} \delta^{\nu_j\nu_k} {C}^{\rho}_{\rho} {}^{\mu\nu_1\cdots\widehat{\nu_j}\cdots\widehat{\nu_k}\cdots\nu_l} \Bigg]\, .
\end{multline}


Before proving this last formula, we point out that, at the special point $u = \ii\frac{l+1}{2}$,
the matrix $\R_{1,l}(u)$ reduces to the orthogonal projector $\mathbb{P}_{1,l}^{(l+1)}$ onto $\V_{l+1}\subset\V_1\otimes\V_{l}$. This fact justifies why the fusion procedure gives new solutions of the Yang--Baxter relation. Its proof goes as follows: first, one notices from \eqref{R1l explicit} that  $\R_{1,l}\left(\ii\frac{l+1}{2}\right)C$ is symmetric traceless in all $l+1$ indices. After that, it is enough to remark that its contraction with any other symmetric traceless tensor $C'\in\V_{l+1}$ is given by $C'\cdot C$.

The proof is made by induction: the property \eqref{R1l explicit} clearly holds for $l=1$ so we assume that it holds for some $l\geqslant 1$. Let us show it for $l+1$, where the fusion procedure states that
\beq
\mathbb{P}_{1',l}^{(l+1)} \R_{1,1'}\left(u-\frac{\ii l}{2}\right) \R_{1,l}\left(u+\frac{\ii}{2}\right) \mathbb{P}_{1',l}^{(l+1)} = \R_{1,l+1}(u)\, .
\eeq
We remind that, due to the Yang--Baxter equation, the left projector could be removed. Consequently, applying the left-hand side to $C\in\V_{1}\otimes\V_{l+1}\subset\V_{1}\otimes\V'_1\otimes\V_{l}$ gives
\begin{multline}
\left[\R_{1,1'}\left(u-\frac{\ii l}{2}\right) \R_{1,l}\left(u+\frac{\ii}{2}\right) C\right]^{\mu\nu_1\cdots\nu_{l+1}} = \frac{1}{u+\ii\frac{2-l}{2}}\Bigg[\left(u-\frac{\ii l}{2}\right)\left[\R_{1,l}\left(u+\frac{\ii}{2}\right) C\right]^{\mu\nu_1\cdots\nu_{l+1}}\\
+ \ii \left[\R_{1,l}\left(u+\frac{\ii}{2}\right) C\right]^{\nu_1\mu\nu_2\cdots\nu_{l+1}} - \ii \frac{u-\frac{\ii l}{2}}{u + \ii\frac{d-2-l}{2}} {\left[\R_{1,l}\left(u+\frac{\ii}{2}\right) C\right]}{^{\rho\rho\nu_2\cdots\nu_{l+1}}}\delta^{\mu\nu_1}\Bigg]\, .
\end{multline}
We now use equation \eqref{R1l explicit} to write the second term in the right-hand side as
\begin{multline}
\left[\R_{1,l}\left(u+\frac{\ii}{2}\right) C\right]^{\nu_1\mu\nu_2\cdots\nu_{l+1}} = \frac{1}{u+\ii\frac{l+2}{2}}\Bigg[\left(u - \ii\frac{l-2}{2}\right)C^{\nu_1\mu\nu_2\cdots\nu_{l+1}} + \ii \sum_{j=2}^{l+1}C^{\nu_j\mu\nu_1\cdots\widehat{\nu_j}\cdots\nu_{l+1}}\\
- \ii\frac{u - \ii\frac{l-2}{2}}{u+\ii\frac{d+l-2}{2}}\sum_{j=2}^{l+1}\delta^{\nu_1\nu_j} {C}^{\rho\mu\rho\nu_2\cdots\widehat{\nu_j}\cdots\nu_{l+1}} + \frac{1}{u+\ii\frac{d+l-2}{2}}\sum_{2\leqslant j<k\leqslant l+1} \delta^{\nu_j\nu_k} {C}^{\rho\mu\rho\nu_1\cdots\widehat{\nu_j}\cdots\widehat{\nu_k}\cdots\nu_{l+1}} \Bigg]\, ,
\end{multline}
and, using the fact that $C$ is symmetric traceless in the last $l+1$ indices, the third term is
\begin{equation}
\left[\R_{1,l}\left(u+\frac{\ii}{2}\right) C\right]^{\rho\rho\nu_2\cdots\nu_{l+1}} = \frac{\left(u - \ii\frac{l-2}{2}\right)\left(u+\ii\frac{d-l-2}{2}\right)}{\left(u+\ii\frac{l+2}{2}\right)\left(u+\ii\frac{d+l-2}{2}\right)}C^{\rho\rho\nu_2\cdots\nu_{l+1}}\, .
\end{equation}
Putting everything together we straightforwardly recover \eqref{R1l explicit} for $\R_{1,l+1}(u)$.
\noindent
Turning our attention to the more general case, it suffices to prove \eqref{Rll} for $l_1\leqslant l_2$, which we shall do by induction on $l_1$ for given $l_2$. We have just verified it for $l_1=1$, and assuming it holds for some $l_1\leqslant l_2 - 1$, one just needs to use fusion to compute $\R_{l_1+1,l_2}(u)$:
\begin{multline}
x^{\otimes l_1+1}\otimes y^{\otimes l_2}\cdot\R_{l_1,l_2}\left(u+\frac{\ii}{2}\right) \R_{1,l_2}\left(u-\frac{\ii l_1}{2}\right) \zeta^{\otimes l_1+1}\otimes\eta^{\otimes l_2}\\
= x^{\otimes l_1+1}\otimes y^{\otimes l_2}\cdot\R_{l_1+1,l_2}(u) \zeta^{\otimes l_1+1}\otimes\eta^{\otimes l_2}\, .
\end{multline}
In the previous equation the product of the two R-matrices is taken in $\V_{l_2}$. In order to compute this product, one may insert a resolution of the identity of $\V_{l_2}$ between the two matrices. More explicitly, if $\{C_{j,l}\}_{1\leqslant j\leqslant d_l}$ is an orthonormal basis of $\V_{l}$ (for the inner product $(C,C') = C^*_{\mu_1\dots\mu_l}C'^{\mu_1\dots\mu_l} = C^*\cdot C'$), one can write
\begin{multline}
x^{\otimes l_1+1}\otimes y^{\otimes l_2}\cdot\R_{l_1,l_2}\left(a\right) \R_{1,l_2}\left(b\right) \zeta^{\otimes l_1+1}\otimes\eta^{\otimes l_2}\\
= \sum_{j=1}^{d_{l_2}} x^{\otimes l_1}\otimes y^{\otimes l_2}\cdot\left[\R_{l_1,l_2}(a)
\zeta^{\otimes l_1}\otimes C_{j,l_2}\right] x\otimes C^*_{j,l_2}\cdot\left[\R_{1,l_2}(b)
\zeta\otimes \eta^{\otimes l_2}\right]\, .
\end{multline}
\noindent
Thus, according to the formulas \eqref{R1l} and \eqref{Rll}, we can write
\begin{multline}\label{intermediate step Rll}
x^{\otimes l_1+1}\otimes y^{\otimes l_2}\cdot\R_{l_1,l_2}\left(u+\frac{\ii}{2}\right) \R_{1,l_2}\left(u-\frac{\ii l_1}{2}\right) \zeta^{\otimes l_1+1}\otimes\eta^{\otimes l_2}\\
= \frac{\left(\ii u + \frac{l_2 - l_1 - 1}{2}\right)_{l_1+1}}{\left(\ii u - \frac{l_1 + l_2 + 1}{2}\right)_{l_1+1}} \sum_{k+n \leqslant l_1} \frac{l_1! l_2!}{k! n! (l_1 - k - n)! (l_2 - k - n)!} \frac{ \left(x\cdot y\right)^k (y\cdot\zeta)^n (x\cdot\zeta)^{l_1 - k - n}}{\left(\ii u + \frac{3 - l_1 - l_2 - d}{2}\right)_k\left(-\ii u + \frac{3-l_1-l_2}{2}\right)_n} \\
\sum_{j=1}^{d_{l_2}} (\zeta^{\otimes k}\otimes x^{\otimes n}\otimes y^{\otimes (l_2-k-n)}\cdot C_{j,l_2})\Bigg[x\cdot\zeta\, (C_{j,l_2}^{*}\cdot\eta^{\otimes l_2}) + \frac{l x\cdot\eta\, (C_{j,l_2}^{*}\cdot\zeta\otimes\eta^{\otimes (l_2-1)}) }{-\ii u + \frac{1-l_1-l_2}{2}} \\
+ \frac{l \zeta\cdot\eta\, (C_{j,l_2}^{*}\cdot x\otimes\eta^{\otimes (l_2-1)}) }{\ii u + \frac{3 + l_1 - l_2 - d}{2}} \Bigg]\, .
\end{multline}
\noindent
The only additional formulas needed in order to conclude the proof are
\begin{equation}
\sum_{j=1}^{d_{l_2}} (\zeta^{\otimes k}\otimes x^{\otimes n}\otimes y^{\otimes (l_2-k-n)}\cdot C_{j,l_2})(C_{j,l_2}^{*}\cdot\eta^{\otimes l_2}) = (\eta\cdot\zeta)^k (\eta\cdot x)^n (\eta\cdot y)^{l_2-k-n}\,,
\end{equation}
and
\begin{multline}
\sum_{j=1}^{d_{l_2}} (\zeta^{\otimes k}\otimes x^{\otimes n}\otimes y^{\otimes (l_2-k-n)}\cdot C_{j,l_2})(C_{j,l_2}^{*}\cdot\theta\otimes\eta^{\otimes (l_2-1)})
= \frac{(\eta\cdot\zeta)^{k-1} (\eta\cdot x)^{n-1} (\eta\cdot y)^{l_2-k-n-1}}{l_2}\\ \times\Bigg[(l_2-k-n) (y\cdot\theta)(x\cdot\eta)(\zeta\cdot\eta) + n (x\cdot\theta)(y\cdot\eta)(\zeta\cdot\eta) + k (\zeta\cdot\theta)(y\cdot\eta)(x\cdot\eta) \\
-\frac{2\theta\cdot\eta}{d+2(l_2-2)}\left[(l_2-k-n)n\, (y\cdot x)(\zeta\cdot\eta) + (l_2-k-n)k (y\cdot\zeta)(x\cdot\eta) + kn (x\cdot\zeta)(y\cdot\eta)\right]\Bigg]\, ,
\end{multline}
where $\theta^2=0$, and we will apply it to $\theta\in\{\zeta,x\}$. The first one is trivial since $\eta_2^{\otimes l_2}\in\V_{l_2}$, and $\{C_{j,l_2}\}_{1\leqslant j\leqslant d_{l_2}}$ is an orthonormal basis of $\V_{l_2}$. The second one is a consequence of
\begin{multline}
\sum_{j=1}^{d_{l_2}} C_{j,l_2}^{\mu_1\cdots\mu_{l_2}} (C_{j,l_2}^{*}\cdot\theta\otimes\eta^{\otimes (l_2-1)}) = \frac{1}{l_2}\Bigg[\sum_{i=1}^{l_2}\theta^{\mu_i}\eta^{\mu_1}\dots
\widehat{\eta^{\mu_i}}\dots\eta^{\mu_{l_2}}\\
-\frac{2}{d+2(l_2-2)} \sum_{1\leqslant i<j\leqslant l_2}\theta\cdot\eta\, \delta^{\mu_i\mu_j} \eta^{\mu_1}\dots\widehat{\eta^{\mu_i}}\dots\widehat{\eta^{\mu_j}}\dots\eta^{\mu_{l_2}}\Bigg]\,,
\end{multline}
which is the orthogonal projection of $\theta\otimes \eta^{\otimes (l_2-1)}\in\V_1\otimes\V_{l_2-1}$ onto $\V_{l_2}\subset\V_1\otimes\V_{l_2-1}$, as explained at the beginning of this section (recall that this projector is nothing else than $\R_{1,l_2-1}\left(\ii\frac{l_2}{2}\right)$).

Using the two additional formulas, we can compute the sum over $d_{l_2}$ appearing in \eqref{intermediate step Rll}:
\begin{multline}
\label{F1}
\sum_{j=1}^{d_{l_2}} (\zeta^{\otimes k}\otimes x^{\otimes n}\otimes y^{\otimes (l_2-k-n)}\cdot C_{j,l_2})\Bigg[x\cdot\zeta\, (C_{j,l_2}^{*}\cdot\eta^{\otimes l_2}) \\
+ \frac{l\,x\cdot\eta\, (C_{j,l_2}^{*}\cdot\zeta\otimes\eta^{\otimes (l_2-1)}) }{-\ii u + \frac{1-l_1-l_2}{2}}  + \frac{l\,\zeta\cdot\eta\, (C_{j,l_2}^{*}\cdot x\otimes\eta^{\otimes (l_2-1)}) }{\ii u + \frac{3 + l_1 - l_2 - d}{2}} \Bigg]\\
= \frac{(\eta\cdot\zeta)^{k} (\eta\cdot x)^{n} (\eta\cdot y)^{l_2-k-n-1}}{\left(-\ii u + \frac{1-l_1-l_2}{2}\right)\left(\ii u + \frac{3 + l_1 - l_2 - d}{2}\right)}\Bigg[(x\cdot\zeta) (\eta\cdot y)\left(-\ii u + \frac{1-l_1-l_2}{2} + n\right)\\
\times\left(\ii u + \frac{3 + l_1 - l_2 - d}{2} + k\right) + (y\cdot\zeta) (\eta\cdot x)  (l_2-k-n) \left(\ii u + \frac{3 + l_1 - l_2 - d}{2} + k\right)\\
+ (y\cdot x) (\eta\cdot\zeta) (l_2-k-n) \left(-\ii u + \frac{1-l_1-l_2}{2} + n\right)\Bigg]\, .
\end{multline}
After plugging \eqref{F1} back into \eqref{intermediate step Rll}, we proceed to rewriting the sum $\sum_{k+n \leqslant l_1}$ into a sum $\sum_{k'+n' \leqslant l_1 + 1}$. The terms contributing to a given pair $(k',n')$ come from $(k,n)\in\lbrace(k',n'),(k'-1,n'),(k',n'-1)\rbrace$. When $(k,n) = (k',n')$, the contribution (without the tensors) is
\begin{equation}
\frac{l_1! l_2!}{k'! n'! (l_1 + 1 - k' - n')! (l_2 - k' - n')!} \frac{l_1+1-k'-n'}{\ii u + \frac{3 + l_1 - l_2 - d}{2}} \frac{\ii u + \frac{3 + l_1 - l_2 - d}{2} + k'}{\left(\ii u + \frac{3-l_1-l_2 - d}{2}\right)_{k'}\left(-\ii u + \frac{1-l_1-l_2}{2}\right)_{n'}} \, ,
\end{equation}
while when $(k,n) = (k'-1,n')$ it is
\begin{equation}
\frac{l_1! l_2!}{k'! n'! (l_1 + 1 - k' - n')! (l_2 - k' - n')!} \frac{k'}{\ii u + \frac{3 + l_1 - l_2 - d}{2}} \frac{1}{\left(\ii u + \frac{3-l_1-l_2 - d}{2}\right)_{k'-1}\left(-\ii u + \frac{1-l_1-l_2}{2}\right)_{n'}} \, ,
\end{equation}
and, when $(k,n) = (k',n'-1)$, it is
\begin{equation}
\frac{l_1! l_2!}{k'! n'! (l_1 + 1 - k' - n')! (l_2 - k' - n')!} \frac{n'}{\ii u + \frac{3 + l_1 - l_2 - d}{2}} \frac{\ii u + \frac{3+l_1-l_2-d}{2} + k'}{\left(\ii u + \frac{3-l_1-l_2-d}{2}\right)_{k'}\left(-\ii u + \frac{1-l_1-l_2}{2}\right)_{n'}} \, .
\end{equation}
The sum of the previous three terms is
\begin{equation}
\frac{(l_1+1)! l_2!}{k'! n'! (l_1 + 1 - k' - n')! (l_2 - k' - n')!} \frac{1}{\left(\ii u + \frac{3-l_1-l_2-d}{2}\right)_{k'}\left(-\ii u + \frac{1-l_1-l_2}{2}\right)_{n'}}\,,
\end{equation}
which concludes the proof of \eqref{Rll} for $(l_1+1,l_2)$.

\paragraph{Extension of \eqref{Rll} to Symmetric Tensors}
We now want to compute $x^{\otimes l_1}\otimes y^{\otimes l_2}\cdot \left[\R_{l_1,l_2}(u)\zeta^{\otimes l_1}\otimes\eta^{\otimes l_2}\right]$ when $\zeta^2 = \eta^2 = 0$, but $x^2\neq 0$ and $y^2\neq 0$. Since $\R_{l_1,l_2}(u)\zeta^{\otimes l_1}\otimes\eta^{\otimes l_2}$ belongs to $\V_{l_1}\otimes\V_{l_2}$, only the symmetric traceless parts of $x^{\otimes l_1}$ and $y^{\otimes l_2}$ are needed. Let us call $X_l$ the symmetric traceless part of $x^{\otimes l}$, it is given by
\begin{equation}
X_{l}^{\mu_1\cdots\mu_l} = \sum_{p=0}^{\lfloor\frac{l}{2}\rfloor} \frac{(x^2)^p}{\left(2-l-\frac{d}{2}\right)_p 2^p}\sum_{\{i_1,j_1\},\dots,\{i_p,j_p\}}\prod_{k=1}^p\delta^{\mu_{i_k}\mu_{j_k}}\prod_{i\notin\{ i_1,j_1,\dots,i_p,j_p\}}x^{\mu_i}\, ,
\end{equation}
where, for a given $p$, we sum over $\frac{l!}{(l-2p)! p! 2^p}$ possible ways of forming $p$ pairs among $l$ elements. We can thus write
\begin{equation}
x^{\otimes l_1}\otimes y^{\otimes l_2}\cdot \left[\R_{l_1,l_2}(u)\zeta^{\otimes l_1}\otimes\eta^{\otimes l_2}\right] = X_{l_1}\otimes Y_{l_2} \cdot \left[\R_{l_1,l_2}(u)\zeta^{\otimes l_1}\otimes\eta^{\otimes l_2}\right]\, ,
\end{equation}
and then use the formula \eqref{Rll}.

To start with, we consider only one vector that is not null : $\alpha^2 = 0$ but $y^2\neq 0$, so that we have
\begin{multline}
\alpha^{\otimes l_1}\otimes y^{\otimes l_2}\cdot \left[\R_{l_1,l_2}(u)\zeta^{\otimes l_1}\otimes\eta^{\otimes l_2}\right] = \frac{\left(\ii u + \frac{l_2 - l_1}{2}\right)_{l_1}}{\left(\ii u - \frac{l_1 + l_2}{2}\right)_{l_1}} \sum_{k,n\geqslant 0} \frac{(-l_1)_{k+n} (-l_2)_{k+n}}{k! n!}\\
\times \frac{(\zeta\cdot \eta)^k(\alpha\cdot\eta)^n(\alpha\cdot\zeta)^{l_2-k-n}}{\left(\ii u + \frac{4 - l_1 - l_2 - d}{2}\right)_k\left(-\ii u + \frac{2-l_1-l_2}{2}\right)_n}\left(Y_{l_2}\cdot \alpha^{\otimes k}\otimes \zeta^{\otimes n}\otimes \eta^{\otimes(l_2-k-n)}\right)\, .
\end{multline}
We then use the explicit expression for $Y_{l_2}$ to compute
\begin{multline}
Y_{l_2}\cdot \alpha^{\otimes k}\otimes \zeta^{\otimes n}\otimes \eta^{\otimes(l_2-k-n)} = \sum_{q=0}^{\lfloor\frac{l_2}{2}\rfloor}\sum_{a+b\leqslant q} \frac{(-k)_{a+b} (-n)_{q-b} (n+k-l_2)_{q-a}}{a! b! (q-a-b)!}\frac{(y^2)^q}{\left(2-l_2-\frac{d}{2}\right)_q 2^q}\\
(\alpha\cdot\zeta)^a (\alpha\cdot\eta)^b (\zeta\cdot\eta)^{q-a-b} (\alpha\cdot y)^{k-a-b} (y\cdot\zeta)^{n+b-q} (y\cdot\eta)^{l_2+a-k-n-q}\, ,
\end{multline}
which implies
\begin{multline}
\alpha^{\otimes l_1}\otimes y^{\otimes l_2}\cdot \left[\R_{l_1,l_2}(u)\zeta^{\otimes l_1}\otimes\eta^{\otimes l_2}\right]\\
= \frac{\left(\ii u + \frac{l_2 - l_1}{2}\right)_{l_1}}{\left(\ii u - \frac{l_1 + l_2}{2}\right)_{l_1}} \sum_{q=0}^{\lfloor\frac{l_2}{2}\rfloor} \sum_{\substack{K,N\geqslant q \\ a+b\leqslant q}} \frac{(-1)^{q+a}(-l_1)_{K+N+a-q} (-l_2)_{K+N}}{(K-q)! (N-q)! a! b! (q-a-b)!}\\
\frac{(\alpha\cdot\zeta)^{l_1+q-N-K} (\alpha\cdot\eta)^{N} (\zeta\cdot\eta)^{K} (y^2)^q (\alpha\cdot y)^{K-q} (y\cdot\zeta)^{N-q} (y\cdot\eta)^{l_2-N-K}}{2^q \left(2-l_2-\frac{d}{2}\right)_q \left(\ii u + \frac{4 - l_1 - l_2 - d}{2}\right)_{K+a+b-q}\left(-\ii u + \frac{2-l_1-l_2}{2}\right)_{N-b}}\, ,
\end{multline}
where we have changed summation indices from $k,n$ to $K = k+q-a-b$ and $N = n+b$. Recalling the Gauss identity
\begin{equation}\label{Gauss hypergeometric}
\sum_{k=0}^{n}\frac{(-n)_k(u)_k}{k! (v)_k} = \frac{(v-u)_n}{(v)_n}\, ,
\end{equation}
one can perform the sums over $a$ and $b$
\begin{multline}
\sum_{a+b\leqslant q} \frac{(-1)^{q+a}(-l_1)_{K+N+a-q}}{a! b! (q-a-b)!}\frac{1}{\left(\ii u + \frac{4 - l_1 - l_2 - d}{2}\right)_{K+a+b-q}\left(-\ii u + \frac{2-l_1-l_2}{2}\right)_{N-b}}\\
= \frac{1}{\left(\ii u + \frac{4 - l_1 - l_2 - d}{2}\right)_{K}\left(-\ii u + \frac{2-l_1-l_2}{2}\right)_{N}}\sum_{a=0}^q \frac{(-l_1)_{K+N+a-q} \left(\frac{d}{2} + l_1 + l_2 - K - N - 1\right)_{q-a}}{a! (q-a)!}\\
= \frac{(-l_1)_{K+N-q}}{q!}\frac{(-1)^q \left(2-l_2-\frac{d}{2}\right)_q}{\left(\ii u + \frac{4 - l_1 - l_2 - d}{2}\right)_{K}\left(-\ii u + \frac{2-l_1-l_2}{2}\right)_{N}}\, ,
\end{multline}
and eventually get
\begin{multline}
\alpha^{\otimes l_1}\otimes y^{\otimes l_2}\cdot \left[\R_{l_1,l_2}(u)\zeta^{\otimes l_1}\otimes\eta^{\otimes l_2}\right] = \frac{\left(\ii u + \frac{l_2 - l_1}{2}\right)_{l_1}}{\left(\ii u - \frac{l_1 + l_2}{2}\right)_{l_1}} \sum_{q=0}^{\lfloor\frac{l_2}{2}\rfloor}\sum_{K,N\geqslant q} \frac{(-l_1)_{K+N-q} (-l_2)_{K+N} (-N)_q}{q! (K-q)! N!}\\
\frac{(y^2)^q (\alpha\cdot\zeta)^{l_1+q-N-K} (\alpha\cdot\eta)^{N} (\zeta\cdot\eta)^{K} (\alpha\cdot y)^{K-q} (y\cdot\zeta)^{N-q} (y\cdot\eta)^{l_2-N-K}}{2^q \left(\ii u + \frac{4 - l_1 - l_2 - d}{2}\right)_{K}\left(-\ii u + \frac{2-l_1-l_2}{2}\right)_{N}}\, .
\end{multline}

\begin{figure}[H]
\begin{center}
\includegraphics[scale=1.1]{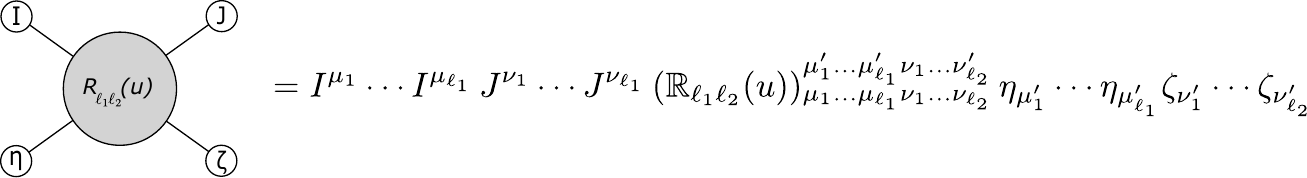}
\end{center}
\caption{Graphical representation of a matrix element of the $O(d)$-invariant $\mathbb{R}$-matrix $\R_{l_1,l_2}(u)$.}
\label{finite_dim_R}
\end{figure}

The same procedure allows to compute $x^{\otimes l_1}\otimes y^{\otimes l_2}\cdot \left[\R_{l_1,l_2}(u)\zeta^{\otimes l_1}\otimes\eta^{\otimes l_2}\right]$. In this case we start from the expression
\begin{multline}
X_{l_1}\cdot y^{\otimes (K-q)}\otimes \zeta^{\otimes (l_1+q-N-K)}\otimes \eta^{\otimes N}\\
= \sum_{p=0}^{\lfloor\frac{l_1}{2}\rfloor}\sum_{a+b+c\leqslant p} \frac{(-K+q)_{2p - a - b - 2c} (N+K-l_1-q)_{b+c} (-N)_{a+c} }{2^{p-a-b-c} a! b! c! (p-a-b-c)!} \frac{(x^2)^p (y^2)^{p-a-b-c}}{\left(2-l_1-\frac{d}{2}\right)_p 2^p}\\
\times (y\cdot\eta)^a (y\cdot\zeta)^b (\zeta\cdot\eta)^{c} (x\cdot y)^{K-q+2c+a+b-2p} (x\cdot\zeta)^{l_1+q-N-K-b-c} (x\cdot\eta)^{N-a-c}\, ,
\end{multline}
and, after the change of summation indices $q' = q+p-a-b-c$, $k=K+c$, $n=N+p-a-c$, the sums over $a$, $b$, and $c$ can be performed via the repeated application of \eqref{Gauss hypergeometric}. One eventually obtains

\begin{multline}\label{Rll general}
x^{\otimes l_1}\otimes y^{\otimes l_2}\cdot \left[\R_{l_1,l_2}(u)\zeta^{\otimes l_1}\otimes\eta^{\otimes l_2}\right]\\
= \frac{\left(\ii u + \frac{l_2 - l_1}{2}\right)_{l_1}}{\left(\ii u - \frac{l_1 + l_2}{2}\right)_{l_1}} \sum_{p=0}^{\lfloor\frac{l_1}{2}\rfloor} \sum_{q=0}^{\lfloor\frac{l_2}{2}\rfloor} \sum_{\substack{k\geqslant p+q\\ n\geqslant 0}} \frac{(-l_1)_{k+n-q} (-l_2)_{k+n-p} (-n)_p (-n)_q}{p! q! (k-p-q)! n! }\\
\times\left(\frac{x^2}{2}\right)^p
\left(\frac{y^2}{2}\right)^q\frac{(x\cdot\zeta)^{l_1+q-n-k} (x\cdot\eta)^{n-p} (\zeta\cdot\eta)^{k} (x\cdot y)^{k-p-q} (y\cdot\zeta)^{n-q} (y\cdot\eta)^{l_2+p-n-k}}{\left(\ii u + \frac{4 - l_1 - l_2 - d}{2}\right)_{k}\left(-\ii u + \frac{2-l_1-l_2}{2}\right)_{n}}\, .
\end{multline}
In what follows, we shall use the graphical representation of the R-matrix shown in Fig.\ref{finite_dim_R}.

\subsection{Spectral Decomposition}

The spectral decomposition of the R-matrix was computed by N.~MacKay
\cite{MacKay:1990mp,MacKay:1991bj}. Since it is clear from our expression \eqref{Rll} that the completely symmetric traceless tensors are eigenvectors with eigenvalue $1$, the normalisation is fixed and MacKay's result reads
\beq
\R_{l_1,l_2}(u) = \sum_{0\leqslant m\leqslant n\leqslant \min(l_1,l_2)} \prod_{p=1}^m \frac{u-\ii\frac{d+l_1+l_2-2-2q}{2}}{u+\ii\frac{d+l_1+l_2-2-2q}{2}} \prod_{q=1}^n \frac{u-\ii\frac{l_1+l_2+2-2q}{2}}{u+\ii\frac{l_1+l_2+2-2q}{2}} \mathbb{P}_{l_1,l_2}^{(l_1+l_2-2n,n-m)}\, ,
\eeq
where $\mathbb{P}_{l_1,l_2}^{(n_1,n_2)}$ is the projector onto the subrepresentation with highest weight $n_1\omega_1 + n_2\omega_2$ of $\V_{l_1}\otimes\V_{l_2}$, the $\omega_a$'s being fundamental weights ($\V_{l}$ has highest weight $l\omega_1$).
When one of the spins is equal to one, the previous decomposition reads
\beq\label{spectral R1l}
\R_{1,l}(u) =  \mathbb{P}_{1,l}^{(l+1,0)} + \frac{u-\ii\frac{l + 1}{2}}{u+\ii\frac{l + 1}{2}} \mathbb{P}_{1,l}^{(l-1,1)}+  \frac{u-\ii\frac{d+l-3}{2}}{u+\ii\frac{d+l-3}{2}} \frac{u-\ii\frac{l + 1}{2}}{u+\ii\frac{l + 1}{2}} \mathbb{P}_{1,l}^{(l-1,0)}\, .
\eeq
Let us check that this coincides with the expression \eqref{R1l explicit} for the R-matrix. We first introduce some operators $\mathcal{P}$, $\mathcal{K}_1$, and $\mathcal{K}_2$, in terms of which the R-matrix reads
\beq\label{R1l explicit bis}
\R_{1,l}(u) = \frac{1}{u+\ii\frac{l+1}{2}}\left[\left(u - \ii\frac{l-1}{2}\right)\mathrm{Id} + \ii \mathcal{P} - \ii\frac{u - \ii\frac{l-1}{2}}{u+\ii\frac{d+l-3}{2}}\mathcal{K}_1+ \frac{1}{u+\ii\frac{d+l-3}{2}}\mathcal{K}_2 \right]\, .
\eeq
We have already explained that $\mathbb{P}_{1,l}^{(l+1,0)} = \R_{1,l}\left(\ii\frac{l+1}{2}\right)$, and in terms of the new operators this reads
\beq
\mathbb{P}_{1,l}^{(l+1,0)} = \frac{1}{l+1} \left[ \mathrm{Id} + \mathcal{P} -\frac{2}{d+2l-2}\left(\mathcal{K}_1+\mathcal{K}_2\right)\right]\, .
\eeq
We claim that the other two projectors are given by
\beq
\mathbb{P}_{1,l}^{(l-1,1)}= \frac{1}{l+1} \left[ l\,\mathrm{Id} - \mathcal{P} +\frac{1}{d+l-3}\left(2\mathcal{K}_2 - (l-1)\mathcal{K}_1\right)\right]
\eeq
and
\beq
\mathbb{P}_{1,l}^{(l-1,0)} = \frac{1}{(d+2l-2)(d+l-3)}\left[(d+2l-4)\mathcal{K}_1 - 2\mathcal{K}_2\right]\, .
\eeq
It is clear that $\mathbb{P}_{1,l}^{(l+1,0)} + \mathbb{P}_{1,l}^{(l-1,1)} + \mathbb{P}_{1,l}^{(l-1,0)} = \mathrm{Id}$, and that \eqref{spectral R1l} is equivalent to \eqref{R1l explicit bis}. It remains to check that they are indeed orthogonal projectors, which leads to a tedious but straightforward computation that we do not show here.

\subsection{Interchange relation and integral representation}

\begin{figure}[H]
\begin{center}
\includegraphics[scale=1.0]{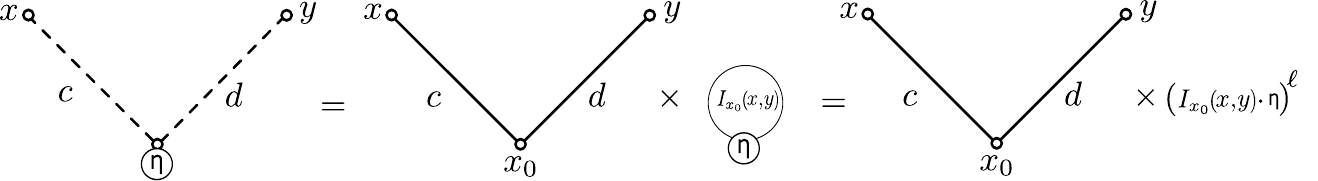}
\end{center}
\caption{Solid lines stand for the usual Feynman diagram notation, that is the square distance of the type $(x-y)^2$ between the extremal points $x$ and $y$, to some power which is written adjacent to the line.  Dashed line notation is illustrated in terms of product of solid lines and a tensor structure. The tensor $I_{x_0}^{\mu}(x,y)=\frac{x^{\mu}-x_0^{\mu}}{(x-x_0)^2}-\frac{y^{\mu}-x_0^{\mu}}{(y-x_0)^2}$ is obtained by a conformal inversion around the point $x_0$ of the vector $x^{\mu}-y^{\mu}$.}
\label{fig:tensor_notations}
\end{figure}

\begin{figure}[H]
\begin{center}
\includegraphics[scale=1.0]{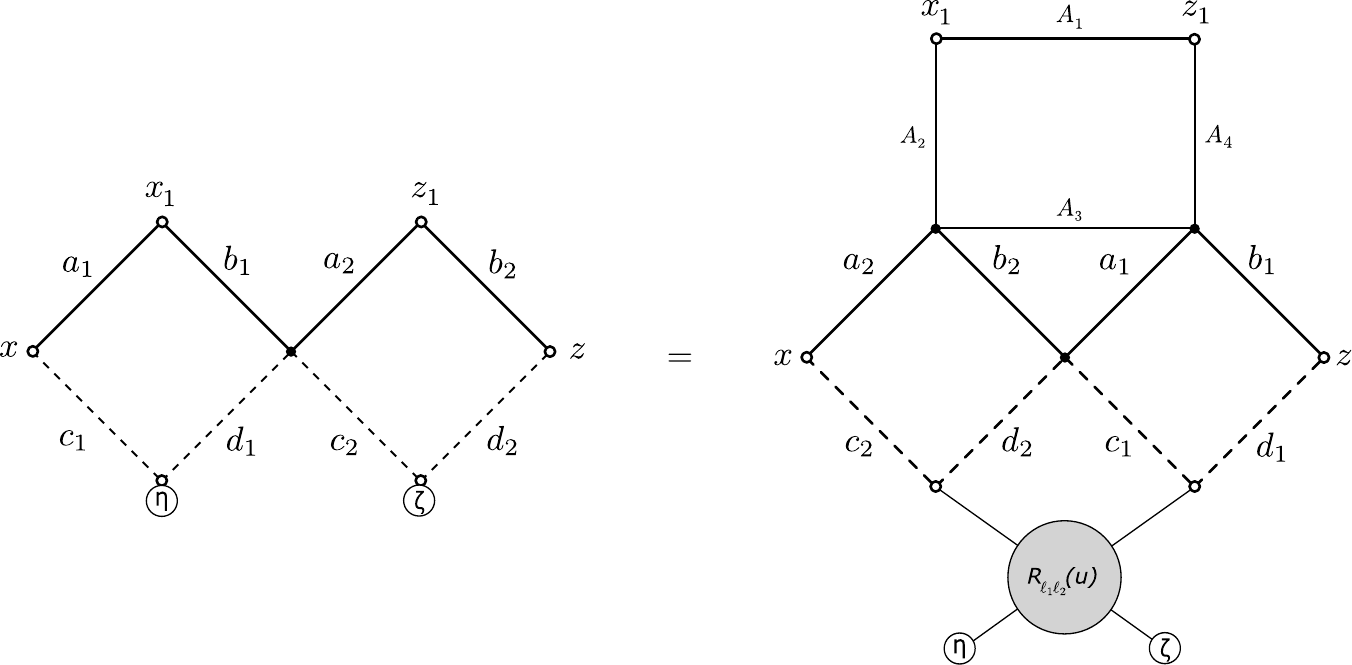}
\end{center}
\caption{Graphical representation of the interchange relation. Black dots are integration points of the diagram, circles are external points. The powers $a_j,b_j,c_j,d_j$, for $j=1,2$, of the solid lines get exchanged between both sides of the identity. }
\label{fig:interchange relation}
\end{figure}

In this section we shall consider the main interchange relation drawn in Fig.\ref{fig:interchange relation} according to the graphical notation of Fig.\ref{fig:tensor_notations}.

\noindent
The interchange relation is equivalent to the explicit integral representation for the
matrix element of the operator $\mathbb{R}_{l_1,l_2}$ and we shall prove
in Appendix \ref{app:main} the equivalence of this integral expression and the expression
\eqref{Rll general}. Here and in the rest of the paper we will use the notation $\tilde{a} \equiv d/2-a$, and we define a few standard functions of $u$:
 \beq\label{alpha beta}
\alpha(u) = \frac{\tdel}{2} - \ii u\, ,\quad\beta(u) = \frac{\tdel}{2} + \ii u\, ,\quad \text{so that}\quad \alpha(u) + \beta(u) + \del = \frac{d}{2}\,,\,\,\, \delta \in i\mathbb{R}\,.
\eeq
We define the powers of solid lines of the two squares in the left-hand side of Fig.\ref{fig:interchange relation} to be
\begin{equation}
a_j = \tilde{\beta}_j-\frac{l_j}{2}\,,\,\, b_j ={\alpha}_j+\frac{d+l_j}{2}-1 \,,\,\, c_j  = {\tilde \alpha}_j +\frac{l_j}{2}\,,\,\, d_j= 1- \tilde{\beta}_j-\frac{l_j}{2}\,,
\end{equation}
and the powers of the square kernel in the right-hand side of Fig.\ref{fig:interchange relation} are
\begin{align}
\begin{aligned}
&A_1=\frac{d}{2}-1 +\tilde{\beta}_1-\tilde{\beta}_2 +\frac{l_1+l_2}{2} \,,\,\, A_2=\frac{d}{2} +\tilde{\beta}_2-\tilde{\beta}_1 +\frac{l_1-l_2}{2} \,,\\&A_3=1-\frac{d}{2}+\tilde{\beta}_1-\tilde{\beta}_2 -\frac{l_1+l_2}{2} \,,\,\, A_4=\frac{d}{2}+\tilde{\beta}_2-\tilde{\beta}_1 +\frac{l_2+l_1}{2} \,.
\end{aligned}
\end{align}

\begin{figure}[H]
\begin{center}
\includegraphics[scale=1.0]{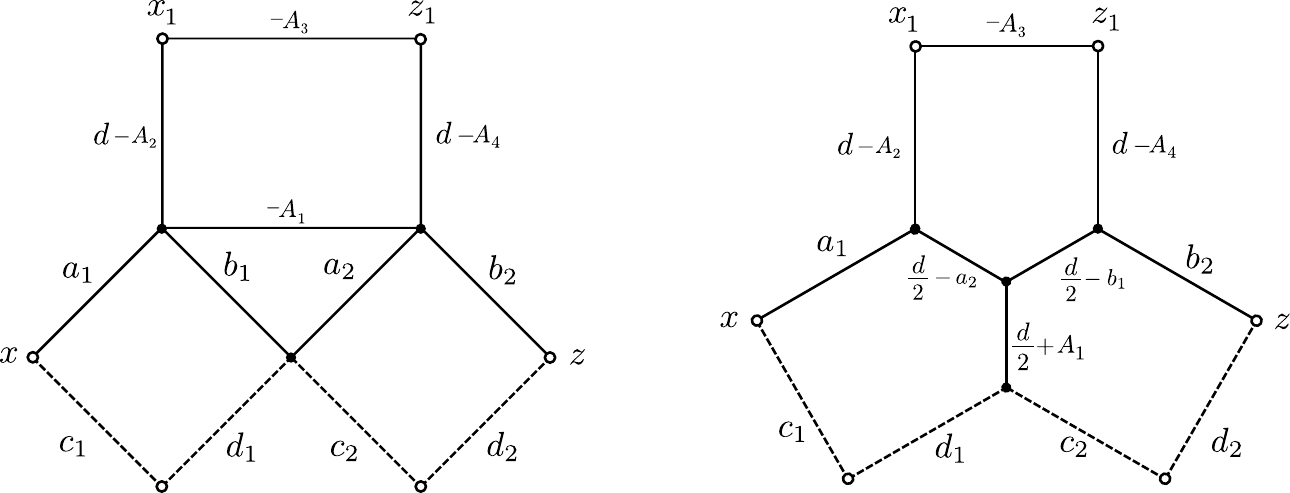}\vspace{15px}
\includegraphics[scale=1.0]{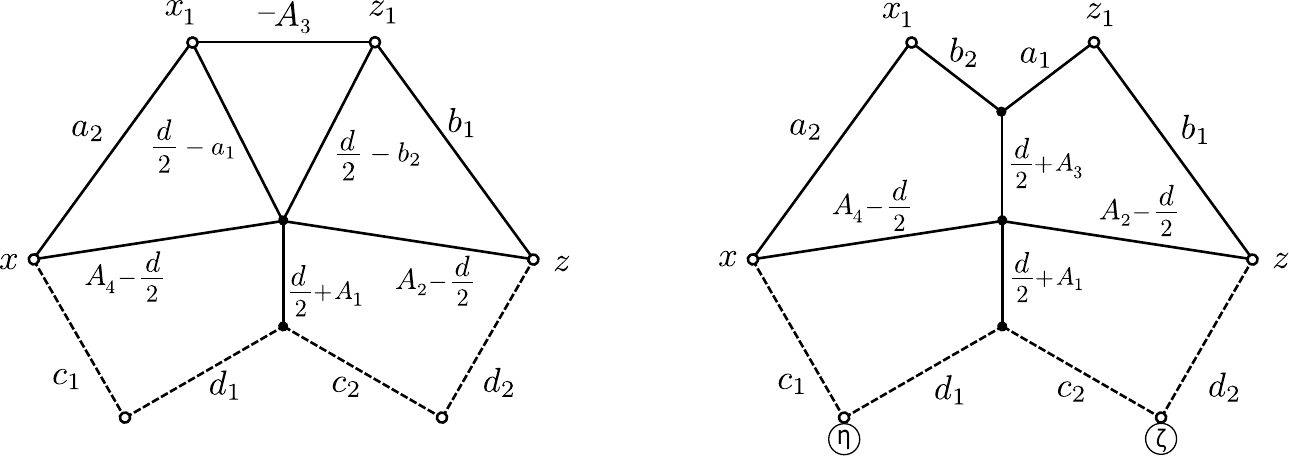}\vspace{15px}
\includegraphics[scale=0.88]{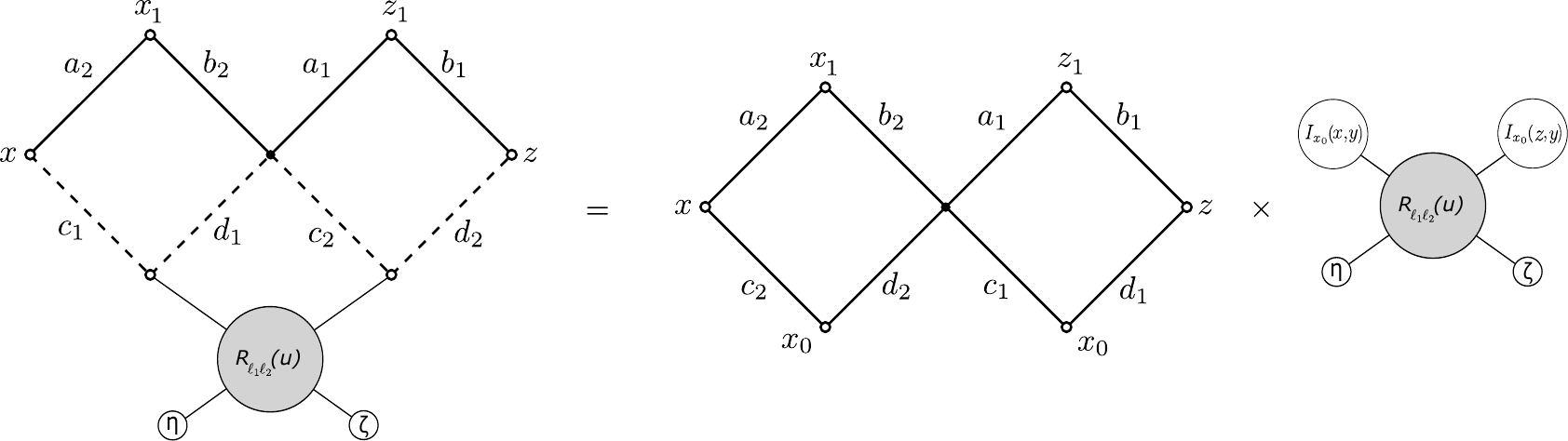}
\end{center}
\caption{\textbf{Up left:} l.h.s. of the interchange relation after multiplication by the inverse of the square kernel both in the r.h.s. and in the l.h.s. \textbf{Up right:} the central triangle $(-A_1,a_2,b_1)$ is transformed into a star integral. \textbf{Middle left:} result of star-triangle identity in the vertices $(a_1,d-A_2,d/2-a_2)$ and $(b_2,d-A_4,d/2-a_1)$. \textbf{Middle right:} the triangle $(-A_3,d/2-a_1,d/2-b_2)$ is transformed into a star integral. \textbf{Down:} result of the integration of the vertex $(d_1,c_2,d/2+A_1)$. The tensors $I_{x_0}(x,y)^{\otimes {l_1}}$ and $I_{x_0}(z,y)^{\otimes {l_2}}$ get mixed by a non-trivial operator acting on $\mathbb{V}^{l_1}\otimes \mathbb{V}^{l_2}$, the matrix $\mathbb{R}_{l_1l_2}(u)$.}
\label{fig:intrel to intrep}
\end{figure}

\begin{figure}[H]
\begin{center}
\includegraphics[scale=1.0]{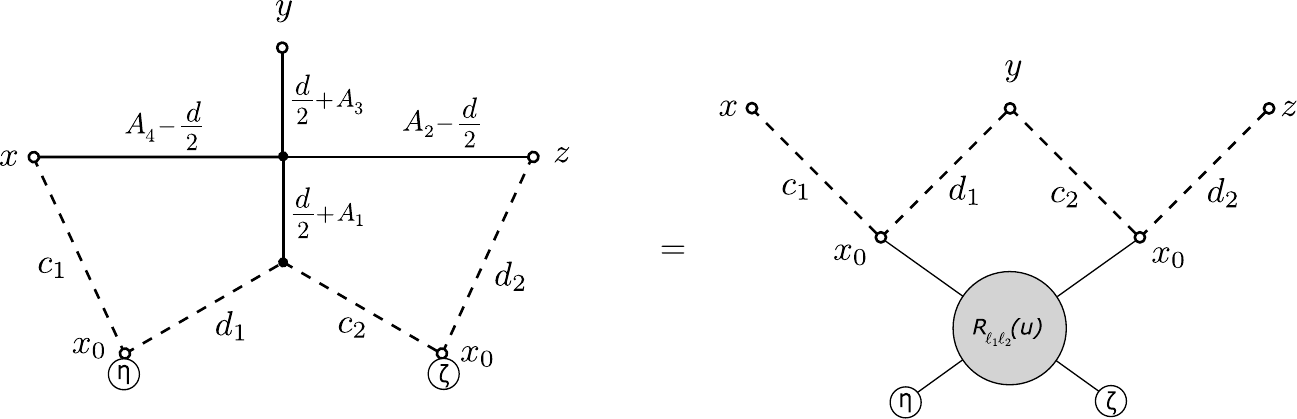}
\end{center}
\caption{Integral representation of a matrix element of the fused R-matrix $\R_{l_1,l_2}(u)$, as encoded into the interchange relation.}
\label{fig:fin form intrep}
\end{figure}

We show in Fig. \ref{fig:intrel to intrep} the chain of equivalent transformations which allows to derive, from the interchange relation, the integral representation for the elements of the R-matrix.
\noindent
The final form of the integral representation for the R-matrix element
is shown in Fig.\ref{fig:fin form intrep} and it follows straightforwardly from the previous chain of relations.
\noindent
The integral representation for the elements of the R-matrix is
\begin{multline}\label{Rll general integral inversion}
\left[\R_{l_1,l_2}(u)\zeta^{\otimes l_1}\otimes\eta^{\otimes l_2}\right]\cdot\left[\left(\frac{x}{x^2}-\frac{w}{w^2}\right)^{\otimes l_1}\otimes \left(\frac{y}{y^2}-\frac{w}{w^2}\right)^{\otimes l_2}\right] = F_{l_1,l_2}(u) \frac{x^{2\left(\ii u+\frac{l_{21}}{2}\right)} y^{2\left(\ii u+\frac{l_{12}}{2}\right)}}{w^{2\left(\ii u+\frac{l_1+l_2}{2}-1\right)}}\\
\times \int \frac{(z-w)^{2\left(\ii u+\frac{l_1+l_2}{2}-1\right)} \left[\zeta\cdot\left(\frac{y}{y^2}-\frac{v}{v^2}\right)\right]^{l_1} \left[\eta\cdot\left(\frac{x}{x^2}-\frac{v}{v^2}\right)\right]^{l_2}}
{(z-x)^{2\left(\ii u+\frac{l_{21}}{2}\right)}
(z-y)^{2\left(\ii u+\frac{l_{12}}{2}\right)}
(z-v)^{2\left(d-1+\frac{l_1+l_2}{2}-\ii u\right)}
v^{2\left(1 - \frac{l_1+l_2}{2} + \ii u\right)}}
\frac{\text{d}^d z}{\pi^{\frac{d}{2}}}
\frac{\text{d}^d v}{\pi^{\frac{d}{2}}}\,,
\end{multline}
where $\zeta^2 = \eta^2 = 0$ but $x^2,y^2$ and $w^2$ are arbitrary, and
\beq
F_{l_1,l_2}(u) = \frac{\Gamma\left(1-\frac{l_1+l_2}{2}-\ii u\right) \Gamma\left(1+\frac{l_1+l_2}{2}+\ii u\right) \Gamma\left(d-1+\frac{l_1+l_2}{2}-\ii u\right)}{\Gamma\left(1+\frac{l_1+l_2}{2}-\ii u\right)
\Gamma\left(\frac{d+l_1+l_2}{2} - 1 -\ii u\right)
\Gamma\left(1-\frac{l_1+l_2+d}{2}+\ii u \right)}\, .
\eeq
The representation \eqref{Rll general integral inversion} is actually equivalent to the main
interchange relation which guarantees the symmetry of eigenvectors w.r.t. the exchange of excitations numbers - explained in the following section.

In the follwing we present the derivation of a few integral
representations for the matrix element \eqref{Rll general integral inversion}. It is natural to perform an inversion of all external vectors $x,y,w$ and
variables of integrations $z\mapsto \frac{z}{z^2}$ and $v\mapsto\frac{v}{v^2}$
in relation~\eqref{Rll general integral inversion}.
After this transformation, one obtains
\begin{multline}\label{Rll general integral w}
\left[\R_{l_1,l_2}(u)\zeta^{\otimes l_1}\otimes\eta^{\otimes l_2}\right]\cdot(x-w)^{\otimes l_1}\otimes (y-w)^{\otimes l_2} \\
= F_{l_1,l_2}(u) \int \frac{(z-w)^{2\left(\ii u+\frac{l_1+l_2}{2}-1\right)} (\zeta\cdot(y-v))^{l_1} (\eta\cdot(x-v))^{l_2}}{(z-x)^{2\left(\ii u+\frac{l_{21}}{2}\right)} (z-y)^{2\left(\ii u+\frac{l_{12}}{2}\right)} (z-v)^{2\left(d-1+\frac{l_1+l_2}{2}-\ii u\right)}} \frac{\text{d}^d z\, \text{d}^d v}{\pi^{d}}\, .
\end{multline}
The integral representation in the right hand side shows manifestly the translation invariance. Thus, for simplicity, we may put $w=0$ without any loss of generality:
\begin{multline}\label{Rll general integral}
\left[\R_{l_1,l_2}(u)\zeta^{\otimes l_1}\otimes\eta^{\otimes l_2}\right]\cdot x^{\otimes l_1}\otimes y^{\otimes l_2} \\
= F_{l_1,l_2}(u) \int \frac{z^{2\left(\ii u+\frac{l_1+l_2}{2}-1\right)} (\zeta\cdot(y-v))^{l_1} (\eta\cdot(x-v))^{l_2}}{(z-x)^{2\left(\ii u+\frac{l_{21}}{2}\right)} (z-y)^{2\left(\ii u+\frac{l_{12}}{2}\right)} (z-v)^{2\left(d-1+\frac{l_1+l_2}{2}-\ii u\right)}} \frac{\text{d}^d z\, \text{d}^d v}{\pi^{d}}\, .
\end{multline}
Despite its simplicity, this integral representation should be used with care because
the integral over $v$ is ill defined. The origin of the problems is our naïve inversion of integration variable
$v\mapsto\frac{v}{v^2}$ in the initial expression, and we illustrate it on the much simpler example of the delta function \eqref{chain relation delta}
\beq
\int\frac{\text{d}^d v}{v^{2a} (z-v)^{2(d-a)}} = A_0(a) A_0(d-a) \pi^{d}\delta^{(d)}(z)\,.
\eeq
If one naïvely performs an inversion $z\mapsto \frac{z}{z^2}$ and $v\mapsto\frac{v}{v^2}$
\beq
\int\frac{\text{d}^d v}{(z-v)^{2(d-a)}} = A_0(a) A_0(d-a) \pi^{d}\frac{\delta^{(d)}\left(\frac{z}{z^2}\right)}{z^{2(d-a)}}\, ,
\eeq
in the result neither the l.h.s. nor the r.h.s. are well-defined. In order to obtain a well defined expression for the R-matrix we perform an inversion of all external vectors $x,y,w$ but not of the variables of integrations $z$ and $v$ in~\eqref{Rll general integral inversion}, for which we obtain (here $w=0$)
\begin{multline}\label{Rll general integral true}
\left[\R_{l_1,l_2}(u)\zeta^{\otimes l_1}\otimes\eta^{\otimes l_2}\right]\cdot x^{\otimes l_1}\otimes y^{\otimes l_2}
= F_{l_1,l_2}(u) \int \frac{\text{d}^d z}{\pi^{\frac{d}{2}}}
\frac{1}{\left(1-2z\cdot x+z^2x^2\right)^{\ii u+\frac{l_{21}}{2}}
\left(1-2z\cdot y+z^2y^2\right)^{\ii u+\frac{l_{12}}{2}}}\\
\int \frac{\text{d}^d v}{\pi^{\frac{d}{2}}}
 \frac{(\zeta\cdot(y-\frac{v}{v^2}))^{l_1} (\eta\cdot(x-\frac{v}{v^2}))^{l_2}}
{(z-v)^{2\left(d-1+\frac{l_1+l_2}{2}-\ii u\right)}
v^{2\left(1-\frac{l_1+l_2}{2}+\ii u\right)}}\, .
\end{multline}
The same representation can be obtained via the inversion $z\mapsto \frac{z}{z^2}$ and $v\mapsto\frac{v}{v^2}$ in relation~\eqref{Rll general integral}.
In fact, the integration of $v$ is reduced to the finite sum of derivatives of delta function, therefore the integration of $z$ can be performed easily, so that we can derive a closed expression for the integrals in r.h.s. of \eqref{Rll general integral true}.
The result reads (see Appendix \ref{app:main})
\begin{multline}\label{v-integral}
\int \frac{\text{d}^d v}{\pi^{\frac{d}{2}}}
 \frac{(\zeta\cdot(y-\frac{v}{v^2}))^{l_1} (\eta\cdot(x-\frac{v}{v^2}))^{l_2}}
{(z-v)^{2\left(d-1+\frac{l_1+l_2}{2}-\ii u\right)}
v^{2\left(1-\frac{l_1+l_2}{2}+\ii u\right)}} =
A_{l_1,l_2}(u) \sum_{n,k,p}\frac{l_1! l_2! (-1)^{k+p}\, 2^{-k-p-3n}}
{(l_1-k-n)! (l_2-p-n)! k! p! n!}\\
\frac{(\zeta\cdot\eta)^n (\zeta\cdot y)^{l_1-k-n}
(\eta\cdot x)^{l_2-p-n}}{\left(1-\frac{l_1+l_2}{2}+\ii u\right)_{k+p+2n}
\left(2-\frac{d}{2}-\frac{l_1+l_2}{2}+\ii u \right)_n}\,
\partial_t^k\partial_s^p \left(\partial_{z_\mu}\partial_{z_\mu}\right)^n \,\delta^{(d)}(z-t\zeta-s\eta)
\end{multline}
where we have to put $t=s=0$ after differentiation, and the explicit
expression for $A_{l_1,l_2}(u)$ is given in \eqref{A}.
For simplicity, we showed in the sum the summation indices only.
The sum is finite and the range of summation is dictated by factorials in denominator:
for each $0\leqslant n \leqslant \min(l_1,l_2)$, we have $0\leqslant k \leqslant l_1-n$ and $0\leqslant p \leqslant l_2-n$.
Now the integral in $z$ can be calculated due to the appearance of the delta function,
and we finally obtain
\begin{multline}\label{R-integral-sum}
\left[\R_{l_1,l_2}(u)\zeta^{\otimes l_1}\otimes\eta^{\otimes l_2}\right]
\cdot x^{\otimes l_1}\otimes y^{\otimes l_2} =
\pi^{-\frac{d}{2}}F_{l_1,l_2}(u)A_{l_1,l_2}(u) \sum_{n,k,p}\frac{l_1! l_2! (-1)^{k+p}\, 2^{-k-p-3n}}
{(l_1-k-n)! (l_2-p-n)! k! p! n!}\\
\frac{(\zeta\cdot\eta)^n (\zeta\cdot y)^{l_1-k-n}
(\eta\cdot x)^{l_2-p-n}}{\left(1-\frac{l_1+l_2}{2}+\ii u\right)_{k+p+2n}
\left(2-\frac{d}{2}-\frac{l_1+l_2}{2}+\ii u\right)_n}\,
\partial_t^k\partial_s^p \\
\left.\left(\partial_{z_\mu}\partial_{z_\mu}\right)^n \,
\frac{1}{\left(1-2z\cdot x+z^2x^2\right)^{\ii u+\frac{l_{21}}{2}}
\left(1-2z\cdot y+z^2y^2\right)^{\ii u+\frac{l_{12}}{2}}}\right|_{z=t\zeta+s\eta}\,.
\end{multline}
It seems that the coincidence of expression \eqref{R-integral-sum} and \eqref{Rll general}
is far from obvious. The direct proof of their equivalence is very
technical and is given in Appendix \ref{app:main}.
Note that equivalence \eqref{R-integral-sum} and \eqref{Rll general}
automatically guarantees the validity of the interchange relations
in Fig.\ref{fig:interchange relation}.

\subsection{Properties of the R matrices}

The integral representation \eqref{Rll general integral} is very useful.
For example, it allows to reduce the derivation of some important properties of the R-matrix
to a few simple standard steps: the integral chain rules \eqref{chain relation} and
\eqref{chain relation delta} and the star-triangle relation \eqref{startriangle}.

\paragraph{Integral Formula for Null Vectors} The fact that \eqref{R-integral-sum}, or equivalently \eqref{Rll general integral}, is the same as \eqref{Rll general} is proved in Appendix \ref{app:main}. However, in the case $x^2 = y^2 = 0$ everything is simpler and the integral over $z$ in \eqref{Rll general integral} can be calculated explicitly using Symanzik's trick \cite{Symanzik:1972wj}: if the parameters $a_1,\dots,a_N$ satisfy $\sum_{k=1}^N a_k = d$, then it holds that
\beq
\int  \prod_{k=1}^N
\frac{\Gamma(a_k)}{(z-x_k)^{2a_k}} \frac{\text{d}^d z}{\pi^{\frac{d}{2}}} = \int_{\mathbb{R}_+^N} \frac{\e^{-\frac{\sum_{i,j}\alpha_i\alpha_j (x_i-x_j)^2}{\sum_{k=1}^N\gamma_k a_k}}}{\left(\sum_{k=1}^N\gamma_k a_k\right)^{\frac{d}{2}}} \prod_{k=1}^N \alpha_k^{a_k -1}\text{d}\alpha_k\,,
\eeq
where the parameters $\gamma_1,\dots,\gamma_N$ can be chosen arbitrarily as long as $\gamma_k\geqslant 0$, and they are not all zero. In our case, $N=4$ and we choose three of the parameters to be $0$ whereas the last one is set to 1, we thus obtain
\begin{multline}
\int \frac{\Gamma\left(\ii u+\frac{l_{21}}{2}\right) \Gamma\left(\ii u+\frac{l_{12}}{2}\right) \Gamma\left(1-\frac{l_1+l_2}{2}-\ii u\right) \Gamma\left(d-1+\frac{l_1+l_2}{2}-\ii u\right)}{(z-x)^{2\left(\ii u+\frac{l_{21}}{2}\right)} (z-y)^{2\left(\ii u+\frac{l_{12}}{2}\right)} z^{2\left(1-\frac{l_1+l_2}{2}-\ii u\right)} (z-v)^{2\left(d-1+\frac{l_1+l_2}{2}-\ii u\right)}} \frac{\text{d}^d z}{\pi^{\frac{d}{2}}}\\
= \int_{\mathbb{R}_+^4} \alpha_1^{\ii u+\frac{l_{21}}{2}-1} \alpha_2^{\ii u+\frac{l_{12}}{2} - 1} \alpha_3^{-\frac{l_1+l_2}{2}-\ii u} \alpha_4^{\frac{d}{2}-2+\frac{l_1+l_2}{2}-\ii u} \e^{-\frac{\alpha_1\alpha_2}{\alpha_4}(x-y)^2 - \alpha_1(x-v)^2 - \alpha_2(y-v)^2 - \alpha_3 v^2} \prod_{k=1}^4\text{d}\alpha_k\\
=\frac{\Gamma\left(\frac{d}{2}+l_1-1\right) \Gamma\left(\frac{d}{2}+l_2-1\right) \Gamma\left(1-\frac{l_1+l_2}{2}-\ii u\right) \Gamma\left(1-\frac{d+l_1+l_2}{2}+\ii u\right)}{(y-v)^{2\left(\frac{d}{2}+l_1 - 1\right)} (x-v)^{2\left(\frac{d}{2}+l_2 - 1\right)} v^{2\left(1-\frac{l_1+l_2}{2}-\ii u\right)} (x-y)^{2\left(1-\frac{d+l_1+l_2}{2}+\ii u\right)}}\, .
\end{multline}
As a consequence, when $x$ and $y$ are null vectors, the formula \eqref{Rll general integral} reduces to
\begin{multline}\label{Rll integral}
\left[\R_{l_1,l_2}(u)\zeta^{\otimes l_1}\otimes\eta^{\otimes l_2}\right]\cdot(x^{\otimes l_1}\otimes y^{\otimes l_2}) = \frac{\Gamma\left(\frac{d}{2}+l_1-1\right) \Gamma\left(\frac{d}{2}+l_2-1\right) \left(\ii u + \frac{l_1+l_2}{2}\right) \Gamma\left(\ii u - \frac{l_1+l_2}{2}\right)}{ \Gamma\left(-\ii u + \frac{d-2+l_1+l_2}{2}\right) \Gamma\left(\ii u + \frac{l_1 - l_2}{2}\right) \Gamma\left(\ii u + \frac{l_2 - l_1}{2}\right)}\\
\times (x-y)^{2\left(-\ii u + \frac{d+l_1+l_2-2}{2}\right)}\int \frac{(\zeta\cdot(v-y))^{l_1} (\eta\cdot(v-x))^{l_2}}{v^{2\left(1-\ii u - \frac{l_1+l_2}{2}\right)} (y-v)^{2\left(\frac{d}{2}+l_1 - 1\right)} (x-v)^{2\left(\frac{d}{2}+l_2 - 1\right)}} \frac{\text{d}^d v}{\pi^{\frac{d}{2}}}\, .
\end{multline}
This integral is well-defined. We postpone to Appendix \ref{app:Symanzik} the direct check of the equivalence of this representation to the expression \eqref{Rll}.

\paragraph{Derivative identity and mixing operator
$\mathbb{O}_{l_1,l_2}$}  For $\zeta$ and $\eta$ two null vectors, it holds that
\begin{multline}\label{derivative identity}
(\zeta\cdot\nabla)^{l_1}(\eta\cdot\nabla)^{l_2} x^{2\left(\frac{l_1+l_2+2-d}{2}+\lambda\right)} = \frac{\left(\frac{4-l_1-l_2-d}{2}+\lambda\right)_{l_1+l_2}}{\left(\frac{4-l_1-l_2-d}{2}-\lambda\right)_{l_1+l_2}}(x^2)^{2\lambda}\\
\times \left[\R_{l_1,l_2}(-\ii\lambda)\zeta^{\otimes l_1}\otimes\eta^{\otimes l_2}\right] \cdot \nabla^{\otimes (l_1+l_2)} x^{2\left(\frac{l_1+l_2+2-d}{2}-\lambda\right)}\, .
\end{multline}
In order to prove this identity one needs to compute $y^{\otimes (l_1+l_2)}\cdot \left[\R_{l_1,l_2}(-\ii\lambda)\zeta^{\otimes l_1}\otimes\eta^{\otimes l_2}\right]$
for arbitrary $y$. The details of calculation and the proof of the
relation \eqref{derivative identity} are given in Appendix \ref{app:O}.

Let us define an operator $\mathbb{O}_{l_1,l_2}(u):\mathbb{V}_{l_1}\otimes \V_{l_2}\rightarrow S^{l_1+l_2}(\mathbb{C}^d)$ that takes values in the space of symmetric tensors of rank $l_1+l_2$ in the following way:
\begin{align}\label{definition O}
(\zeta\cdot\nabla)^{l_1}(\eta\cdot\nabla)^{l_2} x^{2\left(\frac{l_1+l_2+2-d}{2}+\ii u\right)} = 2^{l_1+l_2} \left(\frac{4-l_1-l_2-d}{2}+\ii u\right)_{l_1+l_2} \frac{\left[\mathbb{O}_{l_1,l_2}(u)\zeta^{\otimes l_1}\otimes \eta^{\otimes l_2}\right]\cdot x^{\otimes (l_1+l_2)}}{x^{2\left(\frac{l_1+l_2+d-2}{2}-\ii u\right)}}\, ,
\end{align}
or, equivalently, using \eqref{derivatives LHS},
\begin{align}
\left[\mathbb{O}_{l_1,l_2}(u)\zeta^{\otimes l_1}\otimes \eta^{\otimes l_2}\right]\cdot x^{\otimes (l_1+l_2)} = \sum_{p}\, \frac{l_1! l_2! }{p! (l_1-p)! (l_2-p)!} \frac{(x^2 \zeta\cdot \eta)^{p} (\zeta\cdot x)^{l_1-p} (\eta\cdot x)^{l_2-p}}{2^p \left(\frac{4-l_1-l_2-d}{2}+\ii u\right)_{p}}\, .
\end{align}
The property \eqref{derivative identity} we presented above is now written in a concise manner as
\beq
\mathbb{O}_{l_1,l_2}(u) = \mathbb{O}_{l_1,l_2}(-u)\mathbb{R}_{l_1,l_2}(u)\, .
\eeq
The mixing operator $\mathbb{O}_{l_1,l_2}$ naturally arises in the generalisation of the chain relation \eqref{chain relation}:
\begin{multline}\label{chain harmonic 2}
\int \frac{\text{d}^d w}{\pi^{\frac{d}{2}}}\, \frac{C_1\left(\frac{w}{|w|}\right)C_2\left(\frac{w-x}{|w-x|}\right)}{w^{2a}(w-x)^{2b}} = \\ A_{l_1}(a) A_{l_2}(b) A_{l_1+l_2}(d-a-b)
\,\frac{\left[\mathbb{O}_{l_1,l_2}(\ii(a+b+1-d))C_1\otimes C_2\right]\cdot x^{\otimes (l_1+l_2)}}{x^{2\left(a+b+\frac{l_1+l_2-d}{2}\right)}}\,
\end{multline}
and in the expression for the Basso-Dixon diagram \eqref{I2LO}.

\paragraph{Unitarity} The representation \eqref{Rll} clearly shows that the R matrices are symmetric and transforms simply under complex conjugation:
\beq
^t\R_{l_1,l_2} = \R_{l_1,l_2}\, ,\quad \R_{l_1,l_2}(u)^* = \R_{l_1,l_2}(-u^*)\, .
\eeq
From the integral representation \eqref{Rll general integral}, on the other hand, it is easy to see that the inverse is obtained by changing the sign of the spectral parameter:
\beq
\label{unitar_id}
\R_{l_1,l_2}(u)\R_{l_1,l_2}(-u) = \mathrm{Id}_{l_1}\otimes \mathrm{Id}_{l_2}\, .
\eeq
With the help of the two previous relations this amounts to saying that the R-matrix is unitary when $u$ is real.
\begin{figure}[H]
\begin{center}
\includegraphics[scale=1.0]{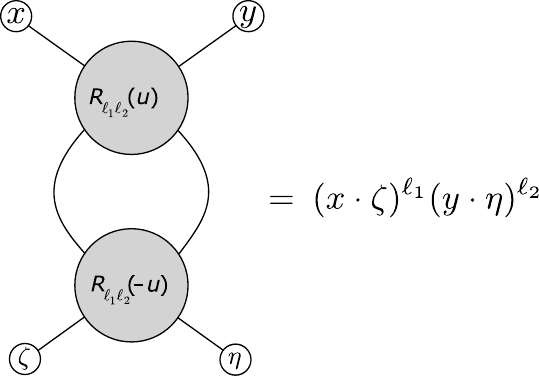}
\end{center}
\caption{Property \eqref{unitar_id} in the notation for the fused $\mathbb{R}_{l_1l_2}(u)$ introduced in Fig.\ref{finite_dim_R}.}
\end{figure}
\noindent
The proof of unitarity goes as follows:
\begin{multline*}
\left[\R_{l_1,l_2}(-u) \R_{l_1,l_2}(u)\zeta^{\otimes l_1}\otimes\eta^{\otimes l_2}\right]\cdot(x^{\otimes l_1}\otimes y^{\otimes l_2})\\
= F_{l_1,l_2}(-u) \int \frac{\left[\R_{l_1,l_2}(u)\zeta^{\otimes l_1}\otimes\eta^{\otimes l_2}\right]\cdot((y-v)^{\otimes l_1}\otimes (x-v)^{\otimes l_2})}{(z-x)^{2\left(-\ii u+\frac{l_{21}}{2}\right)} (z-y)^{2\left(-\ii u+\frac{l_{12}}{2}\right)} z^{2\left(1-\frac{l_1+l_2}{2}+\ii u\right)} (z-v)^{2\left(d-1+\frac{l_1+l_2}{2}+\ii u\right)}} \frac{\text{d}^d z \text{d}^d v}{\pi^d}\\
= F_{l_1,l_2}(-u)F_{l_1,l_2}(u)\int \frac{1}{(z-x)^{2\left(-\ii u+\frac{l_{21}}{2}\right)} (z-y)^{2\left(-\ii u+\frac{l_{12}}{2}\right)} z^{2\left(1-\frac{l_1+l_2}{2}+\ii u\right)} (z-v)^{2\left(d-1+\frac{l_1+l_2}{2}+\ii u\right)}}\\
\frac{(\zeta\cdot(x-v-v'))^{l_1} (\eta\cdot(y-v-v'))^{l_2}}{(z'+v-y)^{2\left(\ii u+\frac{l_{21}}{2}\right)} (z'+v-x)^{2\left(-\ii u+\frac{l_{12}}{2}\right)} z'^{2\left(1-\frac{l_1+l_2}{2}-\ii u\right)} (z'-v')^{2\left(d-1+\frac{l_1+l_2}{2}-\ii u\right)}}\frac{\text{d}^d z' \text{d}^d v' \text{d}^d z \text{d}^d v}{\pi^{2d}}\,.
\end{multline*}
After the natural change of variables $v' \to v'-v$ and $z' \to z'-v$, it is possible to integrate over $v$ explicitly using \eqref{chain relation delta}. One obtains $\delta^{(d)}(z-z')$, and the integration over $z'$ reduces the whole expression to another integral of the type \eqref{chain relation delta}
\begin{multline*}
F_{l_1,l_2}(-u)F_{l_1,l_2}(u)\int \frac{1}{(z-x)^{2\left(-\ii u+\frac{l_{21}}{2}\right)} (z-y)^{2\left(-\ii u+\frac{l_{12}}{2}\right)} z^{2\left(1-\frac{l_1+l_2}{2}+\ii u\right)} (z-v)^{2\left(d-1+\frac{l_1+l_2}{2}+\ii u\right)}}\\
\frac{(\zeta\cdot(x-v'))^{l_1} (\eta\cdot(y-v'))^{l_2}}{(z'-y)^{2\left(\ii u+\frac{l_{21}}{2}\right)} (z'-x)^{2\left(-\ii u+\frac{l_{12}}{2}\right)} (z'-v)^{2\left(1-\frac{l_1+l_2}{2}-\ii u\right)} (z'-v')^{2\left(d-1+\frac{l_1+l_2}{2}-\ii u\right)}}\frac{\text{d}^d z' \text{d}^d v' \text{d}^d z \text{d}^d v}{\pi^{2d}}\\
= \frac{\Gamma\left(1 - \frac{l_1+l_2}{2} + \ii u\right) \Gamma\left(d-1+\frac{l_1+l_2}{2}-\ii u\right)}{\Gamma\left(\frac{d+l_1+l_2}{2} - 1 - \ii u\right) \Gamma\left(1-\frac{l_1+l_2+d}{2}+\ii u\right)} \int \frac{\delta^{(d)}(z-z')}{(z-x)^{2\left(-\ii u+\frac{l_{21}}{2}\right)} (z-y)^{2\left(-\ii u+\frac{l_{12}}{2}\right)} z^{2\left(1-\frac{l_1+l_2}{2}+\ii u\right)}}\\
\times \frac{(\zeta\cdot(x-v'))^{l_1} (\eta\cdot(y-v'))^{l_2}}{(z'-y)^{2\left(\ii u+\frac{l_{21}}{2}\right)} (z'-x)^{2\left(-\ii u+\frac{l_{12}}{2}\right)} (z'-v')^{2\left(d-1+\frac{l_1+l_2}{2}-\ii u\right)}}\frac{\text{d}^d z' \text{d}^d v' \text{d}^d z }{\pi^{d}}\\
= \frac{\Gamma\left(1 - \frac{l_1+l_2}{2} + \ii u\right) \Gamma\left(d-1+\frac{l_1+l_2}{2}-\ii u\right)}{\Gamma\left(\frac{d+l_1+l_2}{2} - 1 - \ii u\right) \Gamma\left(1-\frac{l_1+l_2+d}{2}+\ii u\right)} \int\frac{(\zeta\cdot(x-v'))^{l_1} (\eta\cdot(y-v'))^{l_2}}{z^{2\left(1-\frac{l_1+l_2}{2}+\ii u\right)} (z-v')^{2\left(d-1+\frac{l_1+l_2}{2}-\ii u\right)}} \frac{\text{d}^d v' \text{d}^d z }{\pi^{d}}\\
= \int (\zeta\cdot(x-v'))^{l_1} (\eta\cdot(y-v'))^{l_2} \delta^{(d)}(v')\text{d}^d v' = (\zeta\cdot x)^{l_1} (\eta\cdot y)^{l_2}\, .
\end{multline*}
\paragraph{Crossing Symmetry}
From the explicit representation \eqref{Rll} of the R-matrix one immediately deduces the following crossing property:
\begin{equation}\label{crossing}
^{t_2}\R_{l_1,l_2}\left(\ii\frac{2-d}{2} - u\right) = \frac{\left(-\ii u + \frac{d+l_2-l_1-2}{2}\right)_{l_1}\left(\ii u - \frac{l_1 + l_2}{2}\right)_{l_1}}{\left(-\ii u + \frac{d-l_1-l_2-2}{2}\right)_{l_1}\left(\ii u + \frac{l_2 - l_1}{2}\right)_{l_1}} \R_{l_1,l_2}(u)
\end{equation}
where $t_2$ denotes transposition in $\V_2$ only.

\paragraph{Yang--Baxter Relation} The fusion procedure being a way to construct new solutions of the Yang--Baxter relation, we know that the expresssion \eqref{Rll} satisfies it. It is however also possible to show it directly for the integral representation as we now explain. We want to show that for arbitrary null vectors $\zeta$, $\eta$, and $\theta$ we have
\begin{multline}
\label{YBE_fin_dim}
\R_{l_1,l_2}(\lambda) \R_{l_1,l_3}(\lambda+\mu) \R_{l_2,l_3}(\mu) \zeta^{\otimes l_1}\otimes \eta^{\otimes l_2}\otimes \theta^{\otimes l_3}\\
= \R_{l_2,l_3}(\mu) \R_{l_1,l_3}(\lambda+\mu) \R_{l_1,l_2}(\lambda) \zeta^{\otimes l_1}\otimes \eta^{\otimes l_2}\otimes \theta^{\otimes l_3}\, .
\end{multline}
\begin{figure}[H]
\begin{center}
\includegraphics[scale=1.0]{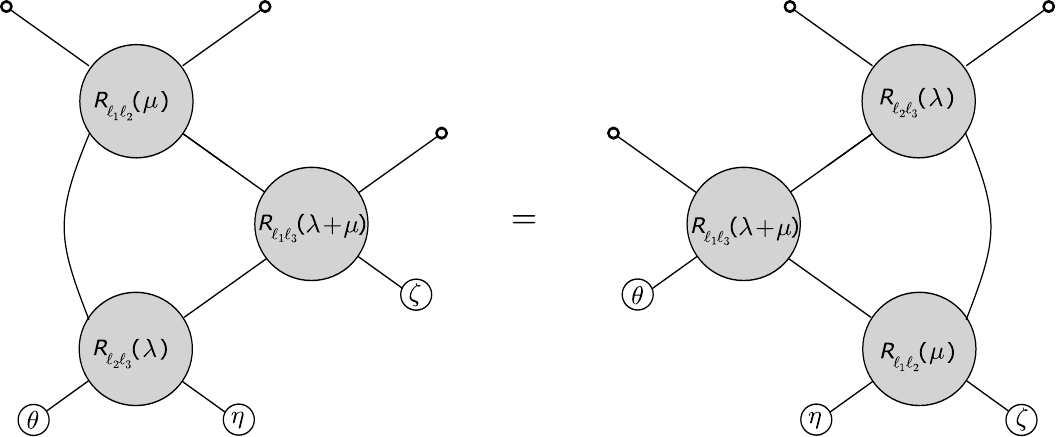}
\end{center}
\caption{Property \eqref{YBE_fin_dim} in the notation for the fused $\mathbb{R}_{l_1l_2}(u)$ introduced in Fig.\ref{finite_dim_R}.}
\end{figure}
\noindent
It suffices to verify that the scalar product with any vector of the form $x^{\otimes l_1}\otimes y^{\otimes l_2} \otimes z^{\otimes l_3}$, for $x$, $y$, and $z$ real, is the same for both sides. After taking the scalar product and using the integral representation (without writing the scalar prefactors $F_{l_i,l_j}$), the left-hand side becomes
\begin{multline*}
[\R_{l_1,l_2}(\lambda) \R_{l_1,l_3}(\lambda+\mu) \R_{l_2,l_3}(\mu) \zeta^{\otimes l_1}\otimes \eta^{\otimes l_2}\otimes \theta^{\otimes l_3}] \cdot x^{\otimes l_1}\otimes y^{\otimes l_2} \otimes z^{\otimes l_3}\\
\propto \int \frac{[\R_{l_1,l_3}(\lambda+\mu) \R_{l_2,l_3}(\mu) \zeta^{\otimes l_1}\otimes \eta^{\otimes l_2}\otimes \theta^{\otimes l_3}]\cdot (y-v)^{\otimes l_1}\otimes (x-v)^{\otimes l_2} \otimes z^{\otimes l_3}}{(w-x)^{2\left(\ii \lambda+\frac{l_{21}}{2}\right)} (w-y)^{2\left(\ii \lambda+\frac{l_{12}}{2}\right)} w^{2\left(1-\frac{l_1+l_2}{2}-\ii \lambda\right)} (w-v)^{2\left(d-1+\frac{l_1+l_2}{2}-\ii \lambda\right)}} \frac{\text{d}^d w \text{d}^d v}{\pi^d}\\
\propto \int \frac{\pi^{-d} \text{d}^d w \text{d}^d v}{(w-x)^{2\left(\ii \lambda+\frac{l_{21}}{2}\right)} (w-y)^{2\left(\ii \lambda+\frac{l_{12}}{2}\right)} w^{2\left(1-\frac{l_1+l_2}{2}-\ii \lambda\right)} (w-v)^{2\left(d-1+\frac{l_1+l_2}{2}-\ii \lambda\right)}}\\
\times \frac{[\R_{l_2,l_3}(\mu) \zeta^{\otimes l_1}\otimes \eta^{\otimes l_2}\otimes \theta^{\otimes l_3}]\cdot (z-v')^{\otimes l_1}\otimes (x-v)^{\otimes l_2} \otimes (y-v-v')^{\otimes l_3}}{(w'-y+v)^{2\left(\ii \lambda+\ii\mu+\frac{l_{31}}{2}\right)} (w'-z)^{2\left(\ii \lambda+\ii\mu+\frac{l_{13}}{2}\right)} w'^{2\left(1-\frac{l_1+l_3}{2}-\ii \lambda-\ii\mu\right)} (w'-v')^{2\left(d-1+\frac{l_1+l_3}{2}-\ii \lambda-\ii\mu\right)}} \frac{\text{d}^d w' \text{d}^d v'}{\pi^{d}}\\
\propto \int \frac{\pi^{-d} \text{d}^d w \text{d}^d v}{(w-x)^{2\left(\ii \lambda+\frac{l_{21}}{2}\right)} (w-y)^{2\left(\ii \lambda+\frac{l_{12}}{2}\right)} w^{2\left(1-\frac{l_1+l_2}{2}-\ii \lambda\right)} (w-v)^{2\left(d-1+\frac{l_1+l_2}{2}-\ii \lambda\right)}}\\
\times \frac{\pi^{-d} \text{d}^d w' \text{d}^d v'}{(w'-y+v)^{2\left(\ii \lambda+\ii\mu+\frac{l_{31}}{2}\right)} (w'-z)^{2\left(\ii \lambda+\ii\mu+\frac{l_{13}}{2}\right)} w'^{2\left(1-\frac{l_1+l_3}{2}-\ii \lambda-\ii\mu\right)} (w'-v')^{2\left(d-1+\frac{l_1+l_3}{2}-\ii \lambda-\ii\mu\right)}}\\
\times \frac{(\zeta\cdot(z-v'))^{l_1} (\eta\cdot(y-v-v'-v''))^{l_2} (\theta\cdot(x-v-v''))^{l_3}\, \pi^{-d} \text{d}^d w'' \text{d}^d v''}{(w''+v-x)^{2\left(\ii\mu+\frac{l_{32}}{2}\right)} (w''+v+v'-y)^{2\left(\ii\mu+\frac{l_{23}}{2}\right)} w''^{2\left(1-\frac{l_2+l_3}{2}-\ii\mu\right)} (w''-v'')^{2\left(d-1+\frac{l_2+l_3}{2}-\ii\mu\right)}}\\
\propto \int \frac{\pi^{-d} \text{d}^d w \text{d}^d v}{(w-x)^{2\left(\ii \lambda+\frac{l_{21}}{2}\right)} (w-y)^{2\left(\ii \lambda+\frac{l_{12}}{2}\right)} w^{2\left(1-\frac{l_1+l_2}{2}-\ii \lambda\right)} (w-v)^{2\left(d-1+\frac{l_1+l_2}{2}-\ii \lambda\right)}}\\
\times \frac{\pi^{-d} \text{d}^d w' \text{d}^d v'}{(w'-y+v)^{2\left(\ii \lambda+\ii\mu+\frac{l_{31}}{2}\right)} (w'-z)^{2\left(\ii \lambda+\ii\mu+\frac{l_{13}}{2}\right)} w'^{2\left(1-\frac{l_1+l_3}{2}-\ii \lambda-\ii\mu\right)} (w'-v')^{2\left(d-1+\frac{l_1+l_3}{2}-\ii \lambda-\ii\mu\right)}}\\
\times \frac{(\zeta\cdot(z-v'))^{l_1} (\eta\cdot(y-v'-v''))^{l_2} (\theta\cdot(x-v''))^{l_3} \, \pi^{-d} \text{d}^d w'' \text{d}^d v''}{(w''-x)^{2\left(\ii\mu+\frac{l_{32}}{2}\right)} (w''+v'-y)^{2\left(\ii\mu+\frac{l_{23}}{2}\right)} (w''-v)^{2\left(1-\frac{l_2+l_3}{2}-\ii\mu\right)} (w''-v'')^{2\left(d-1+\frac{l_2+l_3}{2}-\ii\mu\right)}}\, .
\end{multline*}
At the last step, we have simply performed the change of variables $(w'',v'')\mapsto (w'' - v,v''-v)$ so that now the integral over $v$ is computed by a simple application of the star-triangle identity \eqref{startriangle}. At the same time, we find it convenient to define $\tilde{z} = z - y$ and to perform the change of variables $(w',v')\mapsto (y-w',y-v')$, so that we obtain
\begin{multline*}
[\R_{l_1,l_2}(\lambda) \R_{l_1,l_3}(\lambda+\mu) \R_{l_2,l_3}(\mu) \zeta^{\otimes l_1}\otimes \eta^{\otimes l_2}\otimes \theta^{\otimes l_3}] \cdot x^{\otimes l_1}\otimes y^{\otimes l_2} \otimes z^{\otimes l_3}\\
= F_{l_1,l_2}(\lambda) F_{l_1,l_3}(\lambda+\mu) F_{l_2,l_3}(\mu) A_0\!\left(d-1+\frac{l_1+l_2}{2}-\ii\lambda\right) A_0\!\left(\ii\lambda+\ii\mu+\frac{l_{31}}{2}\right) A_0\!\left(1-\frac{l_2+l_3}{2}-\ii\mu\right)\\
\times \int \frac{\pi^{-\frac{d}{2}} \text{d}^d w}{(w-x)^{2\left(\ii\lambda+\frac{l_2-l_1}{2}\right)} (w-y)^{2\left(\ii\lambda+\frac{l_1-l_2}{2}\right)} w^{2\left(1-\frac{l_1+l_2}{2}-\ii\lambda\right)} (w'-w'')^{2\left(1-\frac{d+l_1+l_2}{2}+\ii\lambda\right)}}\\
\times \frac{\pi^{-d} \text{d}^d w' \text{d}^d v'}{(w'' - w)^{2\left(\frac{d+l_1-l_3}{2} - \ii\lambda - \ii\mu\right)} (w'-\tilde{z})^{2\left(\ii\lambda+\ii\mu+\frac{l_1-l_3}{2}\right)} (w'-y)^{2\left(1-\frac{l_1+l_3}{2}-\ii\lambda-\ii\mu\right)} (w'-v')^{2\left(d-1+\frac{l_1+l_3}{2}-\ii\lambda-\ii\mu\right)}}\\
\times \frac{(\zeta\cdot(\tilde{z}+v'))^{l_1} (\eta\cdot(v'-v''))^{l_2} (\theta\cdot(x-v''))^{l_3} \, \pi^{-d} \text{d}^d w'' \text{d}^d v''}{(w''-x)^{2\left(\ii\mu+\frac{l_3-l_2}{2}\right)} (w''-v')^{2\left(\ii\mu+\frac{l_2-l_3}{2}\right)} (w-w')^{2\left(\frac{d+l_2+l_3}{2}-1+\ii\mu\right)} (w''-v'')^{2\left(d-1+\frac{l_2+l_3}{2}-\ii\mu\right)}}\, .
\end{multline*}

\begin{figure}
\begin{center}
\includegraphics[scale=0.7]{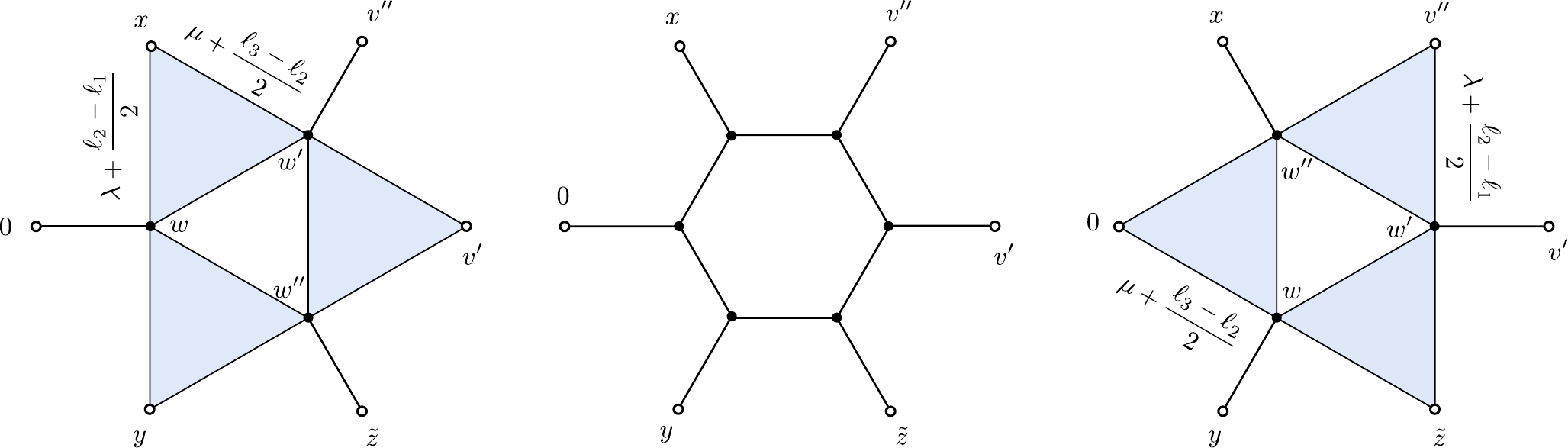}
\end{center}
\caption{The kernels depicted on the left and on the right are related respectively to the l.h.s. and r.h.s. of the Yang-Baxter equation \eqref{YBE_fin_dim} in the integral representation for the R-matrix elements \eqref{Rll}, and the equality of the two kernels is equivalent to the YBE. Both kernels are shown to be proportional to the one in the middle, via star-triangle identity applied to the blue triangles. Keeping track of the proportionality constants shows that l.h.s.$=$r.h.s.}
\label{fig:Yang-Baxter picture}
\end{figure}

Similar manipulations for the right-hand side of the Yang--Baxter relation give
\begin{multline*}
[\R_{l_2,l_3}(\mu) \R_{l_1,l_3}(\lambda+\mu) \R_{l_1,l_2}(\lambda) \zeta^{\otimes l_1}\otimes \eta^{\otimes l_2}\otimes \theta^{\otimes l_3}] \cdot x^{\otimes l_1}\otimes y^{\otimes l_2} \otimes z^{\otimes l_3}\\
= F_{l_1,l_2}(\lambda) F_{l_1,l_3}(\lambda+\mu) F_{l_2,l_3}(\mu) A_0\!\left(d-1+\frac{l_2+l_3}{2}-\ii\mu\right) A_0\!\left(\ii\lambda+\ii\mu+\frac{l_{13}}{2}\right) A_0\!\left(1-\frac{l_1+l_2}{2}-\ii\lambda\right)\\
\times \int \frac{\pi^{-\frac{d}{2}} \text{d}^d w}{w^{2\left(\ii\mu+\frac{l_3-l_2}{2}\right)} (w-\tilde{z})^{2\left(\ii\mu+\frac{l_2-l_3}{2}\right)} (w-y)^{2\left(1-\frac{l_2+l_3}{2}-\ii\mu\right)} (w'-w'')^{2\left(1-\frac{d+l_2+l_3}{2}+\ii\mu\right)}}\\
\times \frac{\pi^{-d} \text{d}^d w'' \text{d}^d v''}{(w''-x)^{2\left(\ii\lambda+\ii\mu+\frac{l_3-l_1}{2}\right)} (w - w')^{2\left(\frac{d+l_3-l_1}{2} - \ii\lambda - \ii\mu\right)} w''^{2\left(1-\frac{l_1+l_3}{2}-\ii\lambda-\ii\mu\right)} (w''-v'')^{2\left(d-1+\frac{l_1+l_3}{2}-\ii\lambda-\ii\mu\right)}}\\
\times \frac{(\zeta\cdot(\tilde{z}+v'))^{l_1} (\eta\cdot(v'-v''))^{l_2} (\theta\cdot(x-v''))^{l_3} \, \pi^{-d} \text{d}^d w' \text{d}^d v'}{(w'-v'')^{2\left(\ii\lambda+\frac{l_2-l_1}{2}\right)} (w'-\tilde{z})^{2\left(\ii\lambda+\frac{l_1-l_2}{2}\right)} (w-w'')^{2\left(\frac{d+l_1+l_2}{2}-1+\ii\lambda\right)} (w'-v')^{2\left(d-1+\frac{l_1+l_2}{2}-\ii\lambda\right)}}\, .
\end{multline*}

Notice now that the numerators in the integrands of the last two formulas are the same, and that these do not involve $w$, $w'$, or $w''$. Consequently, if we can prove that the integrals over these three variables coincide, then we are done. This is actually a straightforward application of the star-triangle identity \eqref{startriangle}, as depicted in Fig.\ref{fig:Yang-Baxter picture}.

\section{Diagonalisation of Graph-building Operators}
\label{sec:diagonalisation}

\subsection{Construction of the Eigenvectors}

In this section, we will work with the choice $\del\in \ii\mathbb{R}$, corresponding to a representation of the unitary principal series of the conformal group \cite{Tod:1977harm}.

We will eventually restore the fishnet framework $0<\del <\frac{d}{2}$ by analytic continuation. We also introduce, in our computations, a reference point $x_0\in\mathbb{R}^d$ (one could set it to $0$ for instance).
\begin{figure}[H]
\begin{center}
\includegraphics[scale=0.9]{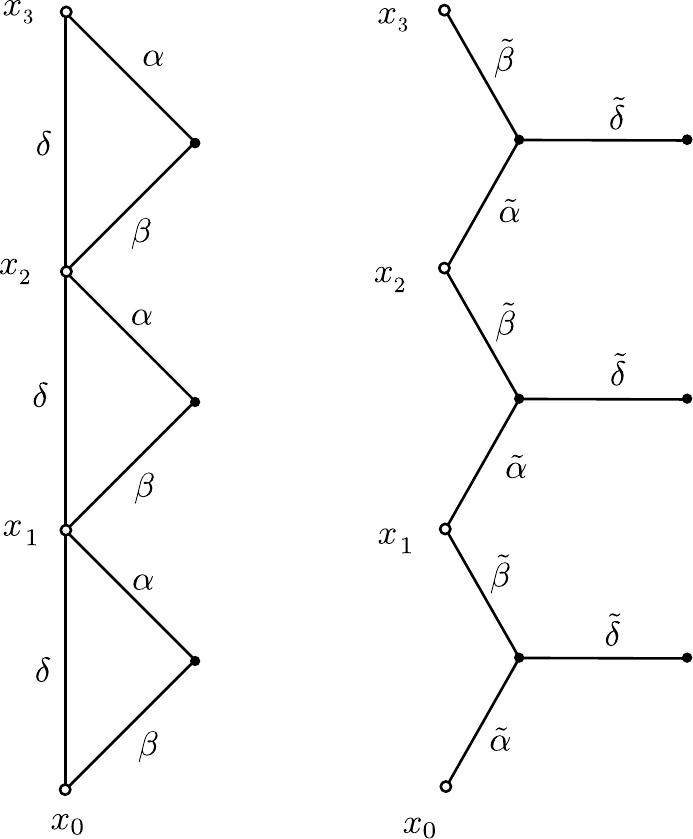}
\end{center}
\caption{Transfer-matrix operator $\mathbf{Q}_3(u)$ as a integral operator with the kernel represented in Feynman diagram notation. The two equivalent forms are related by star-triangle identity.}
\label{fig:Q_op_kernel}
\end{figure}
Let us define the lattice transfer-matrix $\mathbf{Q}_N(u)$  we want to diagonalise, where $N\in\mathbb{N}^*$ is the lattice width and $u \in\mathbb{C}$ is the spectral parameter. The transfeR-matrix $\Qm_N(u)$ is an operator acting on functions $\Phi$ of $N$ points as
\begin{align}
\begin{aligned}
\label{Qmat_any_d}
&\left[\mathbf{\Qm}_N(u)\Phi\right](x_1,\dots ,x_N) = \int \frac{\Phi(y_1, \dots , y_N)}{\prod_{k=1}^N (x_{k k+1})^{2\del} (y_k - x_k)^{2\alpha} (y_k - x_{k-1})^{2\beta}} \prod_{j=1}^N \frac{\text{d}^d y_j}{\pi^{\frac{d}{2}}}\\
&= \left[A_0(\del) A_0(\alpha(u)) A_0(\beta(u))\right]^N \int \frac{\Phi(y_1, \dots , y_N)}{\prod_{k=1}^N (w_k - y_k)^{2\tdel} (w_k - x_k)^{2\tilde{\beta}(u)} (w_k - x_{k-1})^{2\tilde{\alpha}(u)}} \prod_{j=1}^N \frac{\text{d}^d w_j \text{d}^d y_j}{\pi^{d}}\,,
\end{aligned}
\end{align}
where $x_{kk+1} = x_k - x_{k+1}$.
\noindent
The inner product between two functions of $N$ points $\Phi$ and $\Psi$ is defined by
\begin{equation}
\label{inner_product}
\langle{\Phi}\ket{\Psi} = \int \langle{\Phi}\ket{x_1,\dots ,x_N}\langle{x_1,\dots ,x_N}\ket{\Psi} \prod_{k=1}^N \frac{\text{d}^d x_k}{\pi^{\frac{d}{2}}} = \int \Phi^*(x_1,\dots ,x_N)\Psi(x_1,\dots ,x_N) \prod_{k=1}^N \frac{\text{d}^d x_k}{\pi^{\frac{d}{2}}}\, .
\end{equation}
With the definition \eqref{inner_product}, the constant $\pi^{-\frac{d}{2}}$ is included in the integration measure over space-time, i.e. $\ket{x}$ is such that $\langle{x}\ket{y} = \pi^{\frac{d}{2}}\delta(x-y)$.
As a consequence, one can write
\begin{equation}
\bra{x_1,\dots ,x_N} \Qm_N(u) \ket{y_1,\dots ,y_N} = \frac{1}{\prod_{k=1}^N x_{k,k+1}^{2\del} (y_k - x_k)^{2\alpha} (y_k - x_{k-1})^{2\beta}}
\end{equation}
for the kernel of the graph-building operator, which is represented by the diagram of Fig.\ref{fig:Q_op_kernel}.

\begin{figure}[H]
\begin{center}
\includegraphics[scale=0.8]{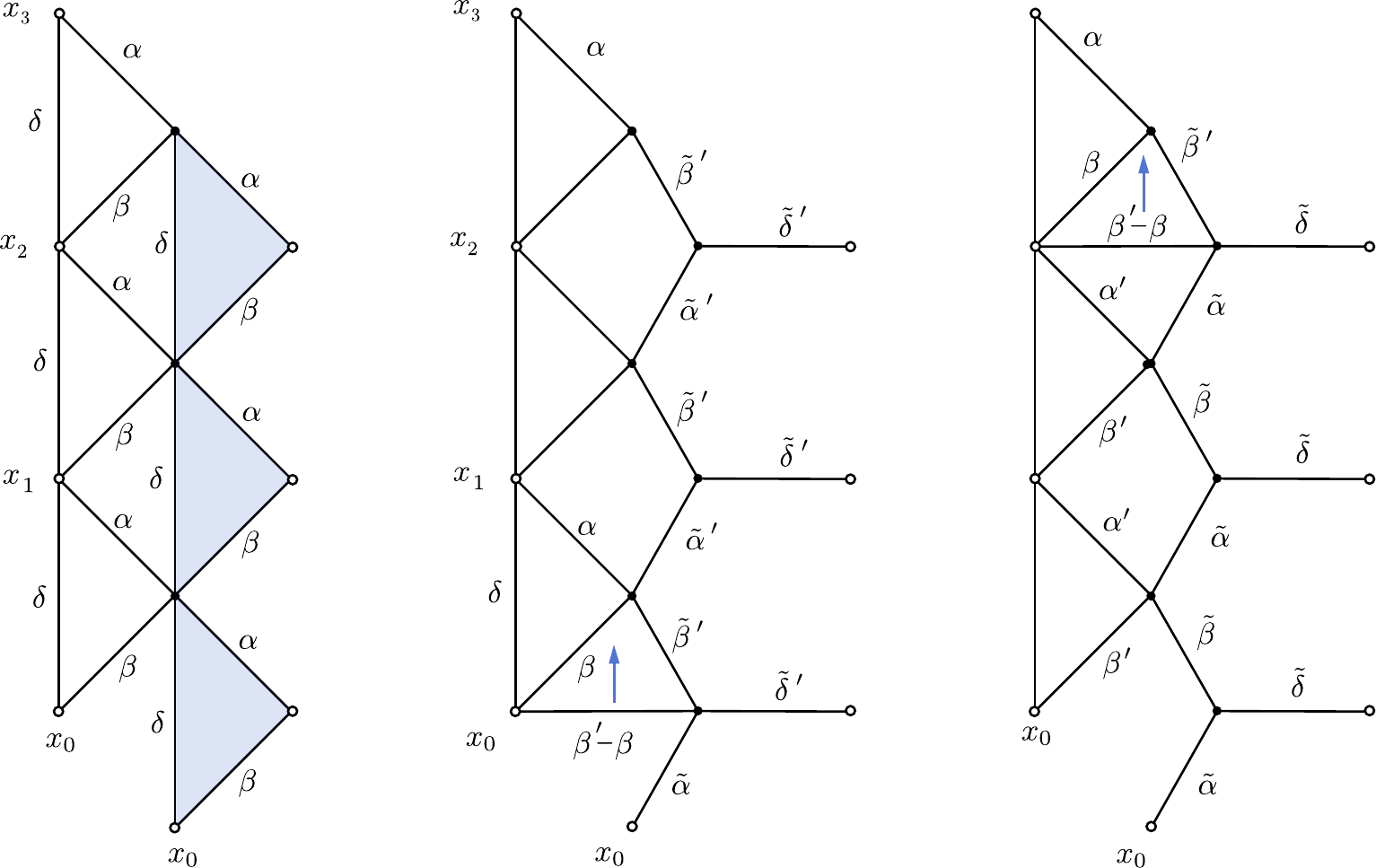}
\includegraphics[scale=0.8]{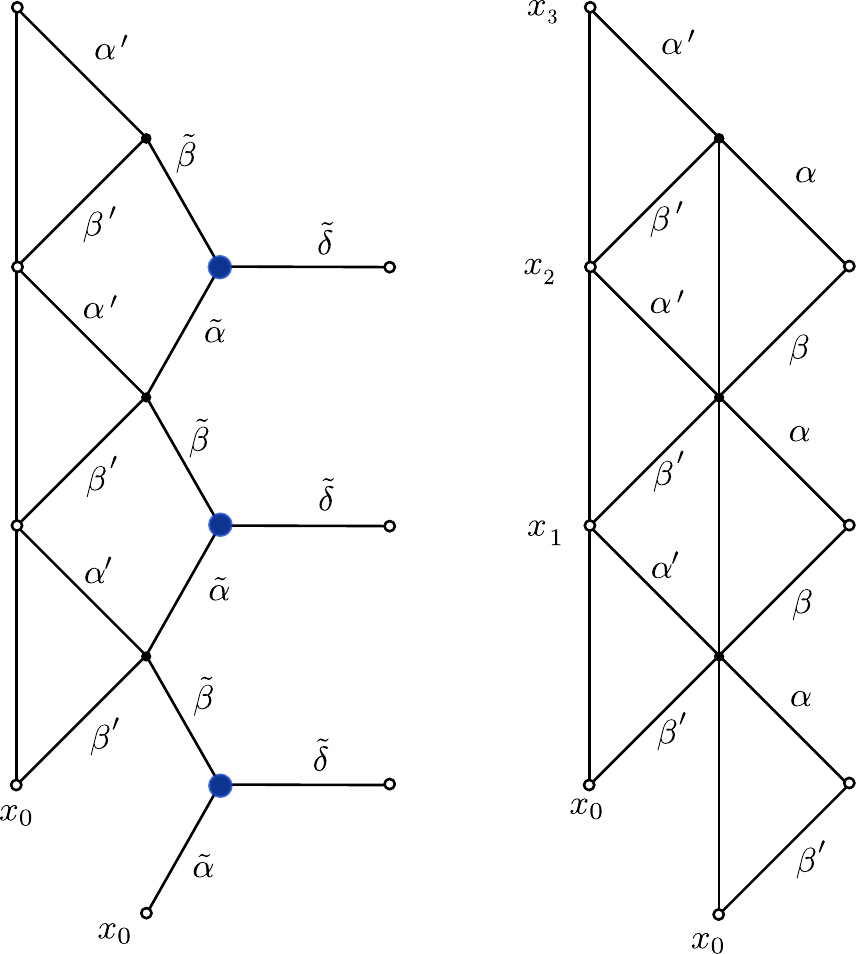}
\end{center}
\caption{Commutation of operators $\mathbf{Q}_3$ computed at different values $u \neq u'$ of the spectral parameter. \textbf{Up left:} The blue triangles are replaced by star integrals. \textbf{Up center:} upwards movement of the horizontal line of power $\beta-\beta'$, by a chain of star-triangle identitites. \textbf{Up right:} The last passage involves star-triangle and chain-rule indentity. \textbf{Down left:} integration of the blue vertices by star-triangle identity. \textbf{Down right:} the final result is equal to the first picture with exchanged $u$ and $u'$, i.e. $(\alpha,\beta) \leftrightarrow (\alpha' , \beta')$.}
\label{fig:Q_comm}
\end{figure}

A particular case of the family of operators \eqref{Qmat_any_d} - for $x_0 = 0$ - is the graph-building operator of the square-lattice fishnet
\beq
\label{graph_fish}
 \mathbf{B}_{N,\tdel}\equiv \Qm_N\left(\ii\frac{\tdel}{2}\right)\, .
\eeq
The operators \eqref{Qmat_any_d} computed at different values of the spectral parameter commute
\beq
[\Qm_N(u),\Qm_N(u')] = 0\,,
\eeq
the proof follows all the steps of the one presented for $d=2$ dimensions in \cite{Derkachov2019,Derkachov:2014gya}, and it is ultimately based on the star-triangle identity \eqref{startriangle}. We show it in Fig.\ref{fig:Q_comm} for completeness. The notation of Feynman diagrams used in Fig.\ref{fig:Q_op_kernel} maps lengthy manipulations of integral kernels into simple moves of lines and vertices, it will therefore be the language of many calculations of this section.

We shall construct iteratively the eigenvectors of $\Qm_N$, starting from $N=1$. Since these operators commute with global rotations and dilations, the eigenvectors of $\Qm_1$ are constrained to be
\beq
\langle{x}\ket{\ub_1;C} = \frac{C(x-x_0)}{(x-x_0)^{2\left(\tilde{\beta}_1 + \frac{l_1}{2}\right)}}\,,\,\,\,\,\,\, \ub_1 = (u_1,l_1)\in\mathbb{C}\times\mathbb{N}\, ,
\eeq
where
\beq
\quad C(y) = C^{\mu_1\dots\mu_{l_1}}y_{\mu_1}\dots y_{\mu_{l_1}}\, ,
\eeq
and $C \in\mathbb{V}_l$ is a symmetric traceless tensors of rank $l_1$. The spectral equation reads
\beq
\Qm_1(u)\ket{\ub_1;C} = Q_{l_1}(u|u_1) \ket{\ub_1;C}\, ,
\eeq
and the eigenvalue, computed using the identity \eqref{chain harmonic}, is
\beq
Q_l(u|u') = A_0(\alpha) A_l(\tilde{\alpha}') A_l(\beta + \tilde{\beta}')\, .
\eeq

\begin{figure}[H]
\begin{center}
\includegraphics[scale=0.71]{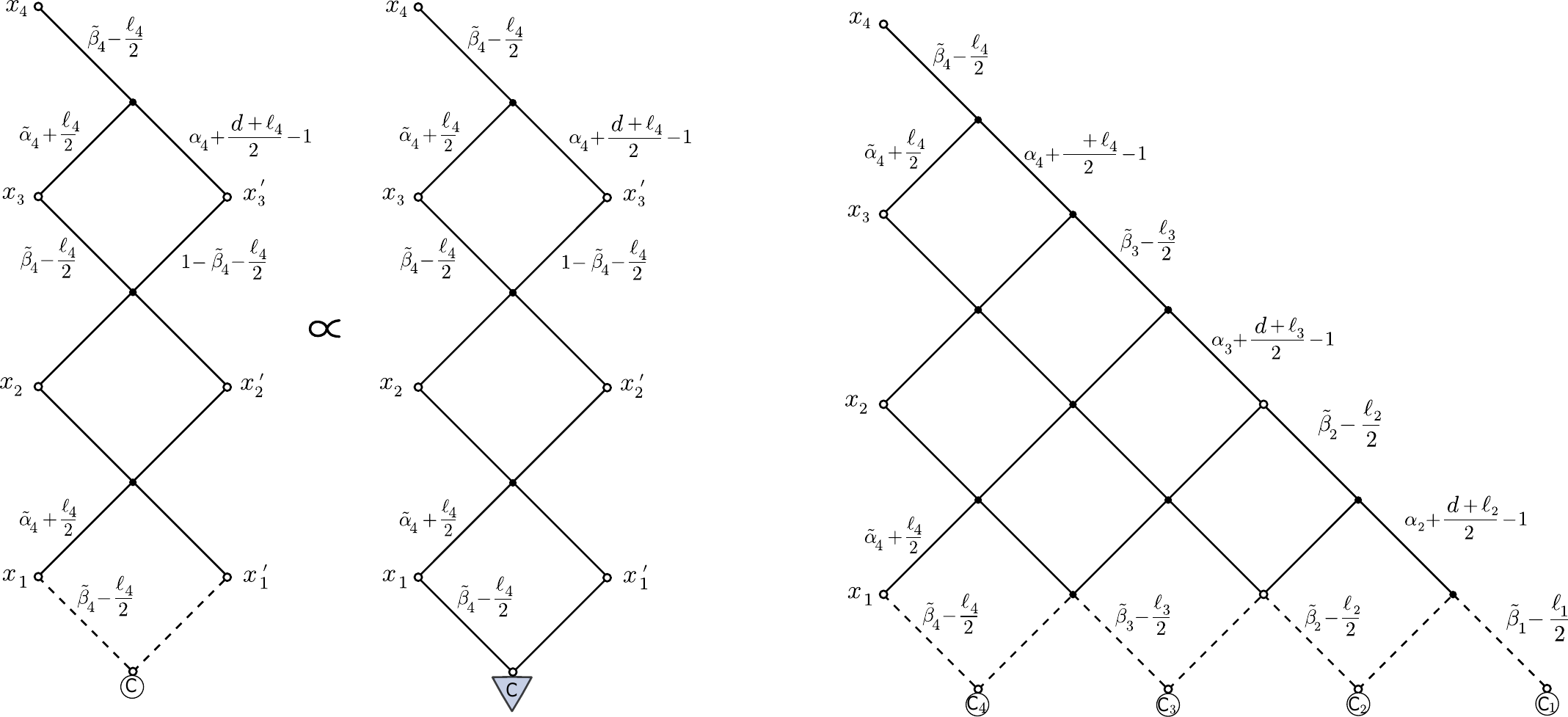}
\end{center}
\caption{\textbf{Left:} graphical representation of the layer operator $C\cdot \mathbf{\Lambda}_4(\ub_4) \propto C(\nabla_{x_0})\cdot {\Lambda}_4(\ub_4)$. The blue triangle stands for the differential operator $C(\nabla_{x_0})$. \textbf{Right:} graphical representation of the eigenvector $\ket{\ub_1,\dots , \ub_4;C}$.}
\label{fig:eigenf}
\end{figure}

For $N>1$, we find the eigenvectors after the definition of a recursive step. For $\ub\in\mathbb{C}\times\mathbb{N}$ and $C\in\mathbb{V}_l$ we introduce the \emph{layer operator} $C\cdot {\mathbf{\Lambda}}_N(\ub)$ acting on functions of $N-1$ points and returning functions of $N$ points:
\begin{multline}
\label{layer_op}
\left[C\cdot {\mathbf{\Lambda}}_N(\ub) \Phi\right](x_1,\dots ,x_N) = \left[A_0\left(1-\alpha-\frac{l}{2}\right)A_0\left(\beta+\frac{l}{2}\right)\right]^{N-1}\\
\times\int \frac{C\left(\frac{x_1-x_0}{(x_1-x_0)^2} - \frac{y_1-x_0}{(y_1-x_0)^2}\right)}{(x_N-w_{N-1})^{2\left(\tilde{\beta} - \frac{l}{2}\right)} \prod_{k=1}^{N-1} (x_k - w_k)^{2\left(\tilde{\alpha} + \frac{l}{2}\right)} (x_k - w_{k-1})^{2\left(\tilde{\beta} - \frac{l}{2}\right)}}
\\
\times\frac{\Phi(y_1, \dots , y_{N-1})}{\prod_{k=1}^{N-1} (y_k - w_k)^{2\left(\alpha + \frac{d+l}{2} - 1\right)} (y_k - w_{k-1})^{2\left(1 - \tilde{\beta} - \frac{l}{2}\right)} }
\prod_{k=1}^{N-1} \frac{\text{d}^d w_k}{\pi^{\frac{d}{2}}}
\frac{\text{d}^d y_k}{\pi^{\frac{d}{2}}}\, ,
\end{multline}
with $w_0 = x_0$. The scalar prefactor in \eqref{layer_op} leads to a convenient normalisation for the eigenvectors, that simplifies the form of their symmetry property and inner products. Strictly speaking, the integrals \eqref{layer_op} are ill-defined if $l>0$ and they should be understood as analytic continuations. Despite that, we can perform on them all the needed manipulations via integral identities presented in Appendix \ref{app:integral relations}. The operator $\mathbf{\Lambda}_N(\ub)$ carries $l$ symmetric traceless tensor indices,
\begin{equation}
{\mathbf{\Lambda}_N}(\ub)^{\mu_1\dots \mu_l}\,,
\end{equation}
and its pairing with the tensor $C$ can be encoded in the action of a differential operator, according to \eqref{harmonic derivative bis}:
\beq
C\cdot \mathbf{\Lambda}_N(\ub) = \frac{C\left(\nabla_{x_0}\right)}{2^l\left(\tilde{\beta}-\frac{l}{2}\right)_l} {\Lambda}_N(\ub)\, .
\eeq
The kernel of this last operator ${\Lambda}_N(\ub)$ is represented by the diagram of Fig.\ref{fig:eigenf} for $N=4$.

\begin{figure}[H]
\begin{center}
\includegraphics[scale=0.8]{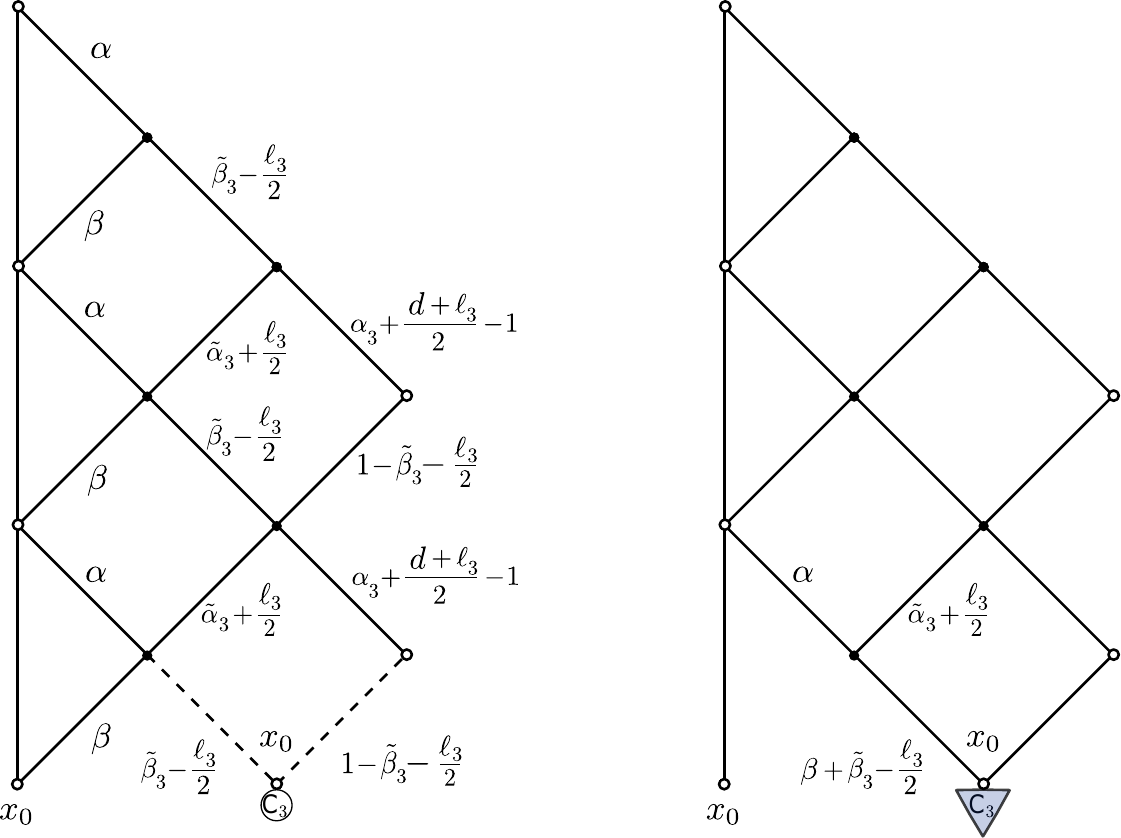}\vspace{15px}
\caption{\textbf{Left:} action of the transfer matrix $\mathbf{Q}_3(u)$ on the layer operator $C_3\cdot \mathbf{\Lambda}_3(u_3)$. \textbf{Right:} the tensor structure is regarded as the action of a differential operator $C_3(\nabla_{x_0})$ as in Equation \eqref{harmonic derivative bis}, and here represented by the blue triangle. The line with power $\beta$ and extreme $x_0$ is pulled under the action of the operator $C_3(\nabla_{x_0})$ by means of the property \eqref{useful property}.}
\includegraphics[scale=0.8]{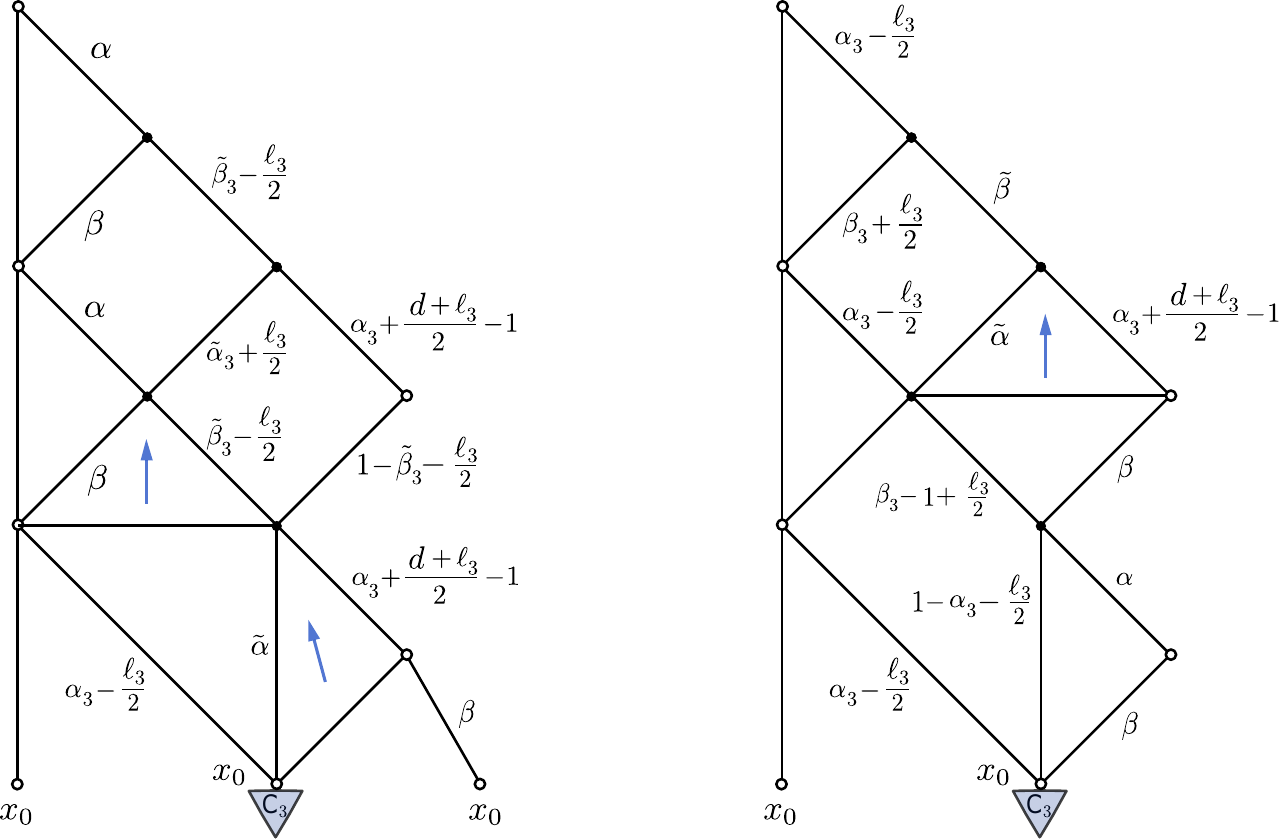}
\end{center}
\caption{Lines are moved in the graphs via several star-triangle and chain-rule identities, illustrated via blue dots, blue triangles, and arrows (see, for details, appendix \ref{app:integral relations}).}
\label{fig:eigenf diag}
\end{figure}

\noindent
The crucial relation satisfied by the layer operator is (see the proof in Fig.\ref{fig:eigenf diag}) is
\begin{equation}
\Qm_N(u) C_N\cdot\mathbf{\Lambda}_N(\ub_N) = q_{l_N}(u|u_N) C_N\cdot\mathbf{\Lambda}_N(\ub_N) \Qm_{N-1}(u)\,,
\label{fundamental commutation}
\end{equation}
and the eigenvectors of $\Qm_N(u)$ are therefore constructed iteratively as
\beq
\label{eigenv_d}
\ket{\ub_1,\dots , \ub_N;C_1\otimes\dots\otimes C_N} = C_N\cdot\mathbf{\Lambda}_N(\ub_N) \cdots C_2\cdot \mathbf{\Lambda}_2(\ub_2) \ket{\ub_1;C_1}\,,
\eeq
with arbitrarily chosen $C_i\in \mathbb{V}_{l_i}$.
The spectral equation for the graph-building transfer-matrix reads
\beq
{\Qm_N(u) \ket{\ub_1,\dots , \ub_N;C} = \prod_{k=1}^{N} q_{l_k}(u|u_k) \ket{\ub_1,\dots , \ub_N;C}}
\eeq
for an arbitrary tensor $C\in\V_{l_1}\otimes\dots\otimes\V_{l_N}$, in agreement with the invariance of $\Qm_N(u)$ under $O(d)$ rotations. The spectrum of the transfer-matrix is factorized into $N$ identical contributions of the type found at $N=1$, each depending on a {rapidity} $u_i$ and a Lorentz {spin} $l_i$, and is symmetric with respect to permutations of these quantum numbers.

\begin{figure}[H]
\begin{center}
\includegraphics[scale=0.8]{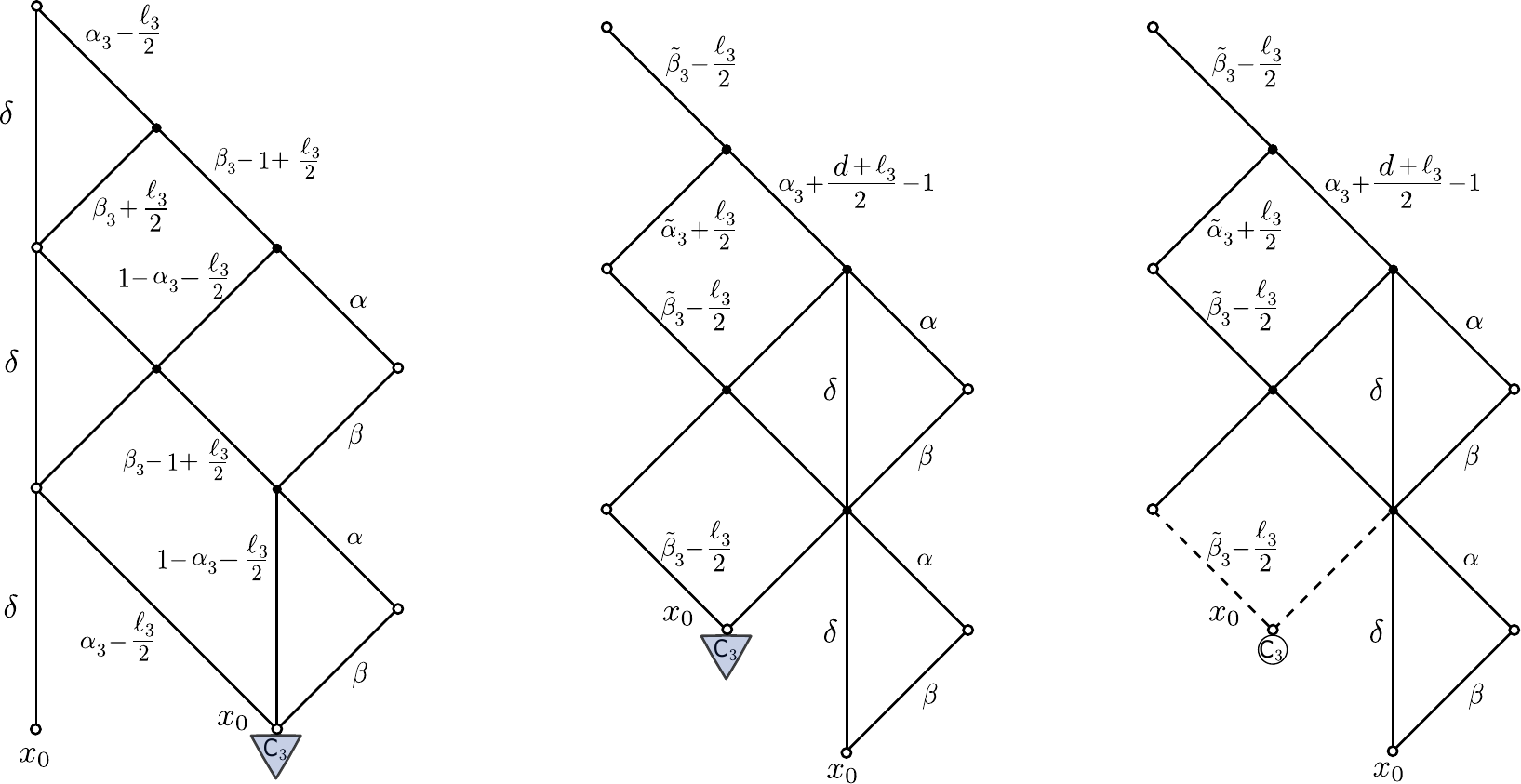}
\end{center}
\caption{Last step of the diagonalisation procedure: the first and second picture differs in the fact that the operator ${C_3}(\nabla_{x_0})$ does not act anymore on the line with power $\beta$, via the application of the property \eqref{useful property}. The last picture corresponds to the expression $C_3 \cdot \mathbf{\Lambda}_3(u_3) \,\mathbf{Q}_2(u)$.}
\end{figure}

\subsection{Symmetry Property}
\begin{figure}[H]
\begin{center}
\includegraphics[scale=0.7]{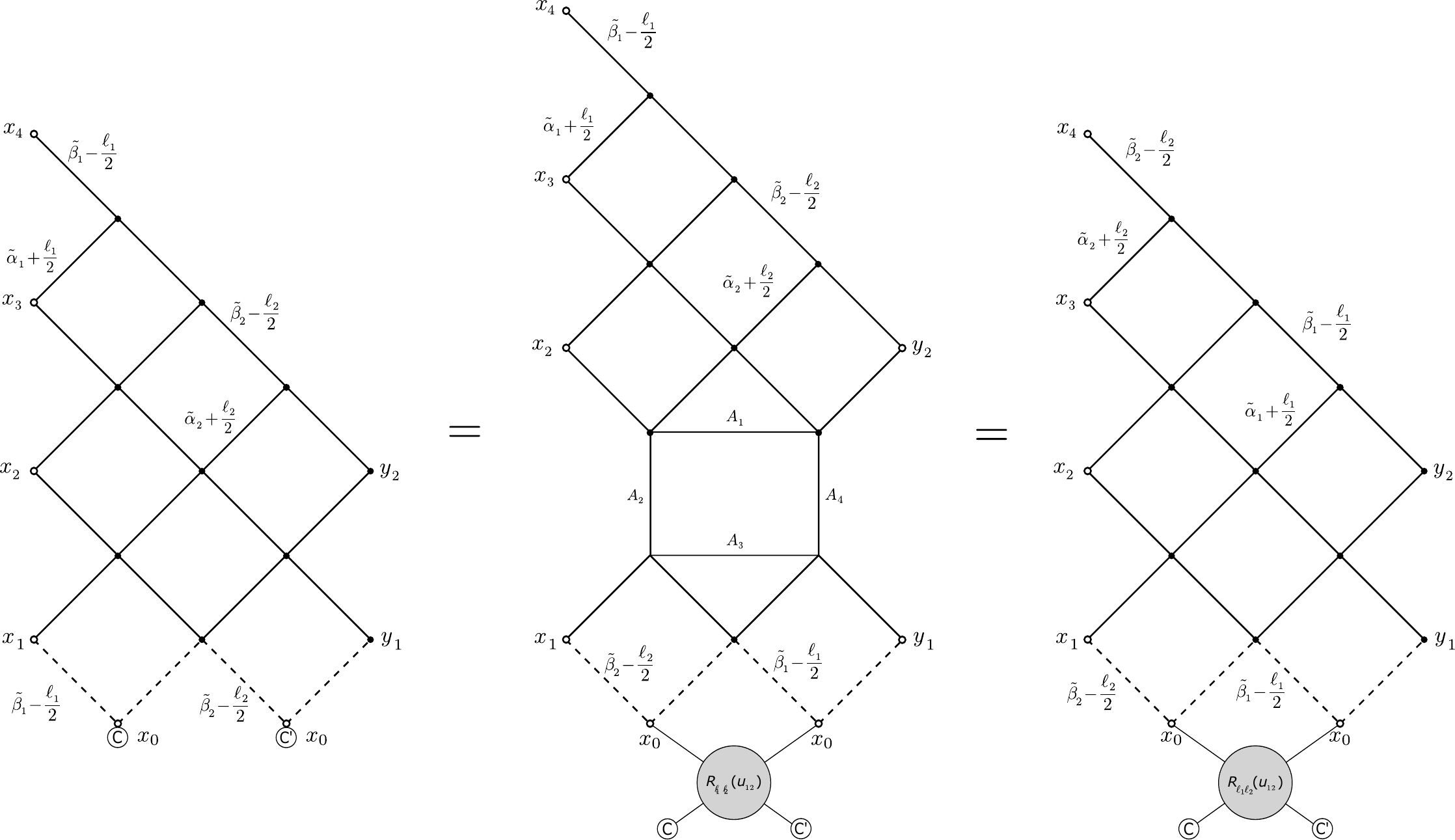}]\vspace{10px}
\includegraphics[scale=0.65]{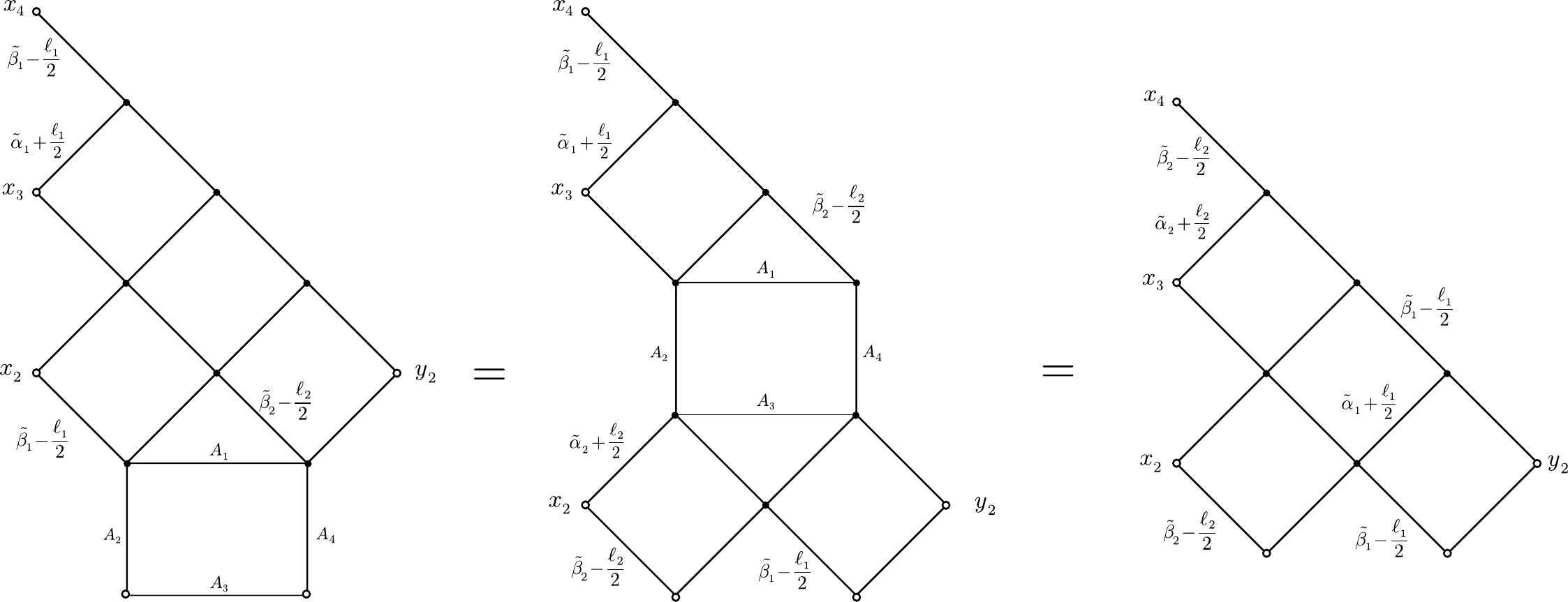}
\end{center}
\caption{Proof of the symmetry of eigenvectors}
\label{fig:symmetry}
\end{figure}

The symmetry of the spectrum of $\Qm_N(u)$ with respect to permutations of quantum numbers $\mathbf{u}_i =(u_i,l_i)$ has its counterpart at the level of the eigenvectors. Using the integral representation \eqref{Rll general integral inversion} of the R-matrix, it is possible to show the following commutation relation:
\begin{equation}\label{ZFexchange}
C\cdot \mathbf{\Lambda}_{N+1}(\ub_1)\otimes \mathbf{\Lambda}_N(\ub_2) = [\Sb_{l_1,l_2}(u_1-u_2)C]\cdot \mathbf{\Lambda}_{N+1}(\ub_2)\otimes \mathbf{\Lambda}_N(\ub_1)\, ,
\end{equation}
where $C\in\V_{l_1}\otimes\V_{l_2}$, and the tensor product notation concerns only the finite-dimensional spaces $\V_{l_1}$ and $\V_{l_2}$. The matrix $\Sb_{l_1,l_2}(u)$ coincides with the fused R-matrix up to a scalar phase:
\begin{equation}\label{S matrix exchange}
\Sb_{l, l'}(u) = (-1)^{l+l'}\frac{\Gamma(1+\tfrac{l+l'}{2}-\ii u)}{\Gamma(1+\tfrac{l+l'}{2}+\ii u)} \frac{\Gamma(\frac{d}{2} - 1 + \tfrac{l+l'}{2} - \ii u)}{\Gamma(\frac{d}{2} - 1 + \tfrac{l+l'}{2} + \ii u)}\,\R_{l, l'}(u)\, .
\end{equation}
The proof of \eqref{ZFexchange} is heavily based on the main interchange relation 
(Fig.\ref{fig:interchange relation}) and presented in Fig.\ref{fig:symmetry}.

The property \eqref{ZFexchange} relates two eigenvectors with exchanged $\mathbf{u}_k$, $\mathbf{u}_{k+1}$. For a generic permutation $\sigma\in\mathfrak{S}_N$,  the matrix $\Sb_{l_k,l_{k+1}}(u_k-u_{k+1})$ is replaced by an operator $\Sb(\ub_1,\dots , \mathbf{u}_N;\sigma)$ acting on $\mathbb{V}_{l_1}\otimes\dots\otimes\mathbb{V}_{l_N}$. First, we set
\beq
\Sb(\mathbf{u}_1,\dots , \mathbf{u}_N;\text{id}) = \mathrm{Id}_{l_1}\otimes\dots\otimes\mathrm{Id}_{l_N}\, ,
\eeq
moreover, if $\sigma$ is the transposition $(k, k+1)$, we impose
\beq
\Sb(\mathbf{u}_1,\dots , \mathbf{u}_N;(k k+1)) = \mathrm{Id}_{l_1}\otimes\dots\otimes\mathrm{Id}_{l_{k-1}}\otimes\Sb_{l_k,l_{k+1}}(u_{k+1}-u_k)\otimes\mathrm{Id}_{l_{k+2}}\otimes\dots\otimes\mathrm{Id}_{l_{N}}\, .
\eeq
Finally, we require the factorisation property
\beq\label{S product}
\Sb(\mathbf{u}_1,\dots , \mathbf{u}_N;(kk+1)\sigma) = \Sb(\mathbf{u}_{\sigma^{-1}(1)},\dots , \mathbf{u}_{\sigma^{-1}(N)};(kk+1)) \Sb(\mathbf{u}_1,\dots , \mathbf{u}_N;\sigma)\,,
\eeq
for any $k\in\{1,\dots,N-1\}$ and any permutation $\sigma$.
Since any permutation can be decomposed into a product of transpositions of the form $(k k+1)$, this is enough to define $\Sb(\mathbf{u_1},\dots , \mathbf{u_N};\sigma)$ for all $\sigma\in\mathfrak{S}_N$. Furthermore, there is no ambiguity in this definition because $\mathbb{S}_{l,l'}$ satisfies the Yang--Baxter equation.

The consequence of \eqref{eigenv_d},\eqref{ZFexchange} and \eqref{S product} on the eigenvectors is the following symmetry property: for any permutation of quantum numbers $\sigma\in\mathfrak{S}_N$, one has
\beq\label{symmetry}
{
\ket{\mathbf{u}_1,\dots , \mathbf{u}_N;C} = \ket{\mathbf{u}_{\sigma^{-1}(1)},\dots , \mathbf{u}_{\sigma^{-1}(N)};\mathbb{P}_\sigma\Sb(\mathbf{u}_1,\dots , \mathbf{u}_N;\sigma)C}}
\eeq
where $\mathbb{P}_\sigma:\V_{l_1}\otimes\dots\otimes\V_{l_N}\rightarrow\V_{l_{\sigma^{-1}(1)}}\otimes\dots\otimes\V_{l_{\sigma^{-1}(N)}}$ is the canonical isomorphism.

We point out that the exchange property \eqref{ZFexchange} is one of the defining properties of the Zamolodchikovs--Faddeev algebra \cite{Zamolodchikov:1978xm,Faddeev:1980zy}. Moreover, the symmetry property \eqref{symmetry} would also be typical for eigenvectors of compact spin chains for instance. However, since we are now considering a model with continuous spectrum, the tensor $C$ and the rapidities can be chosen arbitrarily; there are no (nested) Bethe ansatz equations.

\subsection{Inner Product}

The inner product for eigenvectors of the model of lenght $N=1$ is trivially computed to be
\begin{equation}\label{inner product N=1}
\langle{\ub;C}\ket{\ub';C'} = \int \frac{C^*(x-x_0)C'(x-x_0)}{(x-x_0)^{2\left(\tilde{\beta}^*+\tilde{\beta}' + \frac{l+l'}{2}\right)}}\frac{\text{d}^d x}{\pi^{\frac{d}{2}}}= \delta(\ub-\ub') \langle{C} \ket{C'}\,,
\end{equation}
where $\delta(\ub-\ub') = \delta_{ll'}\delta(u-u')$ and
\beq
\langle{C} \ket{C'} = \pi \int_{\mathbb{S}^{d-1}} C^*(n)C'(n) \frac{\text{d} n}{\pi^{\frac{d}{2}}} = \frac{l!\ 2^{1-l}\pi}{\Gamma\left(\frac{d}{2}+l\right)} C^*_{\mu_1\dots\mu_{l}}C'^{\mu_1\dots\mu_{l}}\,,
\label{inner product}
\eeq
is the inner product we choose on $\V_l$.
The inner product of eigenvectors of length $N>1$ is computed based on the iterative construction via layer operators \eqref{layer_op}. In fact, under the assumption $\ub'\neq\ub$, the overlap of two layer operators of length $N$ is expressed using layers of length $N-1$ via
\begin{multline}\label{contraction N}
C'\cdot \mathbf{\Lambda}^\dagger_{N}(\ub')\, C\cdot \mathbf{\Lambda}_{N}(\ub) =\left(C' \otimes C
\right)\cdot\, \mathbf{\Lambda}^\dagger_{N}(\ub') \otimes \mathbf{\Lambda}_{N}(\ub) \\
= \frac{\left[^{t_2}\Sb_{l,l'}(u-u') C\otimes C'\right]\cdot\, \mathbf{\Lambda}_{N-1}(\ub)\otimes\mathbf{\Lambda}_{N-1}(\ub') }{\left[(u-u')^2 + \frac{(l-l')^2}{4}\right]\left[(u-u')^2 + \frac{(d-2+l+l')^2}{4}\right]}\, ,\quad \text{if}\quad N>2\,,
\end{multline}
and
\begin{multline}\label{contraction_2}
\bra{x} C'\cdot \mathbf{\Lambda}^\dagger_{2}(\ub') C\cdot\mathbf{\Lambda}_{2}(\ub)\ket{y} = \frac{1}{\left[(u-u')^2 + \frac{(l-l')^2}{4}\right]\left[(u-u')^2 + \frac{(d-2+l+l')^2}{4}\right]}\times\\
\times \frac{[^{t_2}\Sb_{l,l'}(u-u') C\otimes C']\cdot(x-x_0)^{\otimes l}\otimes (y-x_0)^{\otimes l'}}{(x-x_0)^{2\left(\tilde{\beta} + \frac{l}{2}\right)} (y-x_0)^{2\left(\beta' + \frac{l'}{2}\right)}}\,.
\end{multline}

\begin{figure}[H]
\begin{center}
\includegraphics[scale=0.7]{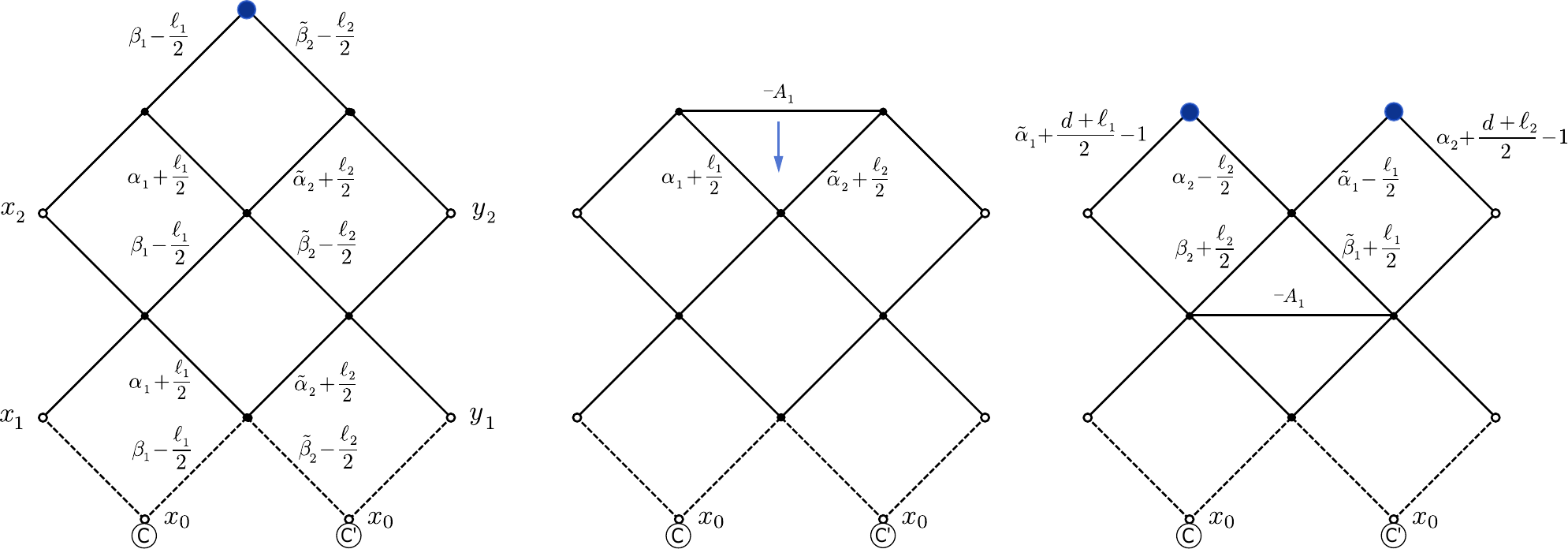}\vspace{10px}
\includegraphics[scale=0.7]{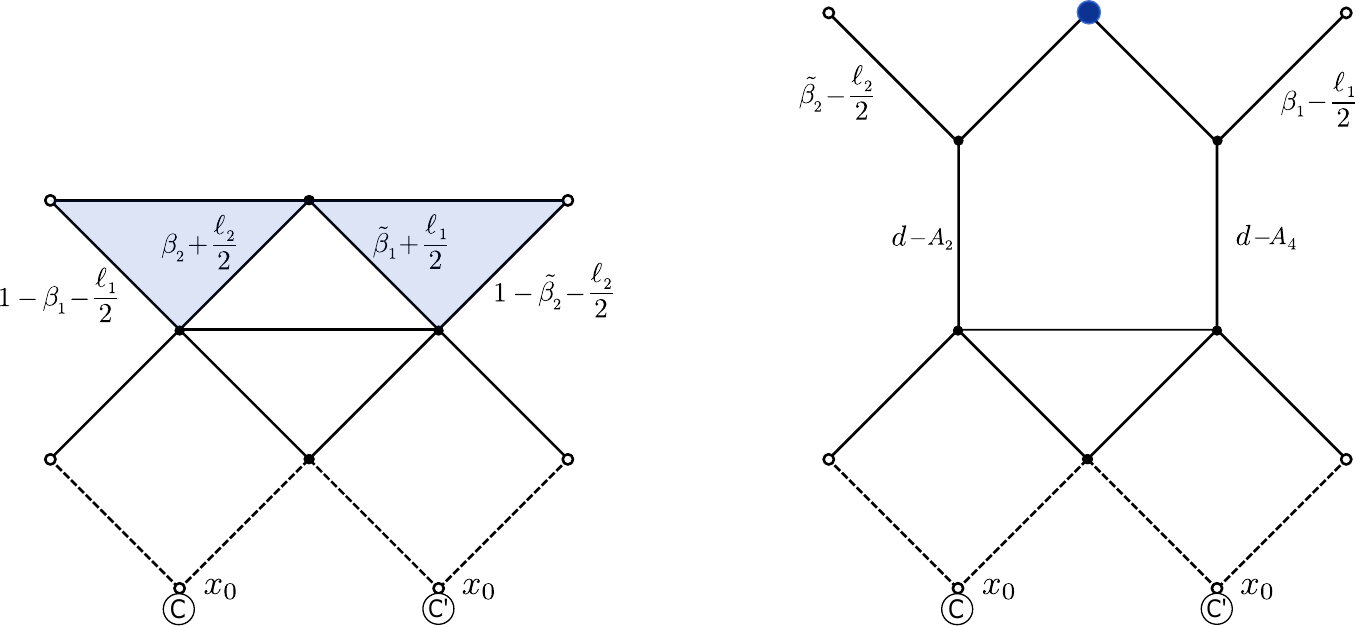}\vspace{10px}
\includegraphics[scale=0.7]{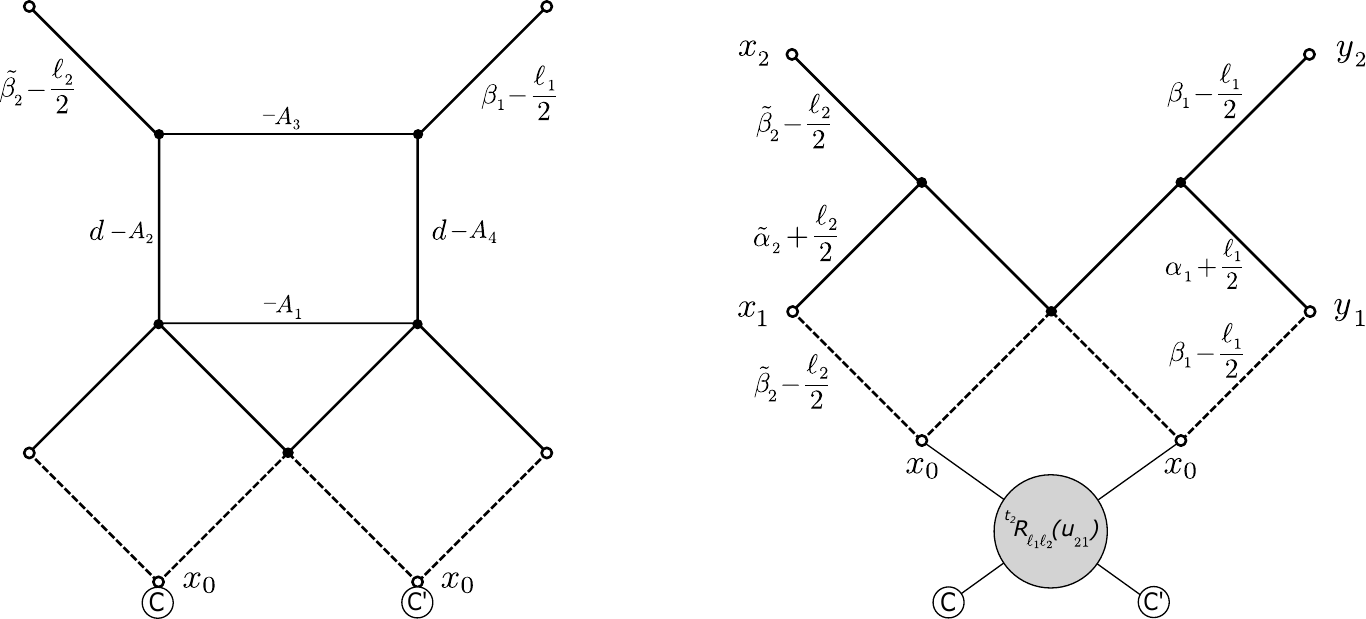}
\end{center}
\caption{Proof of the inner product formula. \textbf{Up left to down right:} the overlap of two layer operators $C'\cdot \mathbf{\Lambda}_3(u_2)$ and  $C\cdot \mathbf{\Lambda}_3(u_1)$ is transformed via application of star-triangle identities and, at the last step, the interchange relation. The notation of blue dots, blue triangles and arrows is the same as in Fig.\ref{fig:Q_comm} and refers to different application of the star-triangle and chain-rule identities. The final expression coincides with the r.h.s. of \eqref{contraction N}. It shows that the overlap of two layers amounts to the scattering of the two excitations $\mathbf{u_1}$ and $\mathbf{u_2}$ across each other, this is expressed by the fused R-matrix $\mathbb{R}_{l_1,l_2}(u_{12})$.}
\label{fig:inner_product}
\end{figure}

The properties \eqref{contraction_2} and \eqref{contraction N} are obtained using the integral representation \eqref{Rll general integral inversion} of the R-matrix and crossing symmetry \eqref{crossing}, and are shown by the diagrams in Fig.\ref{fig:inner_product}.
\noindent
From the iteration of \eqref{contraction N} and the symmetry property \eqref{symmetry}, the overlap of two eigenvectors reads
\begin{equation}\label{inner product N}
\langle{\ub_1,\dots ,\ub_N;C} \ket{\ub'_1,\dots ,\ub'_N;C'} = \frac{\sum_{\sigma\in\mathfrak{S}_N} \prod_{k=1}^N \delta(\ub_{\sigma(k)} - \ub'_{k})\ \langle{C}\ket{\mathbb{P}_\sigma\Sb(\ub'_1,\dots , \ub'_N;\sigma)C'}}{\mu(\ub_1,\dots ,\ub_N)}\,,
\end{equation}
where the measure $\mu$ is defined as
\beq
{\mu(\mathbf{u}_1,\dots , \mathbf{u}_N) = \prod_{1\leqslant j< k\leqslant N} \left[(u_j-u_k)^2 + \frac{(l_j-l_k)^2}{4}\right]\left[(u_j-u_k)^2 + \frac{(d-2+l_j+l_k)^2}{4}\right]}\, .
\eeq
Let us understand this formula with an explicit example. For $N=3$, the inner product is
\begin{align}
\begin{aligned}
&\langle \mathbf{u}_1,\mathbf{u}_2 , \mathbf{u}_3;C_1\otimes C_2\otimes C_3 | \mathbf{u}'_1,\mathbf{u}'_2 , \mathbf{u}'_3;C'_1\otimes C'_2\otimes C'_3 \rangle =\\
&= \bra{\mathbf{u}_1;C_1} C_2\cdot\mathbf{\Lambda}^\dagger_2(\mathbf{u}_2) C_3\cdot\mathbf{\Lambda}^\dagger_3(\mathbf{u}_3) C'_3\cdot\mathbf{\Lambda}_3(\mathbf{u}'_3) C'_2\cdot\mathbf{\Lambda}_2(\mathbf{u}'_2)\ket{\mathbf{u}'_1;C'_1}=\\
&= \int  \langle {\ub_1;C_1}|x\rangle \bra{x} C_2\cdot\mathbf{\Lambda}^\dagger_2(\mathbf{u}_2) C_3\cdot\mathbf{\Lambda}^\dagger_3(\mathbf{u}_3) C'_3\cdot\mathbf{\Lambda}_3(\mathbf{u}'_3) C'_2\cdot\mathbf{\Lambda}_2(\mathbf{u}'_2)\ket{y}  \langle y \ket{\ub_1';C_1'} \frac{\text{d}^d x\text{d}^d y}{\pi^{d}}\, .
\end{aligned}
\end{align}
If we assume that $\ub_3\neq\ub'_3$, $\ub_3\neq\ub'_2$, and $\ub_2\neq\ub'_3$, then thanks the overlap formula \eqref{contraction N} one can write
\begin{multline}\label{over_3}
\langle{\mathbf{u}_1,\mathbf{u}_2 , \mathbf{u}_3;C_1\otimes C_2\otimes C_3}\ket{\mathbf{u}'_1,\mathbf{u}'_2 , \mathbf{u}'_3;C'_1\otimes C'_2\otimes C'_3} \\
\propto \int\left[ ^{t_2}\Sb_{l'_3,l_2}(u'_3-u_2) ^{t_3}\Sb_{l'_2,l_3}(u'_2-u_3) ^{t_3}\Sb_{l'_3,l_3}(u'_3-u_3) C'_2\otimes C'_3\otimes C^*_2\otimes C^*_3\right](z;x;z;y) \\
\times \frac{C_1^*(x-x_0) C'_1(y-x_0)}{x^{2\left(\tilde{\beta}_1^* + \tilde{\beta}'_3 + \frac{l_1+l'_3}{2}\right)} y^{2\left(\tilde{\beta}_3^* + \tilde{\beta}'_1 + \frac{l_3+l'_1}{2}\right)} z^{2\left(\tilde{\beta}_2^* + \tilde{\beta}'_2 + \frac{l_2+l'_2}{2}\right)}}\frac{\text{d}^d x\text{d}^d y\text{d}^d z}{\pi^{\frac{3d}{2}}}\, .
\end{multline}
The integrals over $x$, $y$ and $z$ in \eqref{over_3} are of the form of \eqref{inner product N=1}, and their computation yields
\begin{multline}
\langle{\mathbf{u}_1,\mathbf{u}_2 , \mathbf{u}_3;C_1\otimes C_2\otimes C_3}\ket{\mathbf{u}'_1,\mathbf{u}'_2 , \mathbf{u}'_3;C'_1\otimes C'_2\otimes C'_3}\propto \delta(\ub_1-\ub'_3) \delta(\ub_2-\ub'_2) \delta(\ub_3-\ub'_1)\\
\times \langle{C_3\otimes C_2\otimes C_1}\ket{\Sb_{l'_2,l'_3}(u'_3-u'_2) \Sb_{l'_1,l'_3}(u'_3-u'_1) \Sb_{l'_1,l'_2}(u'_2-u'_1)C'_1\otimes C'_2\otimes C'_3}\, .
\end{multline}

Thanks to the delta functions, the prefactor is actually exactly $\mu(\ub_1,\ub_2,\ub_3)^{-1}$. It remains to notice that
\beq
\Sb_{l'_2,l'_3}(u'_3-u'_2) \Sb_{l'_1,l'_3}(u'_3-u'_1) \Sb_{l'_1,l'_2}(u'_2-u'_1) = \Sb(\ub'_1,\ub'_2,\ub'_3; (12)(23)(12)) = \Sb(\ub'_1,\ub'_2,\ub'_3; (13))
\eeq
because of \eqref{S product}. On the other hand, when $\ub_3\neq\ub'_3$, $\ub_3\neq\ub'_2$, and $\ub_2\neq\ub'_3$, formula \eqref{inner product N} also reduces to
\begin{multline}
\langle{\mathbf{u}_1,\mathbf{u}_2 , \mathbf{u}_3;C_1\otimes C_2\otimes C_3}\ket{\mathbf{u}'_1,\mathbf{u}'_2 , \mathbf{u}'_3;C'_1\otimes C'_2\otimes C'_3} = \mu(\ub_1,\ub_2,\ub_3)^{-1} \\
\times \delta(\ub_1-\ub'_3) \delta(\ub_2-\ub'_2) \delta(\ub_3-\ub'_1) \langle{C_3\otimes C_2\otimes C_1}\ket{ \Sb(\ub'_1,\ub'_2,\ub'_3; (13)) C'_1\otimes C'_2\otimes C'_3}\, .
\end{multline}
The other terms of \eqref{inner product N} appear when requiring, following \eqref{symmetry}, that the full result for the inner product be invariant under
\beq
\ket{\mathbf{u}_1,\mathbf{u}_2,\mathbf{u}_3; C_1\otimes C_2\otimes C_3}\longmapsto \ket{\mathbf{u}_{\sigma^{-1}(1)},\mathbf{u}_{\sigma^{-1}(2)},\mathbf{u}_{\sigma^{-1}(3)}; \mathbb{P}_\sigma\Sb(\mathbf{u}_1,\mathbf{u}_2,\mathbf{u}_3; \sigma)C_1\otimes C_2\otimes C_3}
\eeq
for all the permutations $\sigma\in\mathfrak{S}_3$. This whole procedure is generalized to arbitrary $N$ thanks the iterative form of the property \eqref{contraction N}.

\subsection{Completeness}

Let us fix $\{C_{m,l}\}_{1\leqslant m\leqslant d_l}$ an orthonormal basis of $\V_l$ with respect to the inner product defined in \eqref{inner product} ($d_l$ is the dimension of $\V_l$). We postulate that for any $N$, the following resolution of the identity holds:
\begin{align}\label{completeness}
\sum_{\substack{0\leqslant l_1\leqslant +\infty \\ 1\leqslant m_1\leqslant d_{l_1}}} \dots \sum_{\substack{0\leqslant l_N\leqslant +\infty \\ 1\leqslant m_N\leqslant d_{l_N}}} \int \dots \int \frac{\mu(\mathbf{u}_1,\dots , \mathbf{u}_N)}{N!} \langle{x_1,\dots ,x_N}\ket{\ub_1,\dots ,\ub_N;C_{m_1,l_1}\otimes\dots\otimes C_{m_N,l_N}}\\
\times \langle{\ub_1,\dots ,\ub_N;C_{m_1,l_1}\otimes\dots\otimes C_{m_N,l_N}}\ket{y_1,\dots ,y_N} \prod_{k=1}^N \text{d} u_k  = \prod_{k=1}^N \pi^{\frac{d}{2}}\delta(x_k - y_k)\, .
\end{align}
The power of $\pi$ in the right-hand side comes from the fact that we have defined $\ket{x}$ such that $\langle x\ket{y} = \pi^{\frac{d}{2}}\delta(x-y)$ (see beginning of Section \ref{sec:diagonalisation}). This completeness relation is easily verified in the case $N=1$, as it coincides with the expansion of a radial function id $d$-dimensions in Gegenbauer polynomials on the sphere $S^{d-1}$. We conjecture its validity for $N>1$.

\section{Basso--Dixon Diagrams}
\label{sec:BD diagrams}
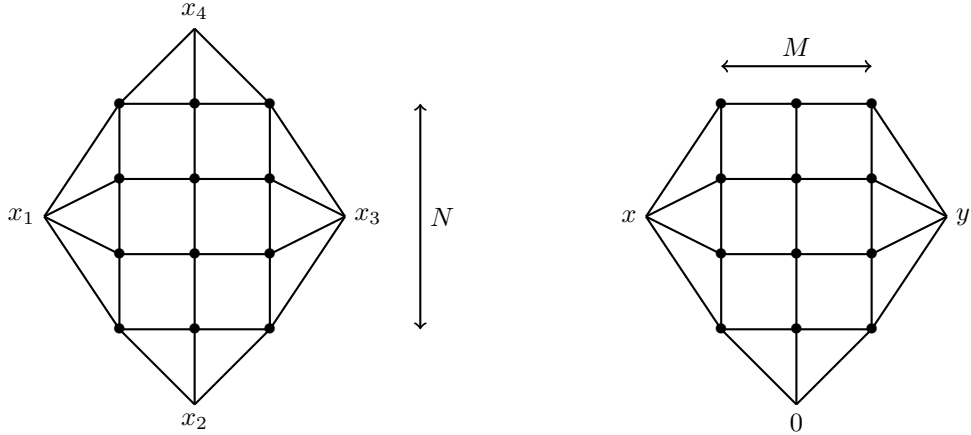
\begin{figure}[H]
\begin{center}
\begin{tikzpicture}

\draw[thick] (-6,0) -- (-5,1.5) -- (-3,1.5) -- (-2,0);
\draw[thick] (-6,0) -- (-5,0.5) -- (-3,0.5) -- (-2,0);
\draw[thick] (-6,0) -- (-5,-0.5) -- (-3,-0.5) -- (-2,0);
\draw[thick] (-6,0) -- (-5,-1.5) -- (-3,-1.5) -- (-2,0);

\draw[thick] (-4,2.5) -- (-5,1.5) -- (-5,-1.5) -- (-4,-2.5);
\draw[thick] (-4,2.5) -- (-4,1.5) -- (-4,-1.5) -- (-4,-2.5);
\draw[thick] (-4,2.5) -- (-3,1.5) -- (-3,-1.5) -- (-4,-2.5);

\draw (-6,0) node[left]{$x_1$};
\draw (-4,-2.5) node[below]{$x_2$};
\draw (-2,0) node[right]{$x_3$};
\draw (-4,2.5) node[above]{$x_4$};

\draw (-5,1.5) node{$\bullet$};
\draw (-5,0.5) node{$\bullet$};
\draw (-5,-0.5) node{$\bullet$};
\draw (-5,-1.5) node{$\bullet$};

\draw (-4,1.5) node{$\bullet$};
\draw (-4,0.5) node{$\bullet$};
\draw (-4,-0.5) node{$\bullet$};
\draw (-4,-1.5) node{$\bullet$};

\draw (-3,1.5) node{$\bullet$};
\draw (-3,0.5) node{$\bullet$};
\draw (-3,-0.5) node{$\bullet$};
\draw (-3,-1.5) node{$\bullet$};

\draw[thick,<->] (-1,1.5) -- (-1,-1.5) node[midway,right]{$N$};
\draw[thick,<->] (3,2) -- (5,2) node[midway,above]{$M$};

\draw[thick] (6,0) -- (5,1.5) -- (3,1.5) -- (2,0);
\draw[thick] (6,0) -- (5,0.5) -- (3,0.5) -- (2,0);
\draw[thick] (6,0) -- (5,-0.5) -- (3,-0.5) -- (2,0);
\draw[thick] (6,0) -- (5,-1.5) -- (3,-1.5) -- (2,0);

\draw[thick] (5,1.5) -- (5,-1.5) -- (4,-2.5);
\draw[thick] (4,1.5) -- (4,-1.5) -- (4,-2.5);
\draw[thick] (3,1.5) -- (3,-1.5) -- (4,-2.5);

\draw (6,0) node[right]{$y$};
\draw (4,-2.5) node[below]{$0$};
\draw (2,0) node[left]{$x$};

\draw (5,1.5) node{$\bullet$};
\draw (5,0.5) node{$\bullet$};
\draw (5,-0.5) node{$\bullet$};
\draw (5,-1.5) node{$\bullet$};

\draw (4,1.5) node{$\bullet$};
\draw (4,0.5) node{$\bullet$};
\draw (4,-0.5) node{$\bullet$};
\draw (4,-1.5) node{$\bullet$};

\draw (3,1.5) node{$\bullet$};
\draw (3,0.5) node{$\bullet$};
\draw (3,-0.5) node{$\bullet$};
\draw (3,-1.5) node{$\bullet$};

\end{tikzpicture}
\end{center}
\caption{Feyman graphs investigated in Section \ref{sec:BD diagrams} in the case $(M,N)=(3,4)$. We call $G^{(d,\del)}_{M,N}(x_1,x_2,x_3,x_4)$ and $I^{(d,\del)}_{M,N}(x,y)$ the integrals represented by the graph on the left and on the right respectively. All the vertical (or ending on $x_2, x_4$ or $0$) segments have weight $\del$ whereas all the horizontal (or ending on $x_1,x_3,x$ or $y$) ones have weight $\tdel$.}
\label{fig:BD diagram}
\end{figure}

In this section, we investigate the application of obtained basis of eigenvectors
and corresponding spectral decomposition of the graph-building operator to the computations of some fishnet Feynman integrals presented in Fig.\ref{fig:BD diagram}. Up to a trivial normalization factor the Feynman graph of the left panel has an interpretation as a four-point correlator in the fishnet theory:
\beq
G^{(d,\del)}_{M,N}(x_1,x_2,x_3,x_4) \propto \left\langle \Tr\left(X^N(x_1)Z^M(x_2)X^{\dagger N}(x_3) Z^{\dagger M}(x_4)\right) \right\rangle\, .
\eeq

Because of the conformal invariance of the integral it is equivalent to compute the integral associated to the right panel of the figure. A simple change of variables indeed shows that
\beq
G^{(d,\del)}_{M,N}(x_1,x_2,x_3,x_4) = \frac{1}{(x_{24}^2)^{M\del} \left(x^2_{14} x^2_{34}\right)^{N\tdel}} I^{(d,\del)}_{M,N}\left(\frac{x_{14}}{x^2_{14}} - \frac{x_{24}}{x^2_{24}},\frac{x_{34}}{x^2_{34}} - \frac{x_{24}}{x^2_{24}}\right)\, .
\eeq

In turn, the integral $I^{(d,\del)}_{M,N}$ is almost a matrix element of the $(M+1)$-th power of the graph-building operator $\mathbf{B}_{N,\tdel} = \Qm_N\left(\ii\frac{\tdel}{2}\right)$, one just has to be careful when sending all the external points to the same point:
\beq\label{IBD}
I^{(d,\del)}_{M,N}(x,y) = \pi^{\frac{Nd}{2}}\bra{x,\dots, x}\left(\prod_{i=1}^N \hat{x}_{i-1,i}^{2\del}\right)\mathbf{B}_{N,\tdel}^{M+1}\ket{y,\dots ,y}\, ,
\eeq
where we have set $x_0 = 0$. It thus seems natural to use the spectral decomposition of the graph-building operator $\Qm_N\left(\ii\frac{\tdel}{2}\right)$ to try to express these integrals in a simpler form. This was successfully achieved in two dimensions in \cite{Derkachov2019} and in four dimensions in \cite{Derkachov:2019tzo,Derkachov:2020zvv}.
For higher dimensions the result is actually more complicated and
we are going to discuss it in a separate paper.
Now we shall consider the first simplest examples to illustrate how the
general scheme works in the case of higher dimensions.

As we have seen above, the eigenvalue of $\Qm_N\left(\ii\frac{\tdel}{2}\right)$ factorises into a product of
\beq
Q_l(u) = Q_l\left(\ii\frac{\tdel}{2}\bigg|u\right)  = \frac{\Gamma(\del)\Gamma\left(\frac{d}{4}-\frac{\del}{2} + \frac{l}{2} - \ii u\right)\Gamma\left(\frac{d}{4}-\frac{\del}{2} + \frac{l}{2} + \ii u\right)}{\Gamma(\tdel)\Gamma\left(\frac{d}{4}+\frac{\del}{2} + \frac{l}{2} + \ii u\right)\Gamma\left(\frac{d}{4}+\frac{\del}{2} + \frac{l}{2} - \ii u\right)}\, .
\label{eigenvalueBD}
\eeq

\subsection{Ladder Diagrams}

We first give the expressions for the so-called ladder diagrams \cite{Usyukina:1993ch,Broadhurst:1985vq,Isaev_2003,Isaev:2007uy} in arbitrary dimension:
\beq
I_{M,1}^{(d,\del)} = \frac{\Gamma\left(\frac{d}{2}\right)}{(x_3^2 x_4^2)^{\frac{\tdel}{2}}}\sum_{l=0}^{+\infty}\frac{2l + d -2}{d-2} C_l^{\left(\frac{d-2}{2}\right)}(\cos\theta) \int \frac{\text{d} u}{2\pi}\, \left(\frac{x_3^2}{x_4^2}\right)^{\ii u} Q_l^{M+1}(u)\, ,
\eeq
where $Q_l$ is given in equation \eqref{eigenvalueBD}, $C_l^{(\mu)}$ are Gegenbauer polynomials and $\cos\theta = x_3\cdot x_4/|x_3||x_4|$. We assume $d\geqslant3$.

The integral is straightforwardly computed by residues but the eigenvalue $Q_l$ generically has infinitely many poles. However, when $\del$ is a positive integer, $Q_l^{-1}$ is a polynomial of degree $2\del$ and there is a finite number of poles. If $\del = 1$, one has $Q_l(u)^{-1} = \Gamma\left(\frac{d-2}{2}\right)\left(u^2 + (l+(d-2)/2)^2/4\right)$ and performing the integral yields
\beq
I_{M,1}^{(d,1)} = \frac{\Gamma\left(\frac{d-2}{2}\right) ^{-M}}{(x_3^2 x_4^2)^{\frac{d-2}{4}}}\sum_{k=M}^{2M}\frac{k!(-2\ln r)^{2M-k}}{M! (k-M)! (2M-k)!} \sum_{l=0}^{+\infty} C_l^{\left(\frac{d-2}{2}\right)}(\cos\theta) \frac{r^{l+ \frac{d-2}{2}}}{\left(l+ \frac{d-2}{2}\right)^k}
\eeq
with $r=\sqrt{x_4^2/x_3^2}$.

When $d$ is even we also have the following property of the Gegenbauer polynomials
\beq
\Gamma\left(\frac{d-2}{2}\right) C_l^{\left(\frac{d-2}{2}\right)}(x) = 2^{\frac{4-d}{2}} \left(\frac{\text{d}}{\text{d} x}\right)^{\frac{d-4}{2}}\,
\left[C^{(1)}_{l+\frac{d-4}{2}}(x)\right]\, .
\label{Gegenbauer derivative}
\eeq
Consequently, for even $d>2$, we can write ($z=r\e^{\ii\theta}$)
\beq
I_{M,1}^{(d,1)} = \frac{2^{\frac{4-d}{2}} \Gamma\left(\frac{d-2}{2}\right) ^{-M-1}}{(x_3^2 x_4^2)^{\frac{d-2}{4}}} \left(\frac{\text{d}}{\text{d}\cos\theta}\right)^{\frac{d-4}{2}} \left[\frac{L_M(z,\bar{z})}{\e^{\ii\theta} - \e^{-\ii\theta}}\right]\, ,
\eeq
where we have introduced the ladder function $L_M$ defined for $M>0$ by
\beq
L_M(z,\bar{z}) = \sum_{k=M}^{2M}\frac{k![-\ln(z\bar{z})]^{2M-k}}{M! (k-M)! (2M-k)!}\left[\mathrm{Li}_k(z) - \mathrm{Li}_k(\bar{z}) \right]\, .
\eeq
with $\mathrm{Li}_k(z) = \sum_{n=1}^{\infty} \frac{z^n}{n^k}$ the polylogarithm.

\subsection{Two-layer Diagrams}

Inserting the resolution of the identity in the expression \eqref{IBD} of $I_{M,2}^{(d,\del)}$, it becomes
\begin{multline}
I_{M,2}^{(d,\del)} = \sum_{\substack{0\leqslant l_1\leqslant +\infty \\ 1\leqslant m_1\leqslant d_{l_1}}} \sum_{\substack{0\leqslant l_2\leqslant +\infty \\ 1\leqslant m_2\leqslant d_{l_2}}}
\int \text{d} u_1 \text{d} u_2\, [Q_{l_1}(u_1) Q_{l_2}(u_2)]^{M+1} \frac{\mu(\ub_1,\ub_2)}{2}\\
\times \bra{x_3,x_3} \hat{x}_{12}^{2\del} \hat{x}_{1}^{2\del}\ket{\ub_1,\ub_2;C_{m_1,l_1}\otimes C_{m_2,l_2}} \bra{\ub_1,\ub_2;C_{m_1,l_1}\otimes C_{m_2,l_2}} {x_4,x_4}\rangle
\end{multline}

One can show that
\begin{align}
\bra{x_3,x_3} \hat{x}_{12}^{2\del} \hat{x}_{1}^{2\del} \ket{\mathbf{u_1},\mathbf{u_2};C} = A_0(\tdel) A_{l_1}(\tilde{\alpha}_1) A_{l_2}(\tilde{\alpha}_2) \frac{4^{1 - \frac{d}{2} + \ii u_{21}}\, \ii^{l_1 + l_2}}{x_3^{2\left(2\alpha_1 - \frac{d}{2} + 1\right)}}
\int \frac{\text{d}^d p}{\pi^{\frac{d}{2}}}\, \frac{C(p,p) \e^{\ii p\cdot x_3}}{p^{2\left(1 + \frac{l_1+l_2}{2} + \ii u_{21}\right)}}\, ,
\label{x3}\\
\langle x_4,x_4 \ket{\mathbf{u_1},\mathbf{u_2};C} = A_0\left(\frac{d}{2} + \del\right) A_{l_1}(\beta_1) A_{l_2}(\beta_2) \frac{4^{1 - \frac{d}{2} + \ii u_{21}}\, \ii^{l_1 + l_2}}{x_4^{2\left(2\tilde{\beta}_1 - \frac{d}{2} + 1\right)}}
\int \frac{\text{d}^d p}{\pi^{\frac{d}{2}}}\,\frac{C(p,p) \e^{\ii p\cdot x_4}}{p^{2\left(1 + \frac{l_1+l_2}{2} + \ii u_{21}\right)}}\, .
\label{x4}
\end{align}
Both of these equations can be rewritten using the operator $\mathbb{O}_{l_1,l_2}$ introduced in \eqref{definition O}, the first one becomes for instance
\begin{align*}
&\bra{x_3,x_3} \hat{x}_{12}^{2\del} \hat{x}_{1}^{2\del} \ket{\mathbf{u_1},\mathbf{u_2};C} = A_0(\tdel) A_{l_1}(\tilde{\alpha}_1) A_{l_2}(\tilde{\alpha}_2)
(-1)^{l_1+l_2} A_{l_1+l_2}(1+\ii u_{21})
\frac{ \left[\mathbb{O}_{l_1,l_2}(u_2 - u_1) C\right]\cdot x_3^{\otimes (l_1+l_2)}}
{x_3^{2\left(\alpha_1 + \alpha_2 + \frac{l_1+l_2}{2}\right)}}\, .
\end{align*}
We have to remark that, necessarily, whatever orthonormal basis of symmetric traceless tensors we chose,
\begin{equation}
\sum_{m=1}^{N(d,l)} C_{m,l}(p) C^*_{m,l}(q) = \frac{\Gamma\left(\frac{d}{2}\right) (2l + d -2)}{2\pi (d-2)} |p|^l |q|^l C_{l}^{(\frac{d-2}{2})}\left(\frac{p\cdot q}{|p||q|}\right)\, .
\label{completeness spherical}
\end{equation}
so that it is possible to rewrite $I_{M,2}^{(d,\del)}$ as
\begin{multline}
I_{M,2}^{(d,\del)} = \frac{\Gamma\left(\frac{d}{2}\right)^2}{2}\sum_{l_1,l_2}
\int \frac{\text{d} u_1\text{d} u_2}{(2\pi)^2}\,
\frac{4^{2-d} [Q_{l_1}(u_1) Q_{l_2}(u_2)]^{M+2}}{x_3^{2\left(2\alpha_1 - \frac{d}{2} + 1\right)} x_4^{2\left(2\beta_1 - \frac{d}{2} + 1\right)}}\\
\times \left[u_{12}^2 + \frac{l_{12}^2}{4}\right] \left[u_{12}^2 + \frac{(l_1+l_2+d-2)^2}{4}\right] \frac{(2l_1 + d - 2) (2l_2 + d - 2)}{(d-2)^2}\\
\times \int \frac{\text{d}^d p\,\text{d}^d q}{\pi^d}\,
\frac{C_{l_1}^{(\frac{d-2}{2})}\left(\frac{p\cdot q}{|p||q|}\right) C_{l_2}^{(\frac{d-2}{2})}\left(\frac{p\cdot q}{|p||q|}\right)}{p^{2\left(1 + \ii(u_2 - u_1)\right)} q^{2\left(1 + \ii(u_1 - u_2)\right)}} \e^{\ii p\cdot x_3 - \ii q\cdot x_4}\, .
\end{multline}

In order to perform the integrals over $p$ and $q$, we may proceed as follows: first one expands the product of two Gegenbauer polynomials according to \cite{vilenkin1978special}
\beq
C_{l_1}^{(\frac{d-2}{2})}\left(\frac{p\cdot q}{|p||q|}\right) C_{l_2}^{(\frac{d-2}{2})}\left(\frac{p\cdot q}{|p||q|}\right) = \sum_{m=0}^{\min (l_1,l_2)} a_{l_1,l_2,m} C_{l_1+l_2-2m}^{(\frac{d-2}{2})}\left(\frac{p\cdot q}{|p||q|}\right)
\eeq
with
\beq
a_{l_1,l_2,m} = \frac{(l_1 + l_2 - 2m + \frac{d-2}{2}) (l_1+l_2-2m)!}{(l_1 + l_2 - m + \frac{d-2}{2}) m! (l_1 - m)! (l_2 - m)!} \frac{\left(\frac{d-2}{2}\right)_{m} \left(\frac{d-2}{2}\right)_{l_1-m} \left(\frac{d-2}{2}\right)_{l_2-m}}{\left(\frac{d-2}{2}\right)_{l_1 + l_2 - m} (d-2)_{l_1+l_2-2m}} (d-2)_{l_1 + l_2 - m}\, .
\eeq
Then one uses the fact that $C_{l}^{(\frac{d-2}{2})}\left(\frac{p\cdot q}{|p||q|}\right)$ is a spherical harmonic with respect to both $p$ and $q$ (see equation \eqref{completeness spherical}) to compute the integrals over these variables using \eqref{fourier}:
\begin{multline}
\int \frac{\text{d}^d p\,\text{d}^d q}{\pi^d}\,
\frac{C_{l}^{(\frac{d-2}{2})}\left(\frac{p\cdot q}{|p||q|}\right)}{p^{2\left(1 - \ii u_{12}\right)} q^{2\left(1 + \ii u_{12}\right)}} \e^{\ii p\cdot x_3 - \ii q\cdot x_4}
 = \Bigg| \frac{\Gamma\left(\frac{l+d-2}{2} + \ii u_{12}\right)}
 {\Gamma\left(\frac{l+2}{2}+\ii u_{12}\right)}\Bigg|^2 \frac{4^{d-2}\,C_{l}^{(\frac{d-2}{2})}\left(\frac{x_3\cdot x_4}{|x_3||x_4|}\right)}{x_3^{2\left(\frac{d}{2} - 1 + \ii u_{12}\right)} x_4^{2\left(\frac{d}{2} - 1 - \ii u_{12}\right)}}\, .
\label{Legendreintegral}
\end{multline}

Consequently, one can write
\begin{multline}
I_{M,2}^{(d,\del)} = \frac{\Gamma\left(\frac{d}{2}\right)^2}{2(x_3^2 x_4^2)^{\tdel}}\sum_{l_1,l_2}
\int \frac{\text{d} u_1\text{d} u_2}{(2\pi)^2} \left(\frac{x_3^2}{x_4^2}\right)^{\ii (u_1 + u_2)} [Q_{l_1}(u_1) Q_{l_2}(u_2)]^{M+2}\\
\times \left[u_{12}^2 + \frac{l_{12}^2}{4}\right] \left[u_{12}^2 + \frac{(l_1+l_2+d-2)^2}{4}\right]\frac{(2l_1 + d - 2) (2l_2 + d - 2)}{(d-2)^2}\\
\times \sum_{m=0}^{\min (l_1,l_2)} a_{l_1,l_2,m} \Bigg| \frac{\Gamma\left(\frac{l_1+l_2-2m+d-2}{2} + \ii u_{12}\right)}{\Gamma\left(\frac{l_1+l_2-2m+2}{2}+\ii u_{12}\right)}\Bigg|^2  C_{l_1+l_2-2m}^{(\frac{d-2}{2})}(\cos \theta)
\label{I2L}
\end{multline}
with $x_3\cdot x_4 = |x_3||x_4|\cos \theta$.
Using the operator $\mathbb{O}_{l_1,l_2}$, this can actually be written in a more concise way:
\begin{multline}
I_{M,2}^{(d,\del)} = \frac{\Gamma\left(\frac{d}{2}\right)^2}{2(x_3^2 x_4^2)^{\tdel}}\sum_{l_1,l_2}
\int \frac{\text{d} u_1\text{d} u_2}{(2\pi)^2}\,
\left(\frac{x_3^2}{x_4^2}\right)^{\ii (u_1 + u_2)} [Q_{l_1}(u_1) Q_{l_2}(u_2)]^{M+2}\\
\times \left[u_{12}^2 + \frac{l_{12}^2}{4}\right] \left[u_{12}^2 + \frac{(l_1+l_2+d-2)^2}{4}\right]
\frac{(2l_1 + d - 2) (2l_2 + d - 2)}{(d-2)^2}\\
\times \Bigg| \frac{\Gamma\left(\frac{l_1+l_2+d-2}{2} + \ii u_{12}\right)}{\Gamma\left(\frac{l_1+l_2+2}{2}+\ii u_{12}\right)}\Bigg|^2  \frac{[\mathbb{O}_{l_1,l_2}(u_{21}) ^t\mathbb{O}_{l_1,l_2}(u_{12}) x_3^{\otimes (l_1+l_2)}]\cdot x_4^{\otimes (l_1+l_2)}}{(|x_3||x_4|)^{l_1+l_2} }\, .
\label{I2LO}
\end{multline}
Notice that since $\mathbb{O}_{l_1,l_2}$ goes from $\V_{l_1}\otimes \V_{l_2}$ to $S^{l_1+l_2}(\mathbb{C}^d)$ we can only multiply it with its transpose. This is what happens here where we need matrix elements of $\mathbb{O}_{l_1,l_2}(u_{21}) ^t\mathbb{O}_{l_1,l_2}(u_{12}):S^{l_1+l_2}(\mathbb{C}^d)\rightarrow S^{l_1+l_2}(\mathbb{C}^d)$.
It seems that expression \eqref{I2LO} is the most natural for the generalization
to the general case $I_{M,N}^{(d,\del)}$.

Let us compare the expressions \eqref{I2L} for integral $I_{M,2}^{(d,\del)}$ in various dimensions.
The limit $d\rightarrow2$ is seemingly singular but one should remember that the Gegenbauer polynomials for $l>0$ tend to $0$ in this limit so that
\beq
\forall l\in\mathbb{N},\quad\frac{2l + d - 2}{d-2} C_{l}^{(\frac{d-2}{2})}\left(\cos\theta\right) \underset{d\rightarrow 2}{\longrightarrow} \frac{2}{1+\delta_{l,0}}\cos l\theta\, .
\eeq
Thus, for $\min(l_1,l_2)>0$, one has
\begin{multline}
\frac{(2l_1 + d - 2)(2l_2 + d - 2)}{(d-2)^2} a_{l_1,l_2,m} C_{l_1+l_2-2m}^{(\frac{d-2}{2})}(\cos\theta)\underset{d\rightarrow 2}{\longrightarrow} \\
2\left[\delta_{m,0}\cos(l_1+l_2)\theta + \delta_{m,\min(l_1,l_2)}\cos(l_1 - l_2)\theta\right]\, .
\end{multline}
In the end, $I_{M,2}^{(2,\del)}$ is finite (as it should be) and, using the additional symmetry $Q_l = Q_{-l}$ valid for $l\in\mathbb{Z}$ when $d=2$, one can extend the sum to $(l_1,l_2)\in\mathbb{Z}^2$ so that
\beq
I_{M,2}^{(2,\del)} = \frac{1}{2(x_3^2 x_4^2)^{\tdel}}\sum_{(l_1,l_2)\in\mathbb{Z}^2} \e^{\ii(l_1 + l_2)\theta} \int \frac{\text{d} u_1\text{d} u_2}{(2\pi)^2} \left(\frac{x_3^2}{x_4^2}\right)^{\ii (u_1 + u_2)} [Q_{l_1}(u_1) Q_{l_2}(u_2)]^{M+2} \left[u_{12}^2 + \frac{l_{12}^2}{4}\right]\, .
\eeq

When $d=4$, the dependence on $u$ of the sum over $m$ disappears and the sum is then simply
\beq
C^{(1)}_{l_1}(\cos\theta) C^{(1)}_{l_2}(\cos\theta)= \frac{(\e^{\ii(l_1+1)\theta} - \e^{-\ii(l_1+1)\theta})(\e^{\ii(l_2+1)\theta} - \e^{-\ii(l_2+1)\theta})}{(\e^{\ii\theta} - \e^{-\ii\theta})^2}\, .
\eeq
Noticing that $Q_l = Q_{-l - 2}$ when $d=4$ we can keep only one of the four terms in the equation above if we extend the summation to $(l_1,l_2)\in\mathbb{Z}^2$. We thus recover the known formula \cite{Basso:2017jwq} (we have also replaced $l_j$ with $a_j = l_j + 1$):
\begin{multline}
I_{M,2}^{(4,\del)} = \frac{1}{2(x_3^2 x_4^2)^{\tdel}}\frac{1}{(\e^{\ii\theta} - \e^{-\ii\theta})^2}\sum_{(a_1,a_2)\in\mathbb{Z}^2} a_1 a_2 \e^{\ii(a_1+a_2)\theta} \\
\times\int \frac{\text{d} u_1\text{d} u_2}{(2\pi)^2} \left(\frac{x_3^2}{x_4^2}\right)^{\ii (u_1 + u_2)} [Q_{a_1-1}(u_1) Q_{a_2-1}(u_2)]^{M+2} \left[u_{12}^2 + \frac{a_{12}^2}{4}\right] \left[u_{12}^2 + \frac{(a_1+a_2)^2}{4}\right]\, .
\end{multline}

The next case we could investigate is $(d,\del)=(6,1)$, the formula \eqref{I2L} then reads
\begin{multline}
I_{M,2}^{(6,1)} = \frac{1}{2(x_3^2 x_4^2)^{2}}\sum_{l_1,l_2}
\int \frac{\text{d} u_1\text{d} u_2}{(2\pi)^2}\, \left(\frac{x_3^2}{x_4^2}\right)^{\ii (u_1 + u_2)} \frac{\left[u_{12}^2 + \frac{l_{12}^2}{4}\right] \left[u_{12}^2 + \frac{(l_1+l_2+4)^2}{4}\right]}{\left[\left(u_1^2 + \frac{(l_1+2)^2}{4}\right)\left(u_2^2 + \frac{(l_2+2)^2}{4}\right)\right]^{M+2}} (l_1+2)(l_2+2) \\
\sum_{m=0}^{\min (l_1,l_2)} \frac{(m+1)(l_1-m+1)(l_2-m+1)(l_1+l_2-m+3)}{(l_1+l_2 - 2m + 1)(l_1+l_2 - 2m + 3)}\,
\left[u_{12}^2 + \frac{(l_1+l_2-2m+2)^2}{4}\right]  C_{l_1+l_2-2m}^{(2)}(\cos \theta)\, .
\end{multline}
The integrals are rather easy, at least when $M$ is not too large,
but the sums seem to be quite tedious to perform.
We hope that this example of the first nontrivial integral
$I_{M,2}^{(d,\del)}$ clearly illustrates the complications
arising in higher dimensions.

\section{Conclusions}

In the present paper, we have constructed the generalised eigenvectors of the graph-building operator for fishnet integrals in $d$ dimensions.
The spectral decomposition of the graph-building operator allowed us to derive a representation for the $d$-dimensional Basso--Dixon diagrams in terms of separated variables, i.e. the rapidities $u_j$ and the bound-state numbers $l_j$ of the fishnet lattice's excitations. According to that, the expression for the Basso--Dixon diagram is an integral over separated variables---with the corresponding Sklyanin measure that we computed for any $d$ from the overlap of eigenvectors, and it reproduces the results of \cite{Basso:2017jwq,Derkachov2019,Derkachov:2019tzo} in two and four dimensions. The integrand is given by the eigenvalues of the graph-building operator and by the reductions of the bra and ket eigenvectors corresponding to pinching their external coordinates to two points $x$ and $y$. The former is factorised into $N$ contributions, each depending on the quantum numbers of one excitation $(u_j,l_j)$, the latter have a more complicated structure for general $d$. In $d=2$ and $d=4$, the eigenvectors are drastically simplified by the reduction, but, as we have demonstrated in the last section, the analogous expression for the general $d$-dimensional situation is more involved.

The present construction of eigenvectors is based on the symmetric
tensor representations of the group $O(d)$. As a consequence, the
corresponding main interchange relation governing the symmetry
of the eigenvectors involves $O(d)$-invariant R-matrix acting
on the tensor product of two symmetric tensor representations.
In this framework, the symmetry of eigenvectors and
scalar product look simple. On the other hand the special reduction of the eigenvector we would
need in order to write the Basso--Dixon diagrams in terms of known functions is quite complicated.

In Appendix \ref{app:spin_basis} we have discussed shortly the construction
of the eigenvectors based on spinor representations of
the group $O(d)$. In some sense these constructions are
complementary and show opposite features: in the spinorial framework the
symmetry properties and scalar products looks more complicated
but the special reductions of eigenvectors is straightforward. A detailed discussion of such duality,
comparison between the spinorial and tensorial frameworks, and
the derivation of the general expression for the reduced
eigenvectors is left to a future paper.

Our results constitute an important step for a formulation of integrability techniques for $n$-point functions  ($n\geqslant 3$) in conformal field theories in dimension $d\neq 4$. This claim is based on the fact that techniques of hexagonalisation \cite{basso2015structure,Eden:2016xvg,Basso_2019} developed in the $4$-dimensional $\mathcal{N}=4$ SYM theory took an important piece information from the knowledge of Fishnet integrals, and can even be derived from first principle for the strongly deformed (Fishnet) theory \cite{Olivucci:2021cfy,Olivucci_hex_II}. At the same time, nothing is known about similar techniques in other dimensions, with the exception of a worldsheet theory without a known field theory dual \cite{Eden:2021xhe}. For example, one can wonder if and how hexagon form factors and octagon functions \cite{Coronado_2019, Coronado_2020} can be computed in $\mathcal{N}=6$ ABJM theory: a rich piece of information would come from an explicit computation of BD diagrams in $3D$ together with the discovery of an analogue of its representation as a determinant of ladder integrals, following the observations in $d=2,4$ \cite{Derkachov2019, Basso:2021omx}.

\begin{acknowledgments}
\label{sec:acknowledgments}
G.F. thanks B.~Basso and V.~Kazakov for numerous fruitful discussions, and D.~Serban and M.~Staudacher for their many useful comments on an early version of the manuscript. The work of S.D. is supported by the Russian Science Foundation project No 19-11-00131.  Research at the Perimeter Institute is supported in part by the Government of Canada through NSERC and by the Province of Ontario through MRI. This work was additionally supported by a grant from the Simons Foundation (Simons Collaboration on the Nonperturbative Bootstrap).

\end{acknowledgments}

\appendix
\section{Basic Integral Relations}
\label{app:integral relations}

We collect in this appendix various formulae which are used
for the calculations of the Feynman diagrams \cite{DEramo:1971hnd,Vasiliev:1982dc,Vasiliev:1981dg,Kazakov:1983ns,
 Kazakov:1984km,Vasil'ev:2004}. We recall that for a complex number $a$ and an integer $l\geqslant 0$ we define
\beq
\tilde{a} = \frac{d}{2} - a\, ,\quad A_l(a) = \frac{\Gamma\left(\tilde{a} + \frac{l}{2}\right)}{\Gamma\left(a + \frac{l}{2}\right)} = \frac{1}{A_l(\tilde{a})}\, ,\quad (a)_l = \frac{\Gamma(a+l)}{\Gamma(a)} = \prod_{k=0}^{l-1} (a+k)\, .
\eeq
If $C$ is a symmetric traceless tensor of rank $l$, i.e. $C\in\V_l$, and $x\in\mathbb{R}^d$ we will also write
\beq
C(x) = C^{\mu_1\cdots\mu_l}x_{\mu_1}\dots x_{\mu_l}\, .
\eeq
Because $C$ is traceless the following two elementary but very useful properties hold for an arbitrary complex number $a$:
\beq\label{harmonic derivative}
\frac{C\left(\frac{x-y}{|x-y|}\right)}{(x-y)^{2a}} = \frac{C(\nabla_x)}{(-2)^l \left(a-\frac{l}{2}\right)_l} \frac{1}{(x-y)^{2\left(a-\frac{l}{2}\right)}}\, ,
\eeq
and
\beq\label{harmonic derivative bis}
\frac{C\left(\frac{x-x_0}{|x-x_0|} - \frac{y-x_0}{|y-x_0|}\right)}{(x-x_0)^{2\left(a - \frac{l}{2}\right)} (y-x_0)^{2\left(1 - a - \frac{l}{2}\right)}} = \frac{C\left(\nabla_{x_0}\right)}{(-2)^l \left(a-\frac{l}{2}\right)_l} \frac{1}{(x-x_0)^{2\left(a - \frac{l}{2}\right)} (y-x_0)^{2\left(1 - a - \frac{l}{2}\right)}}\, .
\eeq
The second property in particular implies that for arbitrary complex numbers $a$ and $b$ one has
\begin{multline}\label{useful property}
C\left(\nabla_{w_0}\right)\left[\frac{1}{(x-x_0)^{2b} (x-w_0)^{2\left(a - \frac{l}{2}\right)} (y-w_0)^{2\left(1 - a - \frac{l}{2}\right)}}\right]\Bigg|_{w_0 = x_0}\\
= \frac{\left(a-\frac{l}{2}\right)_l}{\left(a+b-\frac{l}{2}\right)_l} C\left(\nabla_{w_0}\right)\left[\frac{1}{(x-w_0)^{2\left(a + b - \frac{l}{2}\right)} (y-w_0)^{2\left(1 - a - b - \frac{l}{2}\right)} (y-x_0)^{2b} }\right]\Bigg|_{w_0 = x_0}\, .
\end{multline}

We also recall that, if $\Re(a)>0$, one can write
\beq
\frac{1}{x^{2a}} = \frac{1}{\Gamma(a)}\int_0^{+\infty}\e^{-u x^2}u^{a-1}\text{d} u\, .
\eeq

\paragraph{Fourier transform of a propagator} For $C$ a rank $l$ symmetric traceless tensor,
\beq
\int \frac{C\left(\frac{p}{|p|}\right)}{p^{2 a}} \e^{\ii p\cdot x} \frac{\text{d}^d p}{\pi^{\frac{d}{2}}} = A_l(a)\ii^l 4^{\tilde{a}} \frac{C\left(\frac{x}{|x|}\right)}{x^{2\tilde{a}}}\, .
\label{fourier}
\eeq

\paragraph{Chain relation}
\beq\label{chain relation}
\int\frac{\pi^{-\frac{d}{2}}\text{d}^d z}{(x-z)^{2a} (z-y)^{2b}} = \frac{A_0(a) A_0(b) A_0(d-a-b)}{(x-y)^{2\left(a+b-\frac{d}{2}\right)}}\, .
\eeq


When $a+b = \frac{d}{2}$ the chain relation becomes
\beq\label{chain relation delta}
\int\frac{\pi^{-\frac{d}{2}}\text{d}^d z}{(x-z)^{2a} (z-y)^{2(d-a)}} = A_0(a) A_0(d-a) \pi^{\frac{d}{2}} \delta(x-y)\, .
\eeq

\paragraph{Star-triangle relation} For $a+b+c= d$, one has

\beq\label{startriangle}
\int\frac{\pi^{-\frac{d}{2}}\text{d}^d w}{(w-x)^{2a}(w-y)^{2b}(w-z)^{2c}} = \frac{A_0(a) A_0(b) A_0(c)}{(x-y)^{2\tilde{c}}(y-z)^{2\tilde{a}}(z-x)^{2\tilde{b}}}\, .
\eeq

\begin{figure}[H]
\begin{center}
\includegraphics[scale=1.0]{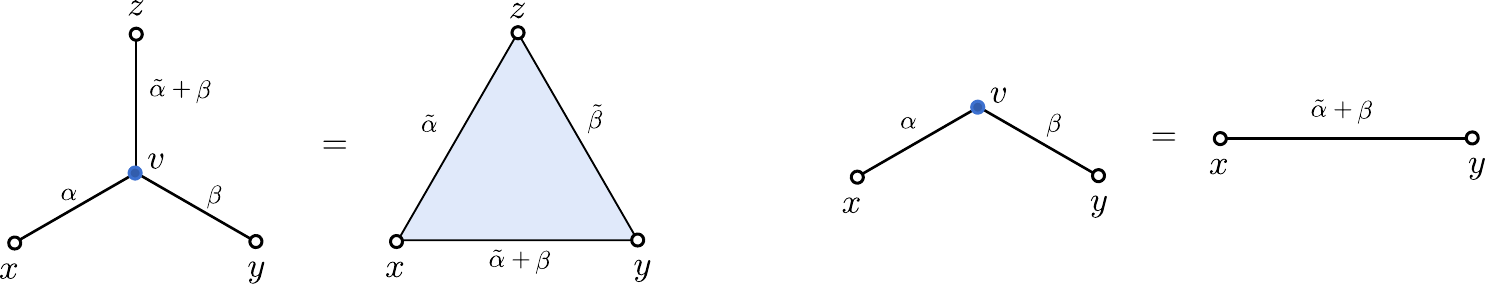}
\end{center}
\caption{\textbf{Left:} Star-triangle identity in graphical form. We recall that $\tilde{x}=\frac{d}{2}-x$. \textbf{Right:} Chain-rule identity in graphical form.}
\end{figure}
\begin{figure}[H]
\begin{center}
\includegraphics[scale=1.0]{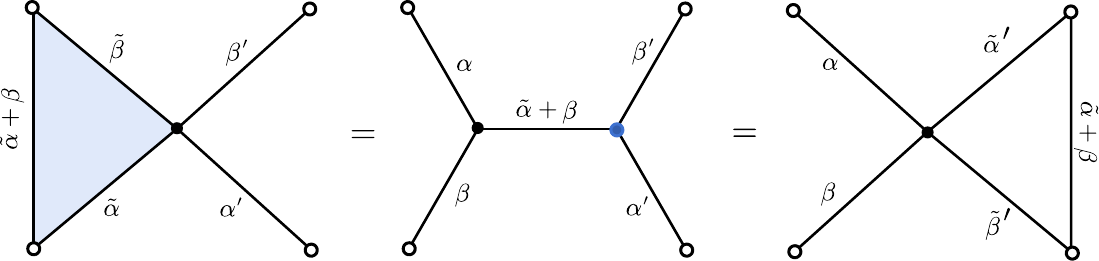}
\end{center}
\caption{The exchange relation corresponding to the notation of blue arrow used in the main text amounts to consecutive triangle-star and star-triangle transformations, and holds under the assumption $\alpha+\beta = \alpha'+\beta'$.}
\end{figure}
\paragraph{Generalization of the chain relation} For $C\in\V_l$, one has
\beq\label{chain harmonic}
\int\frac{C\left(\frac{x-z}{|x-z|}\right)}{(x-z)^{2a} (z-y)^{2b}} \frac{\text{d}^d z}{\pi^{\frac{d}{2}}} = A_l(a) A_0(b) A_l(d-a-b) \frac{C\left(\frac{x-y}{|x-y|}\right)}{(x-y)^{2\left(a+b-\frac{d}{2}\right)}}\, .
\eeq


\section{Equivalence \eqref{Rll general integral true} and \eqref{Rll general}}
\setcounter{equation}{0}
\label{app:main}

\subsection{Derivation of \eqref{v-integral}}

In this section we derive representation \eqref{v-integral} for the v-integral.
For simplicity we shall use the compact notation $\lambda = 1-\frac{l_1+l_2}{2}+\ii u$.
First step is the usual binomial expansion
\begin{multline}
\int \frac{\text{d}^d v}{\pi^{\frac{d}{2}}}
 \frac{(\zeta\cdot(y-\frac{v}{v^2}))^{l_1} (\eta\cdot(x-\frac{v}{v^2}))^{l_2}}
{(z-v)^{2\left(d-\lambda\right)}
v^{2\lambda}} =
\sum_{k,p}\binom{l_1}{k}\binom{l_2}{p}(-1)^{k+p}\\
(\zeta\cdot y)^{l_1-k}
(\eta\cdot x)^{l_2-p}\,
\int \frac{\text{d}^d v}{\pi^{\frac{d}{2}}}
 \frac{(\zeta\cdot\frac{v}{v^2})^{k} (\eta\cdot\frac{v}{v^2})^{p}}
{(z-v)^{2\left(d-\lambda\right)}
v^{2\lambda}}
\end{multline}
On the second step we use representation
\begin{align*}
\frac{(\zeta\cdot\frac{v}{v^2})^{k} (\eta\cdot\frac{v}{v^2})^{p}}
{v^{2\lambda}} = \frac{1}{2^{k+p}(\lambda)_{k+p}}\,\partial_t^k\partial_s^p\,
\frac{1}{(v^2-2t\zeta\cdot v - 2s\eta\cdot v )^{\lambda}}
\end{align*}
and series expansion ($\zeta^2=0$ and $\eta^2 =0$)
\begin{align*}
\frac{1}{(v^2-2t\zeta\cdot v - 2s\eta\cdot v )^{\lambda}} =
\frac{1}{\left((v-t\zeta - s\eta)^2 - 2ts \zeta\cdot\eta \right)^{\lambda}} =
\sum_n \frac{(\lambda)_n}{n!} \frac{(2ts \zeta\cdot\eta)^n}
{\left(v-t\zeta - s\eta\right)^{2(\lambda+n)}}
\end{align*}
to reduce our expression to the sum of the simpler integrals
\begin{multline}
\int \frac{\text{d}^d v}{\pi^{\frac{d}{2}}}
 \frac{(\zeta\cdot(y-\frac{v}{v^2}))^{l_1} (\eta\cdot(x-\frac{v}{v^2}))^{l_2}}
{(z-v)^{2\left(d-\lambda\right)}
v^{2\lambda}} =
\sum_{n,k,p} \binom{l_1}{k}\binom{l_2}{p}(-1)^{k+p}
\frac{2^{n-k-p}(\lambda)_n}{n!(\lambda)_{k+p}}(\zeta\cdot y)^{l_1-k}
(\eta\cdot x)^{l_2-p}(\zeta\cdot\eta)^n
\\
\,\partial_t^k\partial_s^p\, t^n s^n\,
\int \frac{\text{d}^d v}{\pi^{\frac{d}{2}}}
 \frac{1}
{(z-v)^{2\left(d-\lambda\right)}
\left(v-t\zeta - s\eta\right)^{2(\lambda+n)}}
\end{multline}
Next step we reduce remaining integral to the standard form
\begin{align*}
\int \frac{\text{d}^d v}{\pi^{\frac{d}{2}}}
 \frac{1}
{v^{2\left(d-\lambda\right)}
\left(v+z-t\zeta - s\eta\right)^{2\lambda}} =
\pi^{\frac{d}{2}}\frac{\Gamma\left(\lambda-\frac{d}{2}\right)
\Gamma\left(\frac{d}{2}-\lambda\right)}{\Gamma(d-\lambda)\Gamma(\lambda)}
\delta^{(d)}\left(z-t\zeta - s\eta\right)
\end{align*}
using external derivatives
\begin{align*}
\left(\partial_{z_{\mu}}\partial_{z_{\mu}}\right)^n \frac{1}{(z+z_0)^{2\lambda}} =
4^n(\lambda)_n\left(\lambda-\frac{d}{2}+1\right)_n \frac{1}{(z+z_0)^{2(\lambda+n)}}
\end{align*}
\begin{multline}
\int \frac{\text{d}^d v}{\pi^{\frac{d}{2}}}
 \frac{1}
{v^{2\left(d-\lambda\right)}
\left(v+z-t\zeta - s\eta\right)^{2(\lambda+n)}} = \frac{\left(\partial_{z_{\mu}}\partial_{z_{\mu}}\right)^n}{4^n(\lambda)_n\left(\lambda-\frac{d}{2}+1\right)_n}
\int \frac{\text{d}^d v}{\pi^{\frac{d}{2}}}
 \frac{1}
{v^{2\left(d-\lambda\right)}
\left(v+z-t\zeta - s\eta\right)^{2\lambda}} \\
= \frac{\pi^{\frac{d}{2}}\Gamma\left(\lambda-\frac{d}{2}\right)
\Gamma\left(\frac{d}{2}-\lambda\right)}
{4^n(\lambda)_n\left(\lambda-\frac{d}{2}+1\right)_n\Gamma(d-\lambda)\Gamma(\lambda)}
\left(\partial_{z_{\mu}}\partial_{z_{\mu}}\right)^n\,\delta^{(d)}\left(z-t\zeta - s\eta\right)
\end{multline}
We have
\begin{multline}
\int \frac{\text{d}^d v}{\pi^{\frac{d}{2}}}
 \frac{(\zeta\cdot(y-\frac{v}{v^2}))^{l_1} (\eta\cdot(x-\frac{v}{v^2}))^{l_2}}
{(z-v)^{2\left(d-\lambda\right)}
v^{2\lambda}} =
\sum_{n,k,p} \binom{l_1}{k}\binom{l_2}{p}(-1)^{k+p}
\frac{2^{-n-k-p}\pi^{\frac{d}{2}}\Gamma\left(\lambda-\frac{d}{2}\right)
\Gamma\left(\frac{d}{2}-\lambda\right)}
{n!(\lambda)_{k+p}\left(\lambda-\frac{d}{2}+1\right)_n\Gamma(d-\lambda)\Gamma(\lambda)}
\\
(\zeta\cdot y)^{l_1-k}
(\eta\cdot x)^{l_2-p}(\zeta\cdot\eta)^n\,
\partial_t^k\partial_s^p\, t^n s^n\,
\left(\partial_{z_{\mu}}\partial_{z_{\mu}}\right)^n\,
\delta^{(d)}\left(z-t\zeta - s\eta\right)
\end{multline}
The last transformation: using evident formula
\begin{align*}
\left.\partial_t^k\, t^n\,F(t)\right|_{t=0} = \frac{k!}{(k-n)!}\,
\left.\partial_t^{k-n}\,F(t)\right|_{t=0}
\end{align*}
 and similar ones for s-derivative and shifting summation indices $k \to k+n$ and $p \to p+n$
one obtains
\begin{multline}
\int \frac{\text{d}^d v}{\pi^{\frac{d}{2}}}
 \frac{(\zeta\cdot(y-\frac{v}{v^2}))^{l_1} (\eta\cdot(x-\frac{v}{v^2}))^{l_2}}
{(z-v)^{2\left(d-\lambda\right)}
v^{2\lambda}} = \\
\frac{\pi^{\frac{d}{2}}\Gamma\left(\lambda-\frac{d}{2}\right)
\Gamma\left(\frac{d}{2}-\lambda\right)}
{\Gamma(d-\lambda)\Gamma(\lambda)}
\sum_{n,k,p} \binom{l_1}{k+n}\binom{l_2}{p+n}\frac{(k+n)!(p+n)!}{k!p!}
\frac{2^{-3n-k-p}(-1)^{k+p}}
{n!(\lambda)_{k+p+2n}\left(\lambda-\frac{d}{2}+1\right)_n}
\\
(\zeta\cdot y)^{l_1-k-n}
(\eta\cdot x)^{l_2-p-n}(\zeta\cdot\eta)^n\,
\partial_t^k\partial_s^p\,
\left(\partial_{z_{\mu}}\partial_{z_{\mu}}\right)^n\,
\delta^{(d)}\left(z-t\zeta - s\eta\right)
\end{multline}
It is exactly expression \eqref{v-integral} and
\begin{align}\label{A}
A_{l_1,l_2}(u) = \frac{\pi^{\frac{d}{2}}\Gamma\left(\lambda-\frac{d}{2}\right)
\Gamma\left(\frac{d}{2}-\lambda\right)}
{\Gamma(d-\lambda)\Gamma(\lambda)}\ \ ;\ \ \lambda = 1-\frac{l_1+l_2}{2}+\ii u
\end{align}

\subsection{Equivalence}

Now we are going to calculate sum in the right hand side of \eqref{R-integral-sum} and
let us continue to use for simplicity notation $\lambda = 1-\frac{l_1+l_2}{2}+\ii u$.
We have
\begin{align*}
&F_{l_1,l_2}(u) = \frac{\Gamma\left(2-\lambda-l_1-l_2\right)
\Gamma\left(\lambda+l_1+l_2\right)
\Gamma\left(d-\lambda\right)}
{\Gamma\left(2-\lambda\right)
\Gamma\left(\frac{d}{2} -\lambda\right)
\Gamma\left(\lambda -\frac{d}{2}\right)}\ \ \,, \ \
A_{l_1,l_2}(u) = \frac{\pi^{\frac{d}{2}}\Gamma\left(\lambda-\frac{d}{2}\right)
\Gamma\left(\frac{d}{2}-\lambda\right)}
{\Gamma(d-\lambda)\Gamma(\lambda)}\,,\\
&\pi^{-\frac{d}{2}}F_{l_1,l_2}(u)A_{l_1,l_2}(u) =
\frac{\Gamma\left(2-\lambda-l_1-l_2\right)
\Gamma\left(\lambda+l_1+l_2\right)}
{\Gamma\left(2-\lambda\right)\Gamma(\lambda)} =
(-1)^{l_1+l_2}\frac{\lambda-1+l_1+l_2}{\lambda-1}
\end{align*}
so that
\begin{multline}\label{R12}
\left[\R_{l_1,l_2}(u)\zeta^{\otimes l_1}\otimes\eta^{\otimes l_2}\right]
\cdot x^{\otimes l_1}\otimes y^{\otimes l_2} =
\frac{\Gamma\left(2-\lambda-l_1-l_2\right)
\Gamma\left(\lambda+l_1+l_2\right)}
{\Gamma\left(2-\lambda\right)\Gamma(\lambda)} \\
\sum_{n,k,p}\frac{l_1! l_2! (-1)^{k+p}\, 2^{-k-p-3n}}
{(l_1-k-n)! (l_2-p-n)! k! p! n!}
\frac{(\zeta\cdot\eta)^n (\zeta\cdot y)^{l_1-k-n}
(\eta\cdot x)^{l_2-p-n}}{\left(\lambda\right)_{k+p+2n}
\left(\lambda-\frac{d}{2}+1\right)_n}\,\\
\partial_t^k\,\partial_s^p\,
\left.\left(\partial_{z_\mu}\partial_{z_\mu}\right)^n \,
\left(1-2z\cdot x+z^2x^2\right)^{1-\lambda-l_2}
\left(1-2z\cdot y+z^2y^2\right)^{1-\lambda-l_1}\right|_{z=t\zeta+s\eta}
\end{multline}
The first step is the calculation of the expression in the last line.
We introduce Schwinger parameters
\begin{multline*}
\left(1-2z\cdot x+z^2x^2\right)^{1-\lambda-l_2}
\left(1-2z\cdot y+z^2y^2\right)^{1-\lambda-l_1} = \\
\frac{1}{\Gamma(\lambda-1+l_2)\Gamma(\lambda-1+l_1)}
\int_0^{\infty} \text{d}\alpha \alpha^{\lambda-2+l_2}\,e^{-\alpha}
\int_0^{\infty} \text{d}\beta \beta^{\lambda-2+l_1}\,e^{-\beta}
\,e^{2z\cdot(\alpha x+\beta y)-
z^2(\alpha x^2+\beta y^2)}
\end{multline*}
and then calculate z-derivatives using formula
\begin{align}
e^{\gamma\partial_{z_\mu}\partial_{z_\mu}}\, e^{-b z^2+z\cdot c} =
\left(1+4\gamma b\right)^{-\frac{d}{2}}\,e^{\frac{\gamma c^2}{1+4\gamma b}}\,
e^{\frac{-b z^2+z\cdot c}{1+4\gamma b}}
\end{align}
This formula can be easily obtained using Gaussian integral and in our case
$b = \alpha x^2+\beta y^2$ and $c = 2(\alpha x+\beta y)$.
We have
\begin{multline*}
\left(\partial_{z_\mu}\partial_{z_\mu}\right)^n\, e^{-b z^2+z\cdot c} =
\partial_{\gamma}^n\,\left(1+4\gamma b\right)^{-\frac{d}{2}}\,
e^{\frac{\gamma c^2}{1+4\gamma b}}\,
e^{\frac{-b z^2+z\cdot c}{1+4\gamma b}} = \\
4^n b^n \partial_{\gamma}^n\,\left(1+\gamma\right)^{-\frac{d}{2}}\,
e^{\frac{\gamma}{1+\gamma}\frac{c^2}{4b}}\,
e^{\frac{-b z^2+z\cdot c}{1+\gamma}} =
\sum_{m} \frac{1}{m!}\partial_{\gamma}^n\, \gamma^m
\left(1+\gamma\right)^{-\frac{d}{2}-m} 4^{n-m} b^{n-m} \left(c^{2}\right)^m\,
e^{\frac{-b z^2+z\cdot c}{1+\gamma}}
\end{multline*}
Now it is possible to substitute $z=t\zeta+s\eta$ so that
$z^2 = 2 t s \zeta\cdot\eta$ and $z\cdot c = t \zeta\cdot c + s \eta\cdot c $
and calculate t- and s-derivatives:
\begin{multline*}
\partial_t^k\,\partial_s^p\,
e^{\frac{- 2t s b \zeta\cdot\eta+t \zeta\cdot c + s \eta\cdot c}{1+\gamma}} =
(1+\gamma)^{-k} \partial_s^p\, \left(- 2 s b \zeta\cdot\eta + \zeta\cdot c\right)^k
e^{\frac{s \eta\cdot c}{1+\gamma}} = \\
(1+\gamma)^{-k}
\sum_{m_1} \binom{k}{m_1}\frac{p!}{(p-m_1)!} \left(- 2 b \zeta\cdot\eta\right)^{m_1}
\left(\zeta\cdot c\right)^{k-m_1} \partial_s^p\, s^{m_1}\, e^{\frac{s \eta\cdot c}{1+\gamma}} = \\
\sum_{m_1} \binom{k}{m_1} \frac{p!}{(p-m_1)!} (1+\gamma)^{-k-p+m_1}
\left(- 2 b \zeta\cdot\eta\right)^{m_1}
\left(\zeta\cdot c\right)^{k-m_1} \left(\eta\cdot c\right)^{p-m_1}
\end{multline*}
Using formula for $\gamma$-derivative
\begin{align*}
\partial_{\gamma}^n\,\gamma^m\,\left(1+\gamma\right)^{-\frac{d}{2}-m-k-p+m_1} =  (-1)^{n-m}\frac{n!\left(\frac{d}{2}+m+k+p-m_1\right)_{n-m}}{(n-m)!}
\end{align*}
and collecting all terms we obtain
\begin{multline*}
\partial_t^k\,\partial_s^p\,
\left.\left(\partial_{z_\mu}\partial_{z_\mu}\right)^n
\, e^{-b z^2+z\cdot c}\right|_{z=t\zeta+s\eta} =
\sum_{m,m_1} (-1)^{n-m+m_1}\frac{n!k!p!\left(\frac{d}{2}+m+k+p-m_1\right)_{n-m}}
{m!m_1!(n-m)!(k-m_1)!(p-m_1)!}\,4^{n-m}2^{m_1}\,\\
b^{n-m+m_1} \left(c^{2}\right)^m\,
\left(\zeta\cdot\eta\right)^{m_1}
\left(\zeta\cdot c\right)^{k-m_1} \left(\eta\cdot c\right)^{p-m_1}
\end{multline*}
Let us return to our calculation. We have
\begin{multline*}
\partial_t^k\,\partial_s^p\,
\left.\left(\partial_{z_\mu}\partial_{z_\mu}\right)^n
\, e^{2z\cdot(\alpha x+\beta y)-
z^2(\alpha x^2+\beta y^2)}\right|_{z=t\zeta+s\eta} = \\
\sum_{m,m_1} (-1)^{n-m+m_1}\frac{n!k!p!\left(\frac{d}{2}+m+k+p-m_1\right)_{n-m}}
{m!m_1!(n-m)!(k-m_1)!(p-m_1)!}\,4^{n}2^{k+p-m_1}\,\left(\zeta\cdot\eta\right)^{m_1}\\
\left(\alpha x^2+\beta y^2\right)^{n-m+m_1}
\left(\alpha^2x^2 +2\alpha\beta x\cdot y +\beta^2 y^2\right)^m\,
\left(\alpha\zeta\cdot x+\beta\zeta\cdot y\right)^{k-m_1}
\left(\alpha\eta\cdot x+\beta\eta\cdot y\right)^{p-m_1}
\end{multline*}
so that it remains to use binomial expansions
\begin{align*}
\left(\alpha x^2+\beta y^2\right)^{n-m+m_1} =
\sum_{k_1}\binom{n-m+m_1}{k_1} \alpha^{k_1}\beta^{n-m+m_1-k_1}
\,(x^2)^{k_1}(y^2)^{n-m+m_1-k_1}\,,\\
\left(\alpha\zeta\cdot x+\beta\zeta\cdot y\right)^{k-m_1} =
\sum_{k_2}\binom{k-m_1}{k_2} \alpha^{k_2}\beta^{k-m_1-k_2}
\,(\zeta\cdot x)^{k_2}(\zeta\cdot y)^{k-m_1-k_2}\,,\\
\left(\alpha\eta\cdot x+\beta\eta\cdot y\right)^{p-m_1} =
\sum_{k_3}\binom{p-m_1}{k_3} \alpha^{k_3}\beta^{p-m_1-k_3}
\,(\eta\cdot x)^{k_3}(\eta\cdot y)^{p-m_1-k_3}\,,\\
\left(\alpha^2x^2 +2\alpha\beta x\cdot y +\beta^2 y^2\right)^m =
\sum_{s_1,s_2}\frac{2^{s_2}m!\,
\alpha^{2s_1+s_2}\beta^{2m-2s_1-s_2}}{(m-s_1-s_2)!s_1!s_2!}\,
(x^2)^{s_1}\,(x\cdot y)^{s_2}\,(y^2)^{m-s_1-s_2}
\end{align*}
and then calculate $\alpha$- and $\beta$-integrals
\begin{align*}
&\frac{1}{\Gamma(\lambda-1+l_2)}
\int_0^{\infty} \text{d}\alpha \alpha^{\lambda-2+l_2}\,\alpha^{L}\,e^{-\alpha}  =
\left(\lambda-1+l_2\right)_{L}
\\
&\frac{1}{\Gamma(\lambda-1+l_1)}
\int_0^{\infty} \text{d}\beta \beta^{\lambda-2+l_1}\,
\beta^{n+k+p+m-L-m_1}e^{-\beta} =
\left(\lambda-1+l_1\right)_{n+k+p+m-L-m_1}
\end{align*}
where for simplicity we denote $L = k_1+k_2+k_3+2s_1+s_2$.
Collecting all pieces together we obtain the following intermediate result
\begin{multline*}
\partial_t^k\,\partial_s^p\,
\left.\left(\partial_{z_\mu}\partial_{z_\mu}\right)^n \,
\left(1-2z\cdot x+z^2x^2\right)^{1-\lambda-l_2}
\left(1-2z\cdot y+z^2y^2\right)^{1-\lambda-l_1}\right|_{z=t\zeta+s\eta} = \\
\sum \frac{2^{2n+k+p-m_1+s_2}\,(-1)^{n-m+m_1}\,n!k!p!(n-m+m_1)!}
{m_1!(n-m)!(n-m+m_1-k_1)!(k-m_1-k_2)!(p-m_1-k_3)!(m-s_1-s_2)!k_1!k_2!k_3!s_1!s_2!}\\
\left(\frac{d}{2}+m+k+p-m_1\right)_{n-m}\left(\lambda-1+l_2\right)_{L}
\left(\lambda-1+l_1\right)_{n+k+p+m-L-m_1}\\
(x^2)^{k_1+s_1}(y^2)^{n+m_1-k_1-s_1-s_2}
\left(\zeta\cdot\eta\right)^{m_1}(x\cdot y)^{s_2}
(\zeta\cdot x)^{k_2}(\zeta\cdot y)^{k-m_1-k_2}
(\eta\cdot x)^{k_3}(\eta\cdot y)^{p-m_1-k_3}
\end{multline*}
where summation is performed over $m,m_1,k_1,k_2,k_3,s_1,s_2$ and
for simplicity we do not show it explicitly.
Substitution of this expression in \eqref{R12} gives
\begin{multline*}
\left[\R_{l_1,l_2}(u)\zeta^{\otimes l_1}\otimes\eta^{\otimes l_2}\right]
\cdot x^{\otimes l_1}\otimes y^{\otimes l_2} =
\frac{\Gamma\left(2-\lambda-l_1-l_2\right)
\Gamma\left(\lambda+l_1+l_2\right)}
{\Gamma\left(2-\lambda\right)\Gamma(\lambda)} \\
\sum_{n,k,p}\frac{l_1! l_2! (-1)^{k+p}}
{(l_1-k-n)! (l_2-p-n)!}
\frac{1}{\left(\lambda\right)_{k+p+2n}
\left(\lambda-\frac{d}{2}+1\right)_n}\,\\
\sum \frac{(-1)^{n-m+m_1}\,(n-m+m_1)!}
{m_1!(n-m)!(n-m+m_1-k_1)!(k-m_1-k_2)!(p-m_1-k_3)!(m-s_1-s_2)!k_1!k_2!k_3!s_1!s_2!}\\
\left(\frac{d}{2}+m+k+p-m_1\right)_{n-m}\left(\lambda-1+l_2\right)_{L}\,
\left(\lambda-1+l_1\right)_{n+k+p+m-L-m_1}\\
\left(\frac{x^2}{2}\right)^{k_1+s_1}
\left(\frac{y^2}{2}\right)^{n+m_1-k_1-s_1-s_2}
\left(\zeta\cdot\eta\right)^{n+m_1}(x\cdot y)^{s_2}
(\zeta\cdot x)^{k_2}(\zeta\cdot y)^{l_1-n-m_1-k_2}
(\eta\cdot x)^{l_2-p-n+k_3}(\eta\cdot y)^{p-m_1-k_3}
\end{multline*}
Now we are going to show that some sequence of resummations allows to
transform this expression to the form \eqref{Rll general}.
We shall use two variants of Gauss summation formula
\begin{align}
\label{Gauss1}
\sum_{k}\binom{l}{k}(-1)^k \frac{\Gamma(A+k)}{\Gamma(B+k)} & =
\frac{\Gamma(A)\Gamma(B-A+l)}{\Gamma(B-A)\Gamma(B+l)}
\\
\label{Gauss2}
\sum_{k}\binom{l}{k} \frac{1}{\Gamma(A+k)\Gamma(B-k)} & =
\frac{\Gamma(A+B+l-1)}{\Gamma(B)\Gamma(A+B-1)\Gamma(A+l)}
\end{align}
The first step: transformation $p \to p+m_1\,, k \to k+m_1$ and $n \to n-m_1$
\begin{multline*}
\sum\frac{l_1! l_2! (-1)^{k+p}}
{(l_1-k-n)! (l_2-p-n)!}
\frac{1}{\left(\lambda\right)_{k+p+2n}
\left(\lambda-\frac{d}{2}+1\right)_{n-m_1}}\,\\
\frac{(-1)^{n-m}\,(n-m)!}
{m_1!(n-m-m_1)!(n-m-k_1)!(k-k_2)!(p-k_3)!(m-s_1-s_2)!k_1!k_2!k_3!s_1!s_2!}\\
\left(\frac{d}{2}+m+k+p+m_1\right)_{n-m-m_1}\left(\lambda-1+l_2\right)_{L}
\left(\lambda-1+l_1\right)_{n+k+p+m-L}\\
\left(\frac{x^2}{2}\right)^{k_1+s_1}
\left(\frac{y^2}{2}\right)^{n-k_1-s_1-s_2}
\left(\zeta\cdot\eta\right)^{n}(x\cdot y)^{s_2}
(\zeta\cdot x)^{k_2}(\zeta\cdot y)^{l_1-n-k_2}
(\eta\cdot x)^{l_2-p-n+k_3}(\eta\cdot y)^{p-k_3}
\end{multline*}
and then summation over $m_1$ using \eqref{Gauss2}
\begin{align*}
\sum_{m_1} \frac{(n-m)!}
{m_1!(n-m-m_1)!}
\frac{\left(\frac{d}{2}+m+k+p+m_1\right)_{n-m-m_1}}
{\left(\lambda-\frac{d}{2}+1\right)_{n-m_1}} =
\frac{(\lambda)_{k+p+2n}}{(\lambda)_{k+p+n+m}\left(\lambda-\frac{d}{2}+1\right)_{n}}
\end{align*}
leads to expression
\begin{multline*}
\sum\frac{l_1! l_2! (-1)^{k+p}}
{(l_1-k-n)! (l_2-p-n)!}
\frac{1}{\left(\lambda\right)_{k+p+n+m}
\left(\lambda-\frac{d}{2}+1\right)_{n}}\,\\
\frac{(-1)^{n-m}}
{(n-m-k_1)!(k-k_2)!(p-k_3)!(m-s_1-s_2)!k_1!k_2!k_3!s_1!s_2!}\\
\left(\lambda-1+l_2\right)_{L}
\left(\lambda-1+l_1\right)_{n+k+p+m-L}\\
\left(\frac{x^2}{2}\right)^{k_1+s_1}
\left(\frac{y^2}{2}\right)^{n-k_1-s_1-s_2}
\left(\zeta\cdot\eta\right)^{n}(x\cdot y)^{s_2}
(\zeta\cdot x)^{k_2}(\zeta\cdot y)^{l_1-n-k_2}
(\eta\cdot x)^{l_2-p-n+k_3}(\eta\cdot y)^{p-k_3}
\end{multline*}
Now it is possible to perform summation over $m$ using \eqref{Gauss1}
\begin{multline*}
\sum_{m} \frac{(-1)^m \left(\lambda-1+l_1\right)_{n+k+p+m-L}}
{(n-m-k_1)!(m-s_1-s_2)!\left(\lambda\right)_{k+p+n+m}}
= \\
\frac{(-1)^{s_1+s_2}(k_2+k_3+n+s_1-l_1)!\left(\lambda-1+l_1\right)_{n+k+p-k_1-k_2-k_3-s_1}}
{(n-k_1-s_1-s_2)!(L-l_1)!\left(\lambda\right)_{k+p+2n-k_1}}
\end{multline*}
so that one obtains
\begin{multline*}
\sum\frac{l_1! l_2! (-1)^{k+p}}
{(l_1-k-n)! (l_2-p-n)!}
\frac{1}{\left(\lambda\right)_{k+p+2n-k_1}
\left(\lambda-\frac{d}{2}+1\right)_{n}}\,\\
\frac{(-1)^{n+s_1+s_2}(k_2+k_3+n+s_1-l_1)!}
{(n-k_1-s_1-s_2)!(L-l_1)!(k-k_2)!(p-k_3)!k_1!k_2!k_3!s_1!s_2!}\\
\left(\lambda-1+l_2\right)_{L}
\left(\lambda-1+l_1\right)_{n+k+p-k_1-k_2-k_3-s_1}\\
\left(\frac{x^2}{2}\right)^{k_1+s_1}
\left(\frac{y^2}{2}\right)^{n-k_1-s_1-s_2}
\left(\zeta\cdot\eta\right)^{n}(x\cdot y)^{s_2}
(\zeta\cdot x)^{k_2}(\zeta\cdot y)^{l_1-n-k_2}
(\eta\cdot x)^{l_2-p-n+k_3}(\eta\cdot y)^{p-k_3}
\end{multline*}
The second step: transformation $k_1 \to k_1-s_1$ and $p \to p+k_3$
\begin{multline*}
\sum\frac{l_1! l_2! (-1)^{k+p+k_3}}
{(l_1-k-n)! (l_2-p-n-k_3)!}
\frac{1}{\left(\lambda\right)_{k+p+2n-k_1+s_1+k_3}
\left(\lambda-\frac{d}{2}+1\right)_{n}}\,\\
\frac{(-1)^{n+s_1+s_2}(k_2+k_3+n+s_1-l_1)!}
{(n-k_1-s_2)!(k_1+k_2+k_3+s_1+s_2-l_1)!(k-k_2)!p!(k_1-s_1)!k_2!k_3!s_1!s_2!}\\
\left(\lambda-1+l_2\right)_{k_1+k_2+k_3+s_1+s_2}
\left(\lambda-1+l_1\right)_{n+k+p-k_1-k_2}\\
\left(\frac{x^2}{2}\right)^{k_1}
\left(\frac{y^2}{2}\right)^{n-k_1-s_2}
\left(\zeta\cdot\eta\right)^{n}(x\cdot y)^{s_2}
(\zeta\cdot x)^{k_2}(\zeta\cdot y)^{l_1-n-k_2}
(\eta\cdot x)^{l_2-p-n}(\eta\cdot y)^{p}
\end{multline*}
then transformation $k_3\to k_3-s_1$
\begin{multline*}
\sum\frac{l_1! l_2! (-1)^{k+p+k_3}}
{(l_1-k-n)! (l_2-p-n-k_3+s_1)!}
\frac{1}{\left(\lambda\right)_{k+p+2n-k_1+k_3}
\left(\lambda-\frac{d}{2}+1\right)_{n}}\,\\
\frac{(-1)^{n+s_2}(k_2+k_3+n-l_1)!}
{(n-k_1-s_2)!(k_1+k_2+k_3+s_2-l_1)!(k-k_2)!p!(k_1-s_1)!k_2!(k_3-s_1)!s_1!s_2!}\\
\left(\lambda-1+l_2\right)_{k_1+k_2+k_3+s_2}
\left(\lambda-1+l_1\right)_{n+k+p-k_1-k_2}\\
\left(\frac{x^2}{2}\right)^{k_1}
\left(\frac{y^2}{2}\right)^{n-k_1-s_2}
\left(\zeta\cdot\eta\right)^{n}(x\cdot y)^{s_2}
(\zeta\cdot x)^{k_2}(\zeta\cdot y)^{l_1-n-k_2}
(\eta\cdot x)^{l_2-p-n}(\eta\cdot y)^{p}
\end{multline*}
and summation over $s_1$ using \eqref{Gauss2}
\begin{align*}
\sum_{s_1}\frac{1}{s_1!(k_1-s_1)!}\frac{1}{(l_2-p-n-k_3+s_1)!(k_3-s_1)!} = \\
\frac{(l_2-p-n+k_1)!}{k_1!k_3!(l_2-p-n+k_1-k_3)!(l_2-p-n)!}
\end{align*}
finally gives
\begin{multline*}
\sum \frac{l_1! l_2! (-1)^{k+p+k_3}}
{(l_1-k-n)! (l_2-p-n)!}
\frac{1}{\left(\lambda\right)_{k+p+2n-k_1+k_3}
\left(\lambda-\frac{d}{2}+1\right)_{n}}\,\\
\frac{(-1)^{n+s_2}(k_2+k_3+n-l_1)!(l_2-p-n+k_1)!}
{(l_2-p-n+k_1-k_3)!(n-k_1-s_2)!(k_1+k_2+k_3+s_2-l_1)!(k-k_2)!p!k_1!k_2!k_3!s_2!}\\
\left(\lambda-1+l_2\right)_{k_1+k_2+k_3+s_2}
\left(\lambda-1+l_1\right)_{n+k+p-k_1-k_2}\\
\left(\frac{x^2}{2}\right)^{k_1}
\left(\frac{y^2}{2}\right)^{n-k_1-s_2}
\left(\zeta\cdot\eta\right)^{n}(x\cdot y)^{s_2}
(\zeta\cdot x)^{k_2}(\zeta\cdot y)^{l_1-n-k_2}
(\eta\cdot x)^{l_2-p-n}(\eta\cdot y)^{p}
\end{multline*}
Now it is possible to perform summation over $k$ using \eqref{Gauss1}
\begin{multline*}
\sum_{k} \frac{(-1)^{k}}
{(l_1-k-n)! (k-k_2)!} \frac{\left(\lambda-1+l_1\right)_{n+k+p-k_1-k_2}}
{\left(\lambda\right)_{k+p+2n-k_1+k_3}} = \\
\frac{(-1)^{k_2}k_3!\left(\lambda-1+l_1\right)_{n+p-k_1}}
{(l_1-k_2-n)!(n+k_2+k_3-l_1)!\left(\lambda\right)_{p+n+l_1+k_3-k_1}}
\end{multline*}
so that we obtain
\begin{multline*}
\sum \frac{l_1! l_2! (-1)^{k_2+p+k_3}}
{(l_1-k_2-n)! (l_2-p-n)!}
\frac{1}{\left(\lambda\right)_{p+n+l_1+k_3-k_1}
\left(\lambda-\frac{d}{2}+1\right)_{n}}\,\\
\frac{(-1)^{n+s_2}(l_2-p-n+k_1)!}
{(l_2-p-n+k_1-k_3)!(n-k_1-s_2)!(k_1+k_2+k_3+s_2-l_1)!p!k_1!k_2!s_2!}\\
\left(\lambda-1+l_2\right)_{k_1+k_2+k_3+s_2}
\left(\lambda-1+l_1\right)_{n+p-k_1}\\
\left(\frac{x^2}{2}\right)^{k_1}
\left(\frac{y^2}{2}\right)^{n-k_1-s_2}
\left(\zeta\cdot\eta\right)^{n}(x\cdot y)^{s_2}
(\zeta\cdot x)^{k_2}(\zeta\cdot y)^{l_1-n-k_2}
(\eta\cdot x)^{l_2-p-n}(\eta\cdot y)^{p}
\end{multline*}
and arrive to the last step -- summation over $k_3$ using \eqref{Gauss1}
\begin{multline*}
\sum_{k_3} \frac{(-1)^{k_3}}
{(l_2-p-n+k_1-k_3)!(k_1+k_2+k_3+s_2-l_1)!}
\frac{\left(\lambda-1+l_2\right)_{k_1+k_2+k_3+s_2}}
{\left(\lambda\right)_{p+n+l_1+k_3-k_1}} = \\
\frac{\Gamma(\lambda)\Gamma(\lambda-1+l_1+l_2)}
{\Gamma(\lambda-1+l_2)\Gamma(\lambda+l_1+l_2)}
\frac{(-1)^{k_1+k_2+s_2-l_1}}
{(l_1-l_2-2k_1-k_2-s_2+p+n)!(l_2-l_1+2k_1+k_2+s_2-p-n)!}
\end{multline*}
We see that this summation results in the key restriction
$s_2 +2k_1+k_2+l_2 = l_1+p+n$ which fixes right
homogeneity properties of our polynomial as function $x$ and $y$.
The sum now is over four indices and it is easy to check
that after appropriate redefinition of summation
variables one obtains the expression \eqref{Rll general} exactly.

\section{Equivalence \eqref{Rll} and \eqref{Rll integral}}
\setcounter{equation}{0}
\label{app:Symanzik}

When $x$ and $y$ are null vectors, the Symanzik trick allows to reduce representation \eqref{Rll general integral} to the simpler form
\begin{multline}
\left[\R_{l_1,l_2}(u)\zeta^{\otimes l_1}\otimes\eta^{\otimes l_2}\right]\cdot(x^{\otimes l_1}\otimes y^{\otimes l_2}) = \frac{\Gamma\left(\frac{d}{2}+l_1-1\right) \Gamma\left(\frac{d}{2}+l_2-1\right) \left(\ii u + \frac{l_1+l_2}{2}\right) \Gamma\left(\ii u - \frac{l_1+l_2}{2}\right)}{ \Gamma\left(-\ii u + \frac{d-2+l_1+l_2}{2}\right) \Gamma\left(\ii u + \frac{l_1 - l_2}{2}\right) \Gamma\left(\ii u + \frac{l_2 - l_1}{2}\right)}\\
\times (x-y)^{2\left(-\ii u + \frac{d+l_1+l_2-2}{2}\right)}\int \frac{(\zeta\cdot(v-y))^{l_1} (\eta\cdot(v-x))^{l_2}}{v^{2\left(1-\ii u - \frac{l_1+l_2}{2}\right)} (y-v)^{2\left(\frac{d}{2}+l_1 - 1\right)} (x-v)^{2\left(\frac{d}{2}+l_2 - 1\right)}} \frac{\text{d}^d v}{\pi^{\frac{d}{2}}}\, .
\end{multline}
This integral is perfectly well-defined. After having stripped the right-hand side ($RHS$) of the $(x,y)$-independent prefactor it can be represented as (it is important to notice that one can only impose $x^2 = y^2 = 0$ after having taken the derivatives):
\begin{align*}
&RHS = (x-y)^{2\left(-\ii u + \frac{d+l_1+l_2-2}{2}\right)} \frac{(\zeta\cdot\nabla_y)^{l_1} (\eta\cdot\nabla_x)^{l_2}}{2^{l_1+l_2}\left(\frac{d}{2}-1\right)_{l_1} \left(\frac{d}{2}-1\right)_{l_2}} \int \frac{v^{2\left(\ii u + \frac{l_1+l_2}{2} - 1\right)}}{(x-v)^{2\left(\frac{d}{2} - 1\right)} (y-v)^{2\left(\frac{d}{2}- 1\right)}} \frac{\text{d}^d v}{\pi^{\frac{d}{2}}}\\
& = (x-y)^{2\left(-\ii u + \frac{d+l_1+l_2-2}{2}\right)} \frac{\Gamma\left(- \ii u + \frac{d-2-l_1-l_2}{2}\right)(\zeta\cdot\nabla_y)^{l_1} (\eta\cdot\nabla_x)^{l_2}}{2^{l_1+l_2}\Gamma\left(\frac{d}{2}+l_1-1\right) \Gamma\left(\frac{d}{2}+l_2-1\right)\Gamma\left(1-\ii u-\frac{l_1+l_2}{2}\right)}\\
& \int_{[0,1]^3} \alpha_1^{-\ii u-\frac{l_1+l_2}{2}}\left(\alpha_2\alpha_3\right)^{\frac{d}{2} - 2}\left[\alpha_1\alpha_2 y^2 + \alpha_1\alpha_3 x^2 + \alpha_2\alpha_3 (x-y)^2\right]^{\ii u+\frac{2+l_1+l_2-d}{2}}
\delta\left(1-\sum_{k=1}^3\alpha_{k}\right)\prod_{k=1}^3\text{d}\alpha_{k}\\
& = \frac{\Gamma\left(- \ii u + \frac{d-2-l_1-l_2}{2}\right)}{\Gamma\left(\frac{d}{2}+l_1-1\right) \Gamma\left(\frac{d}{2}+l_2-1\right)\Gamma\left(1-\ii u-\frac{l_1+l_2}{2}\right)}\sum_{k}\, \frac{l_1! l_2!}{k! (l_1-k)! (l_2-k)!} \frac{(x-y)^{2k}}{(-2)^k}\\
&\times(\zeta\cdot\eta)^k \frac{\Gamma\left(\ii u + \frac{4-d+l_1+l_2}{2}\right)}{\Gamma\left(\ii u + \frac{4-d-l_1-l_2}{2} + k\right)} \int_{[0,1]^3} (\zeta\cdot(\alpha_1 y+ \alpha_3 (y-x)))^{l_1-k} (\eta\cdot(\alpha_1 x + \alpha_2 (x-y)))^{l_2-k}\\
&\qquad\qquad\qquad\qquad\times \alpha_1^{-\ii u-\frac{l_1+l_2}{2}} \alpha_2^{\ii u + \frac{l_1-l_2}{2}+k-1} \alpha_3^{\ii u + \frac{l_2-l_1}{2}+k-1}\,\delta\left(1-\sum_{k=1}^3\alpha_{k}\right)\prod_{k=1}^3\text{d}\alpha_{k}\, .
\end{align*}
Since $\alpha_1+\alpha_2+\alpha_3 = 1$ one can write
\begin{align*}
(\zeta\cdot(\alpha_1 y+ \alpha_3 (y-x)))^{l_1-k} = \sum_{m=0}^{l_1-k} \binom{l_1-k}{m}((1-\alpha_2)\zeta\cdot y)^m (-\alpha_3\zeta\cdot x)^{l_1-k-m}\,,
\\
(\eta\cdot(\alpha_1 x + \alpha_2 (x-y)))^{l_2-k} = \sum_{n=0}^{l_2-k} \binom{l_2-k}{n} ((1-\alpha_3)\eta\cdot x)^n (-\alpha_2\eta\cdot y)^{l_2-k-n}\, .
\end{align*}
The integral that then appears is of the form
\begin{align*}
&\int_{[0,1]^3} \alpha_1^{a-1} \alpha_2^{b-1} \alpha_3^{c-1} (1-\alpha_2)^m (1-\alpha_3)^n \,\delta(1-\alpha_1-\alpha_2-\alpha_3)\text{d}\alpha_1\text{d}\alpha_2\text{d}\alpha_3\\
&= \Gamma(a)\sum_{p=0}^m \sum_{q=0}^n \binom{m}{p} \binom{n}{q} (-1)^{p+q} \frac{\Gamma(b+p)\Gamma(c+q)}{\Gamma(a+b+c+p+q)}\\
&= \Gamma(a)\Gamma(c)\sum_{p=0}^m  \binom{m}{p}(-1)^p \frac{\Gamma(b+p) (a+b+p)_n}{\Gamma(a+b+c+n+p)}
= \Gamma(a)\Gamma(b)\sum_{q=0}^n  \binom{n}{q}(-1)^q \frac{\Gamma(c+q) (a+c+q)_m}{\Gamma(a+b+c+m+q)}
\end{align*}
where we used the Gauss identity \eqref{Gauss hypergeometric} in the form $\sum_{l=0}^r \binom{r}{l} (-1)^l \frac{\Gamma(A+l)}{\Gamma(B+l)} = \frac{\Gamma(A)}{\Gamma(B+r)} (B-A)_{r}$. In our case the parameters actually are $a = -\ii u + \frac{2-l_1-l_2}{2}$, $b=\ii u + \frac{l_1+l_2}{2} - n$ and $c=\ii u + \frac{l_1+l_2}{2} - m$. In particular $a+b = 1-n$ so that $(a+b+p)_n = (1+p-n)_n = 0$ unless $p\geqslant n$ and one of the formulas above for the integral shows that it vanishes unless $m\geqslant n$. Similarly $a+c = 1-m$ so that we also need $n\geqslant m$. In the end, the integral is
\begin{multline}
\int_{[0,1]^3} \alpha_1^{-\ii u - \frac{l_1+l_2}{2}} \alpha_2^{\ii u + \frac{l_1+l_2}{2} - n - 1} \alpha_3^{\ii u + \frac{l_1+l_2}{2} - m - 1} (1-\alpha_2)^m (1-\alpha_3)^n \,\delta\left(1-\sum_{k=1}^3\alpha_{k}\right)\prod_{k=1}^3\text{d}\alpha_{k}\\
= \delta_{m,n} n! \frac{\Gamma\left(-\ii u + \frac{2-l_1-l_2}{2}\right)\Gamma\left(\ii u + \frac{l_1+l_2}{2}\right)}{\left(\ii u + \frac{l_1+l_2}{2}\right)\left(-\ii u + \frac{2-l_1-l_2}{2}\right)_n}\, .
\end{multline}
Putting everything together yields (we also use $(x-y)^2 = -2x\cdot y$)
\begin{align*}
RHS = \frac{(-1)^{l_1+l_2} \Gamma\left(- \ii u + \frac{d-2-l_1-l_2}{2}\right) \Gamma\left(\ii u + \frac{l_1+l_2}{2}\right)}{\Gamma\left(\frac{d}{2}+l_1-1\right) \Gamma\left(\frac{d}{2}+l_2-1\right) \left(\ii u + \frac{l_1+l_2}{2}\right)} \sum_{k+n\leqslant  \min(l_1,l_2)} \frac{l_1! l_2! (x\cdot y\, \zeta\cdot\eta)^k}{k! n! (l_1 - k - n)! (l_2 - k - n)!}\\
\times \frac{\Gamma\left(\ii u + \frac{4-d+l_1+l_2}{2}\right)}{\Gamma\left(\ii u + \frac{4-d-l_1-l_2}{2} + k\right)} \frac{(\zeta\cdot y\, \eta\cdot x)^n}{\left(-\ii u + \frac{2-l_1-l_2}{2}\right)_n} (\zeta\cdot x)^{l_1-k-n} (\eta\cdot y)^{l_2-k-n}\\
= \frac{(-1)^{l_1+l_2} \Gamma\left(- \ii u + \frac{d-2-l_1-l_2}{2}\right) \Gamma\left(\ii u + \frac{4-d+l_1+l_2}{2}\right) \Gamma\left(\ii u + \frac{l_1 - l_2}{2}\right) \Gamma\left(\ii u + \frac{l_2 - l_1}{2}\right)}{\Gamma\left(\frac{d}{2}+l_1-1\right) \Gamma\left(\frac{d}{2}+l_2-1\right) \left(\ii u + \frac{l_1+l_2}{2}\right) \Gamma\left(\ii u + \frac{4-d-l_1-l_2}{2}\right)  \Gamma\left(\ii u - \frac{l_1+l_2}{2}\right) }\\
\times \left[\R_{l_1,l_2}(u)\zeta^{\otimes l_1}\otimes\eta^{\otimes l_2}\right]\cdot (x^{\otimes l_1}\otimes y^{\otimes l_2})\\
= \frac{\Gamma\left(-\ii u + \frac{d-2+l_1+l_2}{2}\right) \Gamma\left(\ii u + \frac{l_1 - l_2}{2}\right) \Gamma\left(\ii u + \frac{l_2 - l_1}{2}\right)}{\Gamma\left(\frac{d}{2}+l_1-1\right) \Gamma\left(\frac{d}{2}+l_2-1\right) \left(\ii u + \frac{l_1+l_2}{2}\right) \Gamma\left(\ii u - \frac{l_1+l_2}{2}\right) }
\,\left[\R_{l_1,l_2}(u)\zeta^{\otimes l_1}\otimes\eta^{\otimes l_2}\right]\cdot (x^{\otimes l_1}\otimes y^{\otimes l_2})\, .
\end{align*}

\section{Derivative identity}
\setcounter{equation}{0}
\label{app:O}

For $\zeta$ and $\eta$ two null vectors, it holds that
\begin{multline*}
(\zeta\cdot\nabla)^{l_1}(\eta\cdot\nabla)^{l_2} x^{2\left(\frac{l_1+l_2+2-d}{2}+\lambda\right)} = \frac{\left(\frac{4-l_1-l_2-d}{2}+\lambda\right)_{l_1+l_2}}{\left(\frac{4-l_1-l_2-d}{2}-\lambda\right)_{l_1+l_2}}(x^2)^{2\lambda}\\
\times \left[\R_{l_1,l_2}(-\ii\lambda)\zeta^{\otimes l_1}\otimes\eta^{\otimes l_2}\right] \cdot \nabla^{\otimes (l_1+l_2)} x^{2\left(\frac{l_1+l_2+2-d}{2}-\lambda\right)}\, .
\end{multline*}
In order to prove it one first needs to compute $y^{\otimes (l_1+l_2)}\cdot \left[\R_{l_1,l_2}(-\ii\lambda)\zeta^{\otimes l_1}\otimes\eta^{\otimes l_2}\right]$ for arbitrary $y$. We use equation \eqref{Rll general integral} to write (after having performed the integral over $z$ using the star-triangle relation)
\begin{multline}
y^{\otimes (l_1+l_2)}\cdot \left[\R_{l_1,l_2}(-\ii\lambda)\zeta^{\otimes l_1}\otimes\eta^{\otimes l_2}\right] = \frac{\Gamma\left(\frac{d}{2}-2\lambda\right) \Gamma\left(\frac{d+l_1+l_2}{2}-1+\lambda\right) \Gamma\left(1+\frac{l_1+l_2}{2}+\lambda\right)}{\Gamma(2\lambda) \Gamma\left(1+\frac{l_1+l_2}{2}-\lambda\right) \Gamma\left(\frac{d+l_1+l_2}{2} - 1 -\lambda\right)}\\
\quad\times y^{2\left(\frac{d+l_1+l_2}{2} - 1 -\lambda\right)}\int \frac{(\zeta\cdot(y-v))^{l_1} (\eta\cdot(y-v))^{l_2}}{v^{2\left(\frac{d}{2}-2\lambda\right)} (v-y)^{2\left(\frac{d+l_1+l_2}{2}+\lambda-1\right)}} \frac{\text{d}^d v}{\pi^{\frac{d}{2}}}\\
= \sum_{k}\, \frac{l_1! l_2!}{k! (l_1-k)! (l_2-k)!} \frac{(2\lambda)_k (y^2 \zeta\cdot\eta)^{k} (y\cdot\zeta)^{l_1-k} (y\cdot\eta)^{l_2-k}}{2^k \left(\lambda -\frac{l_1 + l_2}{2}\right)_k \left(\lambda + \frac{4 - l_1 - l_2 - d}{2}\right)_{k}}\, .
\end{multline}
Returning to the proof of \eqref{derivative identity}, we can write
\begin{multline*}
\left[\R_{l_1,l_2}(-\ii\lambda)\zeta^{\otimes l_1}\otimes\eta^{\otimes l_2}\right] \cdot \nabla^{\otimes (l_1+l_2)} x^{2\left(\frac{l_1+l_2+2-d}{2}-\lambda\right)} = \sum_{k}\, \frac{l_1! l_2!}{k! (l_1-k)! (l_2-k)!}(2 \zeta\cdot \eta)^k \\
\times \frac{(2\lambda)_k  \left(\lambda+\frac{d-2-l_1-l_2}{2}\right)_{k}}{\left(\lambda + \frac{4 - l_1 - l_2 - d}{2}\right)_{k}} (\zeta\cdot\nabla)^{l_1-k} (\eta\cdot\nabla)^{l_2-k} x^{2\left(\frac{l_1+l_2+2-d}{2}-k -\lambda\right)}\\
= \sum_{k,j}\, \frac{l_1! l_2! 2^{l_1+l_2-k-j}}{k! j! (l_1-k-j)! (l_2-k-j)!}(\zeta\cdot \eta)^{k+j} (\zeta\cdot x)^{l_1-k-j} (\eta\cdot x)^{l_2-k-j}\\
\times x^{2\left(\frac{2-l_1-l_2-d}{2}+k+j -\lambda\right)} \frac{(2\lambda)_k  (-1)^k}{\left(\lambda + \frac{4 - l_1 - l_2 - d}{2}\right)_{k}} \left(\frac{4-l_1-l_2-d}{2}+k+j-\lambda\right)_{l_1+l_2-k-j}\\
= \sum_{p}\, \frac{l_1! l_2! 2^{l_1+l_2-p}}{(l_1-p)! (l_2-p)!} (\zeta\cdot \eta)^{p} (\zeta\cdot x)^{l_1-p} (\eta\cdot x)^{l_2-p} x^{2\left(\frac{2-l_1-l_2-d}{2}+p -\lambda\right)}\\
\times \left(\frac{4-l_1-l_2-d}{2}+p-\lambda\right)_{l_1+l_2-p} \sum_{k=0}^p \frac{1}{k! (p-k)!} \frac{(2\lambda)_k  (-1)^k}{\left(\lambda + \frac{4 - l_1 - l_2 - d}{2}\right)_{k}}\\
= \left(\frac{4-l_1-l_2-d}{2}-\lambda\right)_{l_1+l_2}\sum_{p}\, \frac{l_1! l_2! 2^{l_1+l_2-p}}{p! (l_1-p)! (l_2-p)!}\,\frac{(\zeta\cdot \eta)^{p} (\zeta\cdot x)^{l_1-p} (\eta\cdot x)^{l_2-p}}{\left(\lambda + \frac{4 - l_1 - l_2 - d}{2}\right)_{p}}x^{2\left(\frac{2-l_1-l_2-d}{2}+p -\lambda\right)} \, .
\end{multline*}
On the other hand, one has
\begin{multline}\label{derivatives LHS}
(\zeta\cdot\nabla)^{l_1}(\eta\cdot\nabla)^{l_2} x^{2\left(\frac{l_1+l_2+2-d}{2}+\lambda\right)} = \sum_{p} \frac{l_1! l_2! 2^{l_1+l_2-p}}{p! (l_1-p)! (l_2-p)!} \left(\frac{4-l_1-l_2-d}{2}+p+\lambda\right)_{l_1+l_2-p} \\
\times (\zeta\cdot \eta)^{p} (\zeta\cdot x)^{l_1-p} (\eta\cdot x)^{l_2-p} x^{2\left(\frac{2-l_1-l_2-d}{2}+p +\lambda\right)}
\end{multline}
and since
\begin{equation}
\left(\frac{4-l_1-l_2-d}{2}+p+\lambda\right)_{l_1+l_2-p} = \frac{\left(\frac{4-l_1-l_2-d}{2}+\lambda\right)_{l_1+l_2}}{\left(\frac{4-l_1-l_2-d}{2}+\lambda\right)_{p}}
\end{equation}
equation \eqref{derivative identity} does hold.

\section{Spinor Basis}
\label{app:spin_basis}
The eigenvectors of the graph-building operator \eqref{graph_fish} for the square-lattice fishnet have been first constructed in $d=2,4$ for any number of sites $L$ in \cite{Derkachov2019,Derkachov:2019tzo}, according to the iterative formula \begin{equation}
\label{spinor_eigen}
\Psi(x_1,\dots,x_L) = \mathbf{\Lambda}_L(u_L,l_L)\mathbf{\Lambda}_L(u_2,l_2)\cdots \mathbf{\Lambda}_1(u_1,l_1)\,,
\end{equation}
where the \emph{layer operator} $\mathbf{\Lambda}_k(u,n)$ acts on $k-1$ coordinates $x_1,\dots,x_{k-1}$ and is defined by its integral kernel in $d=2r$ spacetime
 \begin{align}
\begin{aligned}
\label{Layer_inhom}
&\mathbf{\Lambda}_1(u,n)= \frac{U_n(x-x_0)}{(x-x_0)^{2\left(r-\frac{\tilde{\delta}}{2} +i u \right)}}\,,\\&
\mathbf \Lambda_k(u,n) =\mathbb{T}^{(n)}_{12}\left(i u\right)\mathbb{T}^{(n)}_{23}\left(i u\right)\cdots \mathbb{T}^{(n)}_{k-1k}\left(i u\right)\frac{U_n(x_k-x_0)}{(x_k-x_0)^{2\left(r-\frac{\tilde{\delta}}{2}+iu \right)}}\,,\end{aligned}
\end{align}
and the elementary building blocks in $d=2r$ dimensions are
\begin{align}
\begin{aligned}
&[\mathbb{T}^{(n)}_{ij}(w)] \Phi(x_i,x_j) = \int d y \,{T}^{(n)}_w (x_i,x_j|y)  \Phi(y,x_j) \\
&{T}^{(n)}_w (x_i,x_j|y) = \frac{U_n(x_1-y)U_n(y-x_2)^{\dagger}}{(x_i-x_j)^{2(r-\tilde{\delta})}(x_i-y)^{2\left(w +\frac{\tilde{\delta}}{2} \right)}(y-x_j)^{2\left(-w+\frac{\tilde{\delta}}{2}\right)}} \,.
\end{aligned}
\end{align}

The matrices $U_{n}(x)$ belong to the $n$-symmetric representation of the unitary groups $U(1)$ for $2d$ and $SU(2)$ for $4d$. For $n=1$ they defined respectively as
\begin{equation}
U^{(2)}(x) = \frac{x_1+ i x_2}{x_1-ix_2} = e^{i\phi}\,,\, U^{(4)}(x) = \frac{x_{\mu} \boldsymbol{\sigma}^{\mu}}{x^2}= \hat x_{\mu} \boldsymbol{\sigma}^{\mu} \,,
\end{equation}
where $\boldsymbol{\sigma}_k=i\sigma_k$ for $k=1,2,3$ and $\boldsymbol{\sigma}_4=\mathbbm{1}$, and the matrix $U_n(x)$ is the symmetrization of $n$-fold tensor products
\begin{equation}
U_n(x) =\text{Sym}\left[ U(x) \otimes \cdots \otimes U(x)\right]\,,
\end{equation}
namely
\begin{equation}
\label{defSU}
U^{(2)}_n(x) =e^{i\phi n} \,,\,U^{(4)}_n(x) =\frac{ x_{\mu_1}  \cdots x_{\mu_n} }{n!} \sum_{\pi} (-1)^{\sigma(\pi)}\boldsymbol{\sigma}^{\mu_{\pi(1)}} \otimes \cdots \otimes  \boldsymbol{\sigma}^{\mu_{\pi(n)}}\,.
\end{equation}
The definitions \eqref{defSU} can actually be extended to any even dimension $d=2r$, for a unitary matrix $U_n^{(2r)}(x)$ in the $n$-fold symmetric representation of the group $SU(2^{r-1})$
\begin{equation}
U^{(2r)}(x)  = \frac{x_{\mu} \boldsymbol{\Sigma}^{\mu}}{x^2}= \hat x_{\mu} \boldsymbol{\Sigma}^{\mu}\,,
\end{equation}
where the matrices $\boldsymbol{\Sigma}_{\mu}$ and ${\boldsymbol{\overline \Sigma}}_{\mu} = \boldsymbol{\Sigma}_{\mu}^{\dagger} =\boldsymbol{\Sigma}_{\mu}^{-1}$ realize the Weyl spinor representation of Clifford algebra in $2r$ dimensions
\begin{equation}
\Gamma_{\mu}^{(r)} = \begin{pmatrix}
0 &&\boldsymbol{\Sigma}^{(r)}_{\mu}\\
\boldsymbol{\overline \Sigma}^{(r)}_{\mu}&& 0
\end{pmatrix}\,,\,\,\,\, \{\Gamma_{\mu},\Gamma_{\nu}\}=2 \delta_{\mu\nu}\mathbbm{1}_{2^{r}}\,,
\end{equation}
that is
\begin{equation}
\boldsymbol{\Sigma}_{\mu} \boldsymbol{\overline \Sigma}_{\nu}+\boldsymbol{\Sigma}_{\nu} \boldsymbol{\overline \Sigma}_{\mu}=\boldsymbol{\overline \Sigma}_{\mu} \boldsymbol{ \Sigma}_{\nu}+\boldsymbol{\overline \Sigma}_{\nu} \boldsymbol{ \Sigma}_{\mu}= 2 \delta_{\mu\nu}\mathbbm{1}_{2^{r-1}}\,.
\end{equation}
The concrete definition of matrices $\boldsymbol{\Sigma}_{\mu}$  and $\boldsymbol{\overline \Sigma}_{\mu}$ can be done recursively starting from $r=2$, according to the recipe
\begin{align}
\begin{aligned}
&\Gamma_{\mu}^{(r)} = \begin{pmatrix}
0 && i \\
-i && 0
\end{pmatrix} \otimes \Gamma_{\mu}^{(r-1)}\,,
\,\,\, \mu=1,\dots,2r-2\,, \\ &\Gamma_{2r-1}^{(r)}  = \begin{pmatrix}
0 && i \\
-i && 0
\end{pmatrix} \otimes
 \begin{pmatrix}
1 && 0\\
0 && -1
\end{pmatrix} \otimes \mathbbm{1}_{2^{r-2}}\,,\\&\Gamma_{2r}^{(r)} =  \begin{pmatrix}
0 && 1\\
1 && 0
\end{pmatrix} \otimes \mathbbm{1}_{2^{r-1}}\,,\\
&\Gamma^{(2)}_k = \begin{pmatrix}
0 && \boldsymbol{\sigma}_k\\
\overline{\boldsymbol{\sigma}}_k && 0
\end{pmatrix} \,,\,\mu=1,2,3,4\,.\end{aligned}
\end{align}
It is possible to check that with such definition $\det\left(x_\mu \boldsymbol{\Sigma}^{\mu} \right) =\det\left(x_\mu \overline{\boldsymbol{\Sigma}}^{\mu} \right) = (x_{\mu}x^{\mu})^{2^{r-2}}$, and for a normalized vector $\hat x_{\mu}$ the matrices belong to the special unitary group.
The definitions of layer operators in $2d,\,4d$ provide a concrete realization of a symmetric and traceless tensor in the coordinates $x^{\mu}$ as it follows from their definition and the Fierz identity
\begin{equation}
\boldsymbol{\sigma}_{\mu} \otimes \boldsymbol{\sigma}^{\mu} = 2\mathbbm{1} - 2\mathbb{P}\,.
\end{equation}
For the general $d=2r$ situation, the same identity for the matrices $\boldsymbol{\Sigma}_{\mu}^{(r)}$ does not hold, and the layers need to be projected over specific subset of spinor components. To start with we pair each layer's $SU(2^{r-1})$ indices with generic complex vectors
\begin{equation}
\langle \alpha|\mathbf \Lambda_k(u,n)|\beta\rangle= (\alpha^{*})^a \mathbf\Lambda_k(u,n)^{\,b}_{a} \beta_{b} \,\,\,\,\,\,\,\,\text{i.e.}\,\,\,\,\,\,\,\,\, \langle \alpha|U(x)|\beta\rangle\,.
\end{equation}
For $k=1$ the condition of symmetric traceless tensor is mapped to the null vector condition
\begin{equation}
\label{pure_spinor}
\boldsymbol{\Sigma}_{\mu} \otimes \boldsymbol{\Sigma}^{\mu} |\beta\rangle \otimes |\beta\rangle = \underline 0 \,,
\end{equation}
which imposes a constraint on the components of $|\beta\rangle$. For $d=4,6$ there is no need of any such condition while for $d=2r\geq 8$ we need to impose $N(r)$ \emph{pure spinor} conditions, i.e. solve a quadratic system of $N(r)$ independent equations in the vector components.
For example, $N(4)=1$ and the constraint reads
\begin{equation}
{\beta_2\beta_5-\beta_1 \beta_6 +\beta_4 \beta_7}-\beta_8\beta_3=0\,,
\end{equation}
while $N(5)=5$ and the system of constraints read
\begin{align}
\begin{aligned}
\begin{cases}
   {\beta_1 \beta_6-\beta_4 \beta_7+\beta_3 \beta_8}-{\beta_5}\beta_2= 0\,,\\{\beta_6 \beta_9+\beta_8
   \beta_{11}-\beta_7 \beta_{12}}-{\beta_5}\beta_{10}=0\,,\\{\beta_3 \beta_{12}+\beta_6 \beta_{13}}-{\beta_5}\beta_{14}-\beta_4 \beta_{11}=0\,,\\{\beta_3 \beta_9-\beta_1 \beta_{11}+\beta_7 \beta_{13}}-{\beta_5}\beta_{15}= 0\,,\\ {\beta_4 \beta_9-\beta_1 \beta_{12}+\beta_8 \beta_{13}}-\beta_{16}{\beta_5}=0\,.
    \end{cases}
\end{aligned}
\end{align}
In general the $2^{r-1}$ components of such spin vectors are subject to $N(r)>0$ linearly-independent quadratic constraints.
Imposing the latter on spin vectors, the traceless condition is valid also for the length-$k$ layer. Indeed, the matrix structure of each layer is that of an $SU(2^{r-1})$ matrix, with respect to which the kernel of the matrix $\boldsymbol{\Sigma}_{\mu} \otimes \boldsymbol{\Sigma}^{\mu}$ is invariant
\begin{equation}
\boldsymbol{\Sigma}_{\mu} \otimes \boldsymbol{\Sigma}^{\mu}|\beta\rangle \otimes |\beta\rangle = \underline 0 \,\Longleftrightarrow \,( U\otimes U)( \boldsymbol{\Sigma}_{\mu} \otimes \boldsymbol{\Sigma}^{\mu}) |\beta\rangle \otimes |\beta\rangle = \underline 0 \,\Longleftrightarrow \,  (\boldsymbol{ \Sigma}_{\mu} \otimes \boldsymbol{ \Sigma}^{\mu}) (U\otimes U)|\beta\rangle \otimes |\beta\rangle = \underline 0\,.
\end{equation}
The proof that \eqref{spinor_eigen} is an eigenvector of the fishnet in $d=2r$ dimensions is based on a star-triangle identity, and leads to the same eigenvalue as \eqref{eigenvalueBD}. Indeed, the basis \eqref{spinor_eigen} and \eqref{eigenv_d} differ only by a very non-trivial rotation in the space of tensorial indices, respect to which the spectrum is degenerate, while for $l_j=0$ they coincide.
 \subsection{Star-triangle relation in $d=2r$}
 The scalar star-triangle identity in $d$-dimensions is well-known  \cite{DEramo:1971hnd,Symanzik:1972wj,Vasiliev:1982dc,Vasiliev:1981dg,Kazakov:1983ns,
 Kazakov:1984km,Vasil'ev:2004} and reads - in its amputated form or \emph{chain rule} - as
 \begin{align}
 \begin{aligned}
\int d^{2r}y \frac{1}{(x-y)^{2a}(x-y)^{2b}} = \pi^{r} \frac{\Gamma\left(r-a\right)\Gamma\left(r-b\right)\Gamma\left(a+b-r\right)}{\Gamma\left(a-n\right)\Gamma\left(b-n'\right)\Gamma\left(2r-a-b+n+n'\right) } \frac{1}{(x-y)^{2(a+b-r)}}\,.
 \end{aligned}
 \end{align}
 We can generalize it by adding an $SU(2^{r-1})$ angular part to the radial functions $x^2$, that is
 \begin{equation}
 \frac{1}{x^{2a}} \to  \frac{\langle \alpha| U_n^{(2r)}(x)|\beta \rangle}{x^{2a}} = \frac{\left( \langle \alpha| \boldsymbol{\Sigma}^{(r)}_{\mu} |\beta \rangle\, \hat{x}^{\mu} \right)^n}{x^{2a}}=\frac{\left( \langle \alpha| \boldsymbol{\Sigma}^{(r)}_{\mu} |\beta \rangle\, \p^{\mu} \right)^n}{\Gamma(a+\frac{n}{2})/\Gamma\left(a-\frac{n}{2}\right)(-2)^n} \frac{1}{x^{2\left(a-\frac{n}{2}\right)}}\,,
 \end{equation}
 and obtain, for $2r$ dimensions,
 \begin{align}
 \begin{aligned}
 \label{diff_str}
&\int d^{2r}y \frac{\langle \alpha| U_n(x-y)|\beta \rangle \langle \alpha' |U_{n'}(y-z)|\beta'\rangle^* }{(x-y)^{2a}(y-z)^{2b}} = \\&= \pi^{r} \frac{\Gamma\left(r-a+\frac{n}{2}\right)\Gamma\left(r-b+\frac{n'}{2}\right)\Gamma\left(a+b-r+\frac{n'-n}{2}\right)}{(-2)^{n}\Gamma\left(a+\frac{n}{2}\right) \Gamma\left(b+\frac{n'}{2}\right)\Gamma\left(2r-a-b+\frac{n+n'}{2}\right)} \left( \langle \alpha| \boldsymbol{\Sigma}^{(r)}_{\mu} |\beta \rangle\, \p^{\mu} \right)^n \frac{\langle \alpha'| U_{n'}(x-z) |\beta '\rangle^*}{(x-z)^{2\left(a+b-r-\frac{n}{2}\right)}}= \\&= \pi^{r}  \frac{\Gamma\left(r-a+\frac{n}{2}\right)\Gamma\left(r-b+\frac{n'}{2}\right)\Gamma\left(a+b-r+\frac{n-n'}{2}\right)}{(-2)^{n'}\Gamma\left(a+\frac{n}{2}\right) \Gamma\left(b+\frac{n'}{2}\right)\Gamma\left(2r-a-b+\frac{n+n'}{2}\right)}  \left( \langle \beta'| \boldsymbol{\bar \Sigma}^{(r)}_{\mu} |\alpha' \rangle\, \p^{\mu} \right)^{n'} \frac{\langle \alpha| U_{n}(x-z) |\beta \rangle}{(x-z)^{2\left(a+b-r-\frac{n'}{2}\right)}}\,.
 \end{aligned}
 \end{align}
The $n$ terms or $n'$ terms resulting from the derivation can eventually be organized in a mixing matrix for the spin vectors, and in the $r=2$ case  the latter coincides with a fused $SU(2)$ invariant solution of the Yang-Baxter equation. For the particular reduction $n'=0$ (or the analogous $n=0$) the formula \eqref{diff_str}
simplifies as
\begin{equation}
 \begin{aligned}
&\int d^{2r}y \frac{\langle \alpha| U_n(x-y)|\beta \rangle}{(x-y)^{2a}(y-z)^{2b}} = \pi^{r} \frac{\Gamma\left(r-a+\frac{n}{2}\right)\Gamma\left(r-b\right)\Gamma\left(a+b-r+\frac{n}{2}\right)}{\Gamma\left(a+\frac{n}{2}\right)\Gamma\left(b\right)\Gamma\left(2r-a-b+\frac{n}{2} \right) } \frac{\langle \alpha| U_n(x-z)|\beta \rangle}{(x-z)^{2(a+b-r)}}\,.
 \end{aligned}
\end{equation}
This kind of equation is what we need in order to prove to prove that for spin vectors subject to the constraints \eqref{pure_spinor} the functions
\begin{equation}
\langle \alpha_L|\mathbf{\Lambda}_L(u_L,l_L)|\beta_L\rangle\cdots  \langle \alpha_2|\mathbf{\Lambda}_2(u_2,l_2)|\beta_2\rangle \cdot \langle \alpha_1|\mathbf{\Lambda}_1(u_1,l_1)|\beta_1\rangle\,.
\end{equation}
diagonalize the fishnet graph-building operator. The proof is identical to the $d=4$ case treated in \cite{Derkachov:2020zvv}, as it relies only on the star-triangle identity \eqref{diff_str}. The main complication arising for general $r$ respect to the case $r=2$, is that the mixing of spinors is captured by a matrix that is not a solution of Yang-Baxter equation, and even contains explicitly a dependence over the coordinates. This fact can be checked already in $6d$, when the pure spinor condition is trivial - i.e. the spin vectors components are not subject to any constraint. The main consequence is that it is not manifest the symmetry of the eigenvectors respect to the permutation of excitations numbers $(u_k,l_k)$, and for this reason we prefer to use the basis of functions \eqref{eigenv_d} which has a much more involved structure of tensorial indices and a complicated behaviour when one or more coordinates get identified.
\bibliographystyle{nb}
\bibliography{BiblioDdim}
\end{document}